\documentclass[twoside,twocolumn,10pt]{article}
\usepackage{extsizes}
\usepackage[super,sort&compress,comma]{natbib} 
\usepackage[version=3]{mhchem}
\usepackage[left=1.5cm, right=1.5cm, top=1.785cm, bottom=2.0cm]{geometry}
\usepackage{mathptmx}
\usepackage{sectsty}
\usepackage{graphicx} 
\usepackage{lastpage}
\usepackage[format=plain,justification=justified,singlelinecheck=false,font={stretch=1.125,small},labelfont=bf,labelsep=space]{caption}
\usepackage{float}
\usepackage{fancyhdr}
\usepackage{fnpos}
\usepackage[english]{babel}
\addto{\captionsenglish}{%
  
}
\usepackage{array}
\usepackage{charter}
\usepackage[T1]{fontenc}
\usepackage[usenames,dvipsnames]{xcolor}
\usepackage{setspace}
\usepackage[compact]{titlesec}

\usepackage{xr-hyper} % To refer to figures in the SI
\usepackage{hyperref}
\usepackage{filecontents}
\usepackage[dvipsnames]{xcolor} % color text

\makeatletter
\newcommand*{\addFileDependency}[1]{% argument=file name and extension
  \typeout{(#1)}
  \@addtofilelist{#1}
  \IfFileExists{#1}{}{\typeout{No file #1.}}
}
\makeatother

\newcommand*{\myexternaldocument}[1]{%
    \externaldocument{#1}%
    \addFileDependency{#1.tex}%
    \addFileDependency{#1.aux}%
}

\myexternaldocument{mlbse_SI}

\usepackage{authblk}
\usepackage{etoolbox}
\usepackage{lmodern}

\makeatletter

\begin{document}
\title{\textbf{Machine Learning Dielectric Screening for the Simulation \\
of Excited State Properties of Molecules and Materials}}

\author[1,2,3]{Sijia S. Dong}
\author[1,2,*]{Marco Govoni}
\author[2,1,$\dagger$]{Giulia Galli}
\affil[1]{Materials Science Division and Center for Molecular Engineering, Argonne National Laboratory, Lemont, IL 60439, USA}
\affil[2]{Pritzker School of Molecular Engineering, the University of Chicago, Chicago, IL 60637, USA}
\affil[3]{Current Address: Department of Chemistry and Chemical Biology, Northeastern University, Boston, MA 02115, USA}

\affil[*]{mgovoni@anl.gov}
\affil[$\dagger$]{gagalli@uchicago.edu}
\date{}

\twocolumn[
\begin{@twocolumnfalse}
\maketitle
\normalsize
\begin{abstract}
Accurate and efficient calculations of absorption spectra of molecules
and materials are essential for the understanding
and rational design of broad classes of systems. Solving the Bethe-Salpeter
equation (BSE) for electron-hole pairs usually yields accurate predictions
of absorption spectra, but it is computationally expensive, especially
if thermal averages of spectra computed for multiple configurations
are required. We present a method based on machine learning to evaluate a key quantity  entering the definition of absorption spectra: the dielectric screening. We show that our approach yields a model for the screening that is transferable  between multiple configurations sampled during first principles molecular dynamics simulations; hence it leads to a substantial improvement in the efficiency of calculations of finite temperature spectra. We obtained computational gains of one
to two orders of magnitude for systems with 50 to 500 atoms, including
liquids, solids, nanostructures, and solid/liquid interfaces. Importantly,
the models of dielectric screening derived here may
be used not only in the solution of the BSE but also in developing functionals for
time-dependent density functional theory (TDDFT) calculations of
 homogeneous and heterogeneous systems. Overall, our work provides a strategy
to combine machine learning with electronic structure calculations
to accelerate first principles simulations of excited-state properties.
\end{abstract}
\end{@twocolumnfalse}
]

\section*{Introduction}

Characterization of materials often involves investigating their interaction with light. Optical absorption spectroscopy is one of the key experimental techniques for such characterization, and the simulation of optical absorption spectra is essential for interpreting experimental observations and predicting design rules for materials with desired properties. In recent years, absorption spectra of condensed systems have been successfully predicted by solving the Bethe-Salpeter equation (BSE)\cite{salpeter_relativistic_1951,hedin1965new,hanke1980many,onida1995ab,albrecht1997ab,albrecht1998ab,albrecht1998excitonic,benedict1998optical,rohlfing1998electron,rohlfing2000electron,blase2018bethe} in the framework of many-body perturbation theory (MBPT).\cite{strinati1988application,onida_electronic_2002,martin2016interacting,ping_electronic_2013,govoni_2018,golze2019gw} However, for large and complex systems, the use of MBPT is computationally demanding.\cite{govoni_large_2015,seo2016design,gaiduk2016photoelectron,scherpelz_2016,seo2017designing,mcavoy_phonon_2018,smart_fundamental_2018,gaiduk2018electron,gerosa2018role} It is thus desirable to develop methods that can improve the efficiency of optical spectra calculations, especially  if results at  finite temperature (T) are desired.

Simulation of absorption spectra at finite T can be achieved by performing, e.g., first principles molecular dynamics (FPMD)\cite{car1985unified} and by solving the BSE for uncorrelated snapshots extracted from FPMD trajectories. A spectrum can then be obtained by averaging over the results obtained for each snapshot.\cite{garbuio2006ab,lu2008dielectric,bernasconi2010statistical,nguyen_finite-field_2019}

Several schemes have been proposed in the literature to reduce the computational cost of solving the BSE,\cite{marsili2017large,elliott2019koopmans,henneke2020fast} including an algorithm that avoids the explicit calculation of virtual single particle electronic states, as well as the storage and inversion of large dielectric matrices.\cite{rocca_ab_2010,rocca_solution_2012} Recently, a so-called  finite-field (FF) approach\cite{ma_finite-field_2019,nguyen_finite-field_2019} has been proposed, where the calculation of dielectric matrices is bypassed; rather the key quantities to be evaluated are screened Coulomb integrals, which are obtained by solving the Kohn-Sham (KS) equations\cite{hohenberg1964inhomogeneous,kohn1965self} for the electrons in a finite electric field. The ability to describe dielectric screening through finite field calculations also led to the formulation of  GW\cite{ma_finite-field_2019,ma2020correction} and BSE\cite{nguyen_finite-field_2019} calculations beyond the random phase approximation (RPA), and of a quantum embedding approach\cite{ma_quantum_2020,ma2020first} scalable to large systems. 

From a computational standpoint, one important aspect of solving the Kohn-Sham equations in finite field is that the calculations can be  straightforwardly combined with the recursive bisection algorithm\cite{gygi_compact_2009} and thus,  by harnessing orbital localization, one may greatly reduce the number of screened Coulomb integrals that need to be evaluated. Importantly,  the workload to compute those integrals is of $O(N^4)$, irrespective of whether semilocal or hybrid functionals are used.\cite{nguyen_finite-field_2019} In spite of the improvement brought about by the FF algorithm and the use of the bisection algorithm, the solution of the BSE remains a demanding task. One of the quantities particularly challenging to  evaluate is the dielectric matrix of the system, that describes many-body screening effects between the interacting electrons. Intuitively we can understand a dielectric matrix as a complex filter that connects the bare (i.e., unscreened) Coulomb interaction between the electrons to an effective, screened Coulomb interaction. Such screened interaction is used in MBPT to approximately account for electronic correlation effects, when solving the Dyson equation (GW) and the BSE.  Here we turn to  machine learning (ML), in order to tackle the challenge of evaluating the dielectric matrix.

Specifically, for a chosen atomic configuration of a solid or a molecule, we use ML techniques to derive a mapping from the unscreened to the screened Coulomb interaction, thus deriving  a model of the dielectric screening. Once such a model is available, it can be re-used for multiple configurations sampled in a FPMD at finite temperature, without the need to recompute a complex dielectric matrix for each snapshot. Hence the use of a ML-derived model may greatly improve the efficiency of the calculation of finite T absorption spectra, provided the dielectric screening is weakly dependent on atomic configurations explored as a function of simulation time. We will show below that this assumption is indeed verified for several disordered systems, including liquid water and Si/water interfaces at ambient conditions and silicon clusters. Importantly, the use of ML-derived models leads to a reduction of 1 to 2 orders of magnitude in the computational workload required to obtain the dielectric screening for the simulation of optical absorption spectra at finite temperature. Another important advantage of  the ML-derived dielectric screening is that it provides insight into the approximate screening parameters used in the derivation of  hybrid functionals for time-dependent DFT (TDDFT) calculations, including dielectric-dependent hybrid (DDH) functionals. \cite{shimazaki2009first,skone2014self,gerosa2017accuracy,chen2018nonempirical,sun_low-cost_2020}

We emphasize that the strategy adopted here is different in spirit from strategies that use ML to infer structure-property relationships\cite{montavon2013machine,brockherde2017bypassing, welborn2018transferability,schleder_dft_2019,ryczko2019deep,noe_machine_2020,hase2020designing,sutton2020identifying,bogojeski2020quantum} or relationships between computational and experimental data\cite{stein_machine_2018}.
We do not seek to relate structural properties of a molecule or a solid to its absorption spectrum. Rather, either we consider a known microscopic structure of the system or we determine the structure by carrying out first principles MD (e.g., in the case of liquid water or a solid/liquid interface). Then, for a given atomistic configuration we use ML techniques to obtain the model between the unscreened and the screened Coulomb interaction, and we use such a model in the solution of the BSE for multiple configurations. 

 Hence the method proposed here is  conceptually different from the approaches previously adopted to predict the absorption spectra of molecules or materials using ML.\cite{gastegger2017machine,stein_machine_2018,Ye11612,ghosh_deep_2019,carbone2020machine,xue2020machine}.
For example, Ghosh et al.\cite{ghosh_deep_2019} predicted  molecular excitation spectra 
%(using Kohn-Sham eigenvalues computed at the Perdew–Burke–Ernzerhof (PBE)\cite{perdew_generalized_1996} level of theory, with  van der Waals corrections\cite{tkatchenko_accurate_2009}) 
from the knowledge of molecular structures at zero T, by using neural networks trained with a dataset of 132531 small organic molecules. %Stein et al.\cite{stein_machine_2018} identified a mapping between experimental red-green-blue images of  materials (captured by a scanner) and their experimental optical absorption spectra using variational auto-encoders. 
%All data (the images and the spectra) used by Stein et al. are from high-throughput experiments of 178994 metal oxide samples and naturally include finite temperature effects.
Carbone et al.\cite{carbone2020machine} mapped molecular structures to X-ray absorption spectra using message-passing neural networks, and a dataset of $\sim$134000 small organic molecules. Xue et al.\cite{xue2020machine} focused on two specific molecules and used a kernel ridge regression model trained with a minimum of several hundred molecular geometries and their corresponding  excitation energies and oscillator strengths computed at the TDDFT\cite{runge_density-functional_1984} level; they then used the results to predict the excitation energies and oscillator strengths of an ensemble of geometries and absorption spectra. 

All of these methods seek to relate structure to function (absorption spectra). The method presented here uses instead ML to replace a computationally expensive step in first principles simulations, and as we show below, leads to physically  interpretable results. The rest of the paper is organized as follows. In the next section, we briefly summarize our computational strategy. We then discuss homogeneous systems, including liquid water and periodic solids, followed by results  for heterogeneous and finite systems. We conclude by highlighting the innovation and key results of our work. 

\section*{Methods}

We first briefly summarize the technique used here to solve the BSE, including the use of bisection techniques to improve the efficiency of the method. We then describe the method based on ML to obtain the dielectric screening entering the BSE, including the description of the training set of integrals. These integrals are computed for a chosen configuration of a molecule or a solid.

Using the linearized Liouville equation\cite{walker_efficient_2006,rocca_ab_2010,rocca_solution_2012,nguyen_finite-field_2019} and the Tamm-Dancoff approximation\cite{hirata1999time}, the absorption spectrum of a solid or molecule  can be computed from DFT\cite{hohenberg1964inhomogeneous,kohn1965self} single particles eigenfunctions as: 
\begin{equation}
S(\omega) \propto \sum_{i=1}^3 \sum_{v=1}^{n_{occ}} \langle \psi_{v} | r_i | a_v^i(\omega) \rangle + c.c.
\label{eq:spectrum}
\end{equation}
where $\omega$ is the absorption energy, $r_i$ are the Cartesian components of the dipole operator, $n_{\text{occ}}$ is the total number of occupied orbitals, and $\mid\psi_{v}\rangle$ is the $v$-th occupied orbital of the unperturbed KS Hamiltonian, $\hat{H}^0$, corresponding to the eigenvalue $\epsilon_v$. The functions $\mid a_v^i \rangle$ are obtained from the solution of the following equation:\cite{rocca_ab_2010,rocca_solution_2012,nguyen_finite-field_2019}

\begin{equation}
\sum^{n_{\text{occ}}}_{v^\prime=1} (\omega\delta_{vv^\prime}-D_{vv^\prime}-\mathcal{K}^{1e}_{vv^\prime}+\mathcal{K}^{1d}_{vv^\prime}) \mid a^i_{v^\prime}\rangle = \hat{P}_{c} \hat{r}_{i} \mid \psi_{v} \rangle
\label{eq:linear_eq}
\end{equation}
where
\begin{equation}
D_{vv^\prime}\mid a^i_{v^\prime}\rangle = \hat{P}_{c} (\hat{H}^0-\epsilon_v)\delta_{vv^\prime} \mid a^i_{v^\prime}\rangle,
\label{eq:Da}
\end{equation}
\begin{equation}
\mathcal{K}_{vv^\prime}^{1e}\mid a_{v^\prime}^{i}\rangle=2\hat{P}_{c}\left(\int d\mathbf{r^{\prime}}V_c(\mathbf{r},\mathbf{r^{\prime}})\psi_{v^\prime}^{*}(\mathbf{r^{\prime}})a_{v^\prime}^{i}(\mathbf{r^{\prime}})\right)\psi_{v}(\mathbf{r}),
\label{eq:K1e_a}
\end{equation}
\begin{equation}
\mathcal{K}_{vv^\prime}^{1d}\mid a_{v^\prime}^{i}\rangle=\hat{P}_{c}\tau_{vv^\prime}(\mathbf{r})a^i_{v^\prime}(\mathbf{r}),
\label{eq:K1d_a}
\end{equation}
$\hat{P}_c=1-\sum^{n_\text{occ}}_{v=1}\mid \psi_{v}\rangle\langle \psi_{v}\mid$ is the projector on the unoccupied manifold, and $V_c=\frac{e^2}{\mid\mathbf{r}-\mathbf{r^{\prime}}\mid}$ is the unscreened Coulomb potential. Following the derivation reported by Nguyen et al.,\cite{nguyen_finite-field_2019} we defined screened Coulomb integrals, $\tau_{vv^\prime}$, entering Eq.~\ref{eq:K1d_a}, as: 
\begin{eqnarray}
\tau_{vv^\prime}(\mathbf{r}) &=& \int W(\mathbf{r},\mathbf{r^{\prime}})\psi_{v}(\mathbf{r^{\prime}})\psi_{v^\prime}^{*}(\mathbf{r^{\prime}})d\mathbf{r}'\label{eq:tau0}\\
&=&\tau^u_{vv^\prime}(\mathbf{r})+\Delta\tau_{vv^\prime}(\mathbf{r}),\label{eq:tau}
\end{eqnarray}
where the screened Coulomb interaction $W$ is given by
$W=\epsilon^{-1} V_c$,  and $\epsilon^{-1}$ is the inverse of the dielectric matrix (dielectric screening). Analogously, unscreened Coulomb integrals, $\tau^u_{vv^\prime}$, are defined as: 
\begin{equation}
\tau^u_{vv^\prime}(\mathbf{r})=\int V_c(\mathbf{r,r^{\prime}})\psi_{v}(\mathbf{r^{\prime}})\psi^*_{v^\prime}(\mathbf{r^{\prime}})d\mathbf{r^{\prime}}. \label{eq:tauu}
\end{equation}
By carrying out finite field calculations\cite{ma_finite-field_2019, ma2020correction, nguyen_finite-field_2019}, one can obtain screened Coulomb integrals without an explicit evaluation of the dielectric matrix (Eq.~\ref{eq:tau0}), but rather by adding to the unscreened Coulomb integrals the second term on the right hand side of Eq.~\ref{eq:tau}, which is computed as: 
\begin{equation}
    \Delta\tau_{vv^\prime}(\mathbf{r})=\int V_c(\mathbf{r,r^{\prime}})\frac{\rho^+_{vv^\prime}(\mathbf{r^{\prime}})-\rho^-_{vv^\prime}(\mathbf{r^{\prime}})}{2}d\mathbf{r^{\prime}}.
    \label{eq:delta_tau_delta_rho}
\end{equation}
The densities $\rho_{vv^\prime}^{\pm}$ are obtained by solving the KS equations with the perturbed Hamiltonian $\hat{H}\pm\tau^u_{vv^\prime}$; both indexes $v$ and $v^\prime$ run over  all occupied orbitals. While all potential terms of $\hat{H}$ may be computed self-consistently\cite{nguyen_finite-field_2019}, in this work the exchange-correlation potential was evaluated for the initial unperturbed electronic density and kept fixed during the self-consistent iterations. This amounts to evaluating the dielectric screening within the RPA. The FF-BSE approach has been implemented by coupling the WEST\cite{govoni_large_2015} and Qbox\cite{gygi_architecture_2008} codes in client-server mode.\cite{ma_finite-field_2019,nguyen_finite-field_2019,Govoni2021} 

The maximum number of integrals, $n_{\text{int}}=n_{\text{occ}}(n_{\text{occ}}+1)/2$, is determined by the total number of pairs of occupied orbitals. The actual number of integrals to be evaluated can be greatly reduced by using the recursive bisection method,\cite{gygi_compact_2009} which allows one to localize orbitals and consider only integrals generated by pairs of overlapping orbitals\cite{nguyen_finite-field_2019}. The systems studied in this work contain tens to hundreds of atoms, with hundreds to thousands of electrons. For example, for one of the Si/water interfaces discussed below,  we considered a slab with 420 atoms, 1176 electrons and each single particle state is doubly occupied. Hence, $n_{\text{occ}}$=588, and $n_{\text{int}}=173166$. Using the recursive bisection method  the total number of $vv^\prime$ pairs is reduced to $n_{\text{int}}=5574$ (a reduction factor slightly larger than 30) without compromising accuracy, when a bisection threshold of 0.05 and five bisection levels in each Cartesian direction are adopted\cite{gygi_compact_2009}.  

We note that the Liouville formalism used in this work (Eq.~\ref{eq:spectrum}) only involves summations over occupied states. Such formalism was shown to yield absorption spectra equivalent to solving the BSE with explicit and converged summations over empty states.\cite{rocca_ab_2010,rocca_solution_2012,nguyen_finite-field_2019} The same formalism may also be used to describe absorption spectra within TDDFT\cite{runge_density-functional_1984}, albeit employing a different definition of the $\mathcal{K}^{1e}$ and $\mathcal{K}^{1d}$ terms. \cite{hutter2003,walker_efficient_2006,rocca2008turbo,malciouglu2011turbotddft,ping_electronic_2013,ge2014turbotddft,nguyen_finite-field_2019} 

The key point of our work is the use of ML to generate a model for the calculation of screened Coulomb integrals (Eq.~\ref{eq:tau}) that is transferable to multiple atomic configurations; the goal is to reduce the computational cost in the solution of Eq.~\ref{eq:spectrum}. In particular, we consider the mapping between unscreened Coulomb integrals, $\tau^u_{vv\prime}$, and screened Coulomb integrals, $\Delta \tau_{vv\prime}$. Such transformation is mapping $n_\text{int}$ pairs of a 3D array, i.e., $\{ F: \tau^u_{vv^\prime}\to\Delta\tau_{vv^\prime},\,\forall v,v^\prime\in[1,\cdots,n_\text{occ}]\}$ and is similar to 3D image processing. Our objective is to learn the mapping functions and  hence it is natural here to use convolutional neural networks (CNN), a widely used technique in image classification. CNNs are artificial neural networks with spatial-invariant features. The screened and unscreened Coulomb integrals are related by the dielectric matrix, which describes a linear response function of the system to an external perturbation. Therefore, the mapping we aim to obtain should follow a linear relationship for physical reasons, and one convolutional layer without nonlinear activation functions should be considered. Here, the surrogate model $F$, used to bypass the explicit calculation of Eq.~\ref{eq:delta_tau_delta_rho}, is represented by a single convolutional layer $K$:
\begin{equation}
    \Delta\tau_{vv^\prime}(x,y,z) = (K * \tau^u_{vv^\prime} )(x,y,z)
    \label{eq:CNN}
\end{equation}
where $K$ is the convolutional filter of size $(n_x,n_y,n_z)$ (see the Electronic Supplementary Information (ESI) for details).

The filter, $K$, is determined through an optimization procedure that utilizes $n_\text{int}$ pairs of $\tau^u_{vv^\prime}$ and $\Delta\tau_{vv^\prime}$ as the dataset, obtained for one configuration (i.e., one set of atomic positions) using Eq.~\ref{eq:tauu} and Eq.~\ref{eq:delta_tau_delta_rho}, respectively. Therefore this filter captures features in the dielectric screening that are translationally invariant. When the filter size is reduced to $(1,1,1)$, the training procedure is effectively a linear regression and Eq.~\ref{eq:CNN} amounts to applying a global scaling factor to $\tau^u_{vv^\prime}$, which we label $f^{\text{ML}}$. 

In our calculations, the mapping $F$ corresponds to evaluating the dielectric screening arising from the short-wavelength part (i.e., the body) of the dielectric matrix. The long-wavelength part (i.e., the head of the dielectric matrix) corresponds to the macroscopic dielectric constant $\epsilon_\infty$. The definitions of the head and body of the dielectric matrix are given in Eq.~\ref{eq:dtau_G} of the ESI.

\begin{figure}[htbp]
\centering
\includegraphics[width=\linewidth]{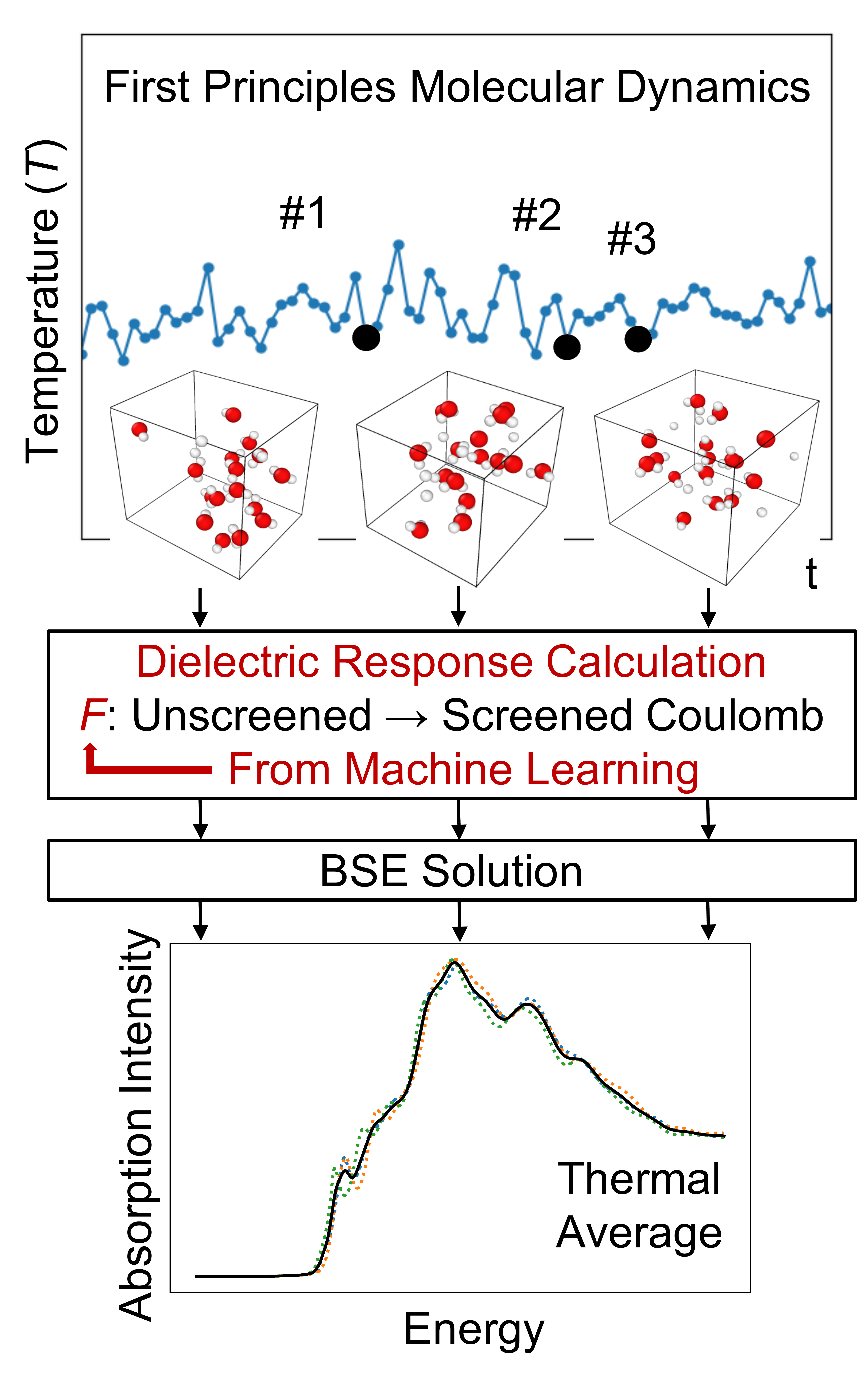}
\caption{Illustration of the strategy to predict absorption spectra at finite temperature based on the solution of the Bethe-Salpeter equation (BSE) and machine learning techniques. $F$ is the mapping obtained by machine learning.}
\label{fig:workflow}
\end{figure}

One of the main advantages of a ML-based model for the screening is that it may be reused for multiple configurations sampled during a FPMD simulation, thus avoiding the calculations of dielectric matrices for each snapshot, as illustrated in Figure \ref{fig:workflow}.The validity of such an approach and its robustness are discussed below for several systems.  In our calculations, we carried out FPMD with the Qbox\cite{gygi_architecture_2008} code and MBPT theory calculations with the WEST\cite{govoni_large_2015} code, coupled in client server mode with Qbox in order to evaluate the screened integrals (Eq.s~\ref{eq:tau}-\ref{eq:delta_tau_delta_rho}), which constitute our training dataset. We implemented an interface between Tensorflow\cite{tensorflow2015-whitepaper} and WEST, including a periodic padding of the data for the convolution in Eq.~\ref{eq:CNN}, in order to satisfy periodic boundary conditions. The computational details of each system investigated here are reported in the ESI.

\section*{Results}

We now turn to present our results for several systems, starting from liquid water.

\subsection*{Liquids}

To establish baseline results with small computational cost, we first considered a water supercell containing 16 water molecules. We tested the accuracy of a single convolutional layer with different filter sizes, from $(1,1,1)$ to $(20,20,20)$. We find that a convolutional model (Eq.~\ref{eq:CNN}) can be used to bypass the calculation of $\Delta\tau$ in Eq.~\ref{eq:delta_tau_delta_rho}, yielding absorption spectra in good agreement with the FF-BSE method. In particular, we find that a filter size of $(1,1,1)$, i.e., a global scaling factor, is sufficient to accurately yield the positions of the lower-energy peaks of the absorption spectra, with an error of only -0.03 eV (see the ESI for a detailed quantification of the error).

We then turned to interpret the meaning of the global scaling factor $f^{\text{ML}}$, and we computed the quantity $\epsilon_f^{\text{ML}}=(1+f^{\text{ML}})^{-1}$. For 20 independent snapshots extracted from a FPMD trajectory of the 16-H$_2$O system, we find that $\epsilon_{f}^{\text{ML}}$ ($1.84\pm0.02$ is the same, within statistical error bars, as that of the PBE\cite{perdew_generalized_1996} macroscopic static dielectric constant computed using the polarizability tensor (as implemented in the Qbox code\cite{gygi_architecture_2008}): $\epsilon_\infty^{\text{PT}}=1.83\pm0.01$. 
Therefore, the global scaling factor that we learned is closely related to the long-wavelength dielectric constant of the system.
Interestingly, we obtained similar scaling factors for a simulation using a larger cell, with 64-molecules, e.g., $\epsilon_f^{\text{ML}}=1.83$ for a given, selected snapshot, for which $\epsilon_\infty^{\text{PT}}=1.86$. To further interpret the factor $f^{\text{ML}}$ obtained by ML,  we computed the average of $\Delta\tau_{vv^\prime}/\tau_{vv^\prime}^{u}$ over all $vv^\prime$. Specifically, we define $f^{\text{Avg}}=\frac{1}{\Omega}\int f^{\text{Avg}}(\mathbf{r}) d\mathbf{r}$, where $f^{\text{Avg}}(\mathbf{r})=\frac{1}{N_{vv^\prime}}\sum_{v,v^\prime} \Delta\tau_{vv^\prime}(\mathbf{r})/\tau_{vv^\prime}^{u}(\mathbf{r})$, $\Omega$ is the volume of the simulation cell, and $N_{vv^\prime}$ is the total number of ${vv^\prime}$ in the summation. Using one snapshot of the 16-H$_2$O system as an example, we find that $\epsilon_{f}^{\text{Avg}}=(1+f^{\text{Avg}})^{-1}=1.79$, similar to $\epsilon_{f}^{\text{ML}}=1.86$ for the same snapshot. 

To evaluate how sensitive the peak positions in the absorption spectra of water are to the value of the global scaling factor, we varied $\epsilon_f$ from 1.67 to 1.92. We find that the position of the lowest-energy peak varies approximately in a linear fashion,  from 8.69 eV to 8.76 eV. This analysis shows that a global scaling factor is sufficient to represent the average effect of the body (i.e., short-wavelength part) of the dielectric matrix and that this factor is approximately equal to the head of the matrix (related to the long-wavelength dielectric constant). Hence, our results show that a diagonal dielectric matrix is a sufficiently good approximation to represent the screening of liquid water and to obtain its optical spectrum by solving the BSE. This simple finding is in fact an important result, leading to a substantial reduction in the computational time necessary to obtain the absorption spectrum of water at the BSE level of theory.

\begin{figure}[h]
\centering
\includegraphics[width=\linewidth]{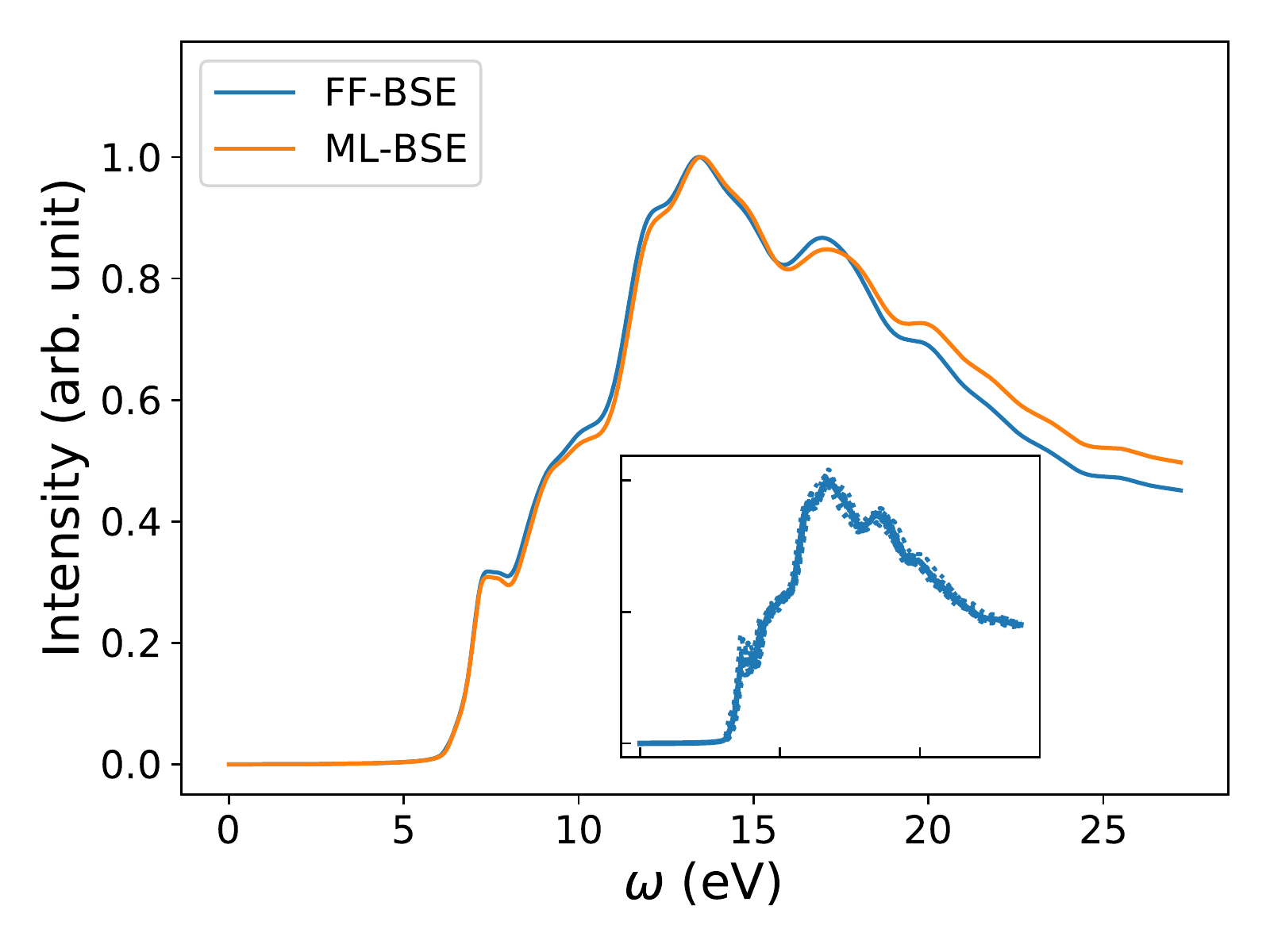}
\caption{Averaged spectra of liquid water obtained by solving the Bethe-Salpeter equation (BSE) in finite field (FF) and using machine learning techniques (ML). Results have been averaged over 10 snapshots obtained from first principles simulations at 400K, using supercells with 64 water molecules. The variability of the FF-BSE spectra within the 10 snapshots is shown in the inset. See also Figure~\ref{fig:wat64_ave_ind} of the ESI for the same variability when using ML-BSE.}
\label{fig:wat64_ave}
\end{figure}

In order to understand how the screening varies over a FPMD trajectory, we applied the global scaling factor $f^{\text{ML}}$ obtained from one snapshot of the 16-H$_2$O system to 10 different snapshots of a  64-H$_2$O system,\cite{dawson_equilibration_2018} at the same T, 400~K, and we computed an average spectrum. As shown in Figure \ref{fig:wat64_ave}, we can accurately reproduce the average spectrum computed with FF-BSE. The RMSE between the two spectra is 0.027. These results show that the global scaling factor is transferable from the 16 to the 64 water cell and that the dependence of the global scaling factor on the 
atomic positions may be neglected, for the thermodynamic conditions considered here. While it was recognized that the dielectric constant of water is weakly dependent on the cell size, it was not known that the average effect of the body of the dielectric matrix is also weakly dependent on the cell size. In addition, our results show that the dielectric screening can be considered independent from atomic  positions for water at ambient conditions. This property of the dielectric screening was not previously recognized; it is not only an important recognition from a physical standpoint, but also from an efficiency standpoint, to improve the efficiency of BSE calculations.

The timing acceleration of ML-BSE compared to FF-BSE is a function of the size of the system (characterized by the number of screened integrals $n_{\text{int}}$ and the number of plane waves (PWs) $n_{\text{pw}}$). We denote by $t_d$ the total number of core hours required to compute the net screening $\Delta\tau$ for all pairs of orbitals. We do not include in $t_d$ the training time, which usually takes only several minutes on one GPU for the systems studied here. Since we perform the training procedure once, we consider the training time to be negligible. We define the acceleration to compute the net effect of the screening as $\alpha_d=t^{\text{FF-BSE}}_d/t^{\text{ML-BSE}}_d$, and we  
find that $\alpha_d$ increases as $n_{\text{int}}$ and $n_{\text{pw}}$ increase. See the ESI for details.

For the 64-H$_2$O system discussed above, we used a bisection threshold equal to 0.05, and a  bisection level of 2 for each of the Cartesian direction. This reduces $n_{\text{int}}$ from $256(256+1)/2=32896$ to 3303. In this case, the gain achieved with our machine learning technique is close to two orders of magnitude: $\alpha_d=87$.

\subsection*{Solids}

\begin{figure}[h]
\centering
\includegraphics[width=\linewidth]{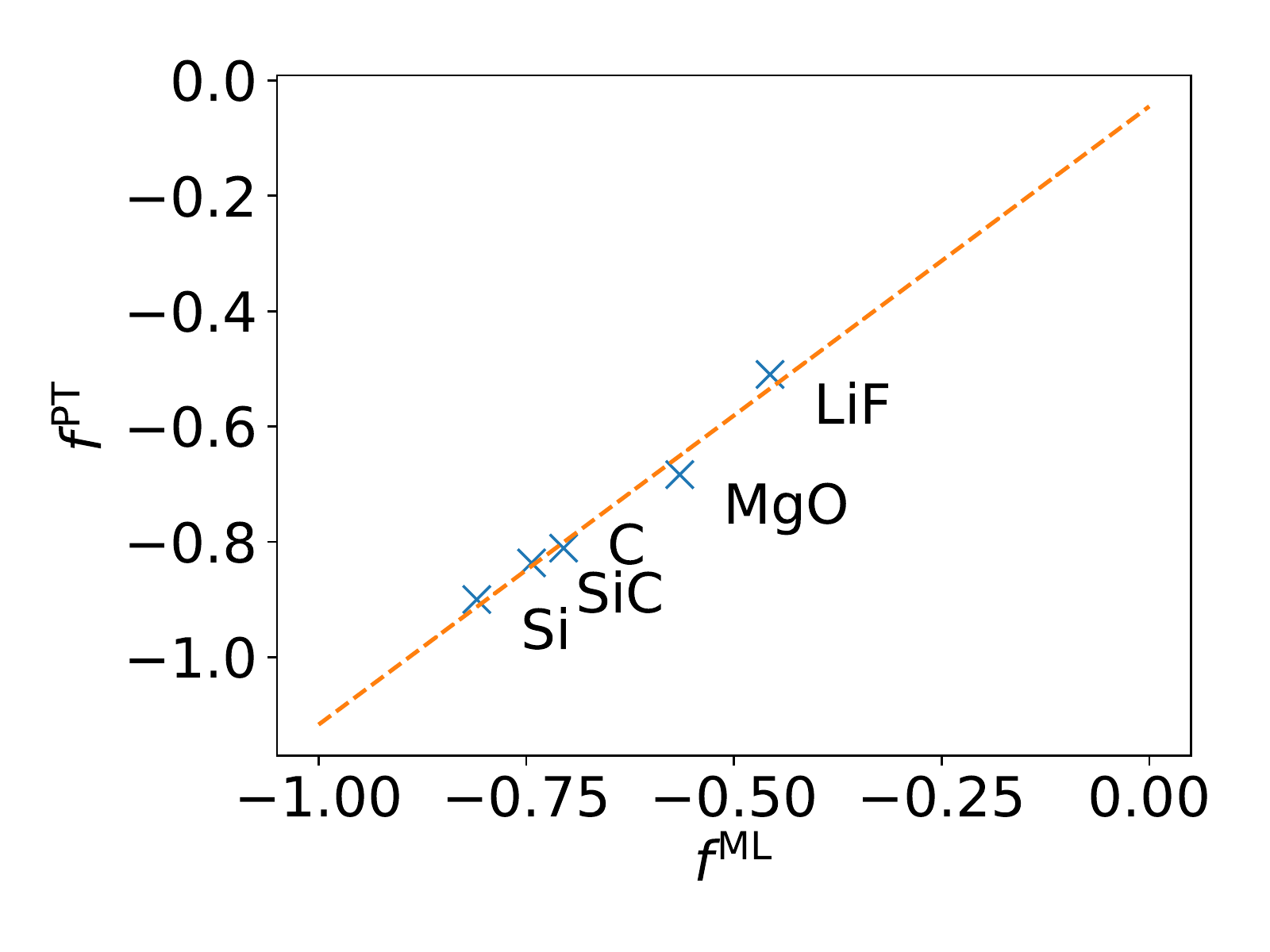}
\caption{Relationship between the scaling factor obtained by machine learning ($f^{\text{ML}}$) and that obtained by computing the dielectric constant at the same level of theory ($f^{\text{PT}}$) (see text).}
\label{fig:solids_epsilon}
\end{figure}

We now turn to discussing the  accuracy of ML-BSE for several solids, including LiF, MgO, Si, SiC, and C (diamond), for which we found again remarkable efficiency gains, ranging from 13 to 43 times for supercells with 64 atoms.  In all cases, we used the experimental lattice constants.\cite{haas2009calculation} Similar to water, we found that a convolutional model (Eq.~\ref{eq:CNN}) can reproduce the absorption spectra of solids at the FF-BSE level, and that global scaling factors, either from linear regression or from averaging $\Delta\tau/\tau^u$ yield similar accuracy (Figures~\ref{fig:si_sf_cnn},\ref{fig:lif_sf_cnn} of the ESI). As shown in Figure~\ref{fig:solids_epsilon}, where we have defined $f^{\text{PT}}=(\epsilon_\infty^{\text{PT}})^{-1}-1$, we found that $f^{\text{ML}}$ is again numerically close to $f^{\text{PT}}$, for $\epsilon_\infty^{\text{PT}}$ computed using the polarizability tensor,\cite{gygi_architecture_2008} and the same level of theory and $k$-point sampling.  
These results show that, for ordered solids, the average effect of the body (short-wavelength part) of the dielectric matrix, $\epsilon_f^{\text{ML}}$, is similar to that of the head (long-wavelength limit) of the matrix and hence a diagonal screening is sufficient to describe the absorption  spectra, similar to the case of water. This is an interesting result that supports the validity of the approximation chosen to derive the DDH functional.\cite{shimazaki2009first,marques2011density,refaely2013gap,skone2014self,skone2016nonempirical,brawand2016generalization,brawand2017performance,pham2017electronic,gerosa2017accuracy}

\begin{figure}[h]
    \centering
    
    \includegraphics[width=\linewidth]{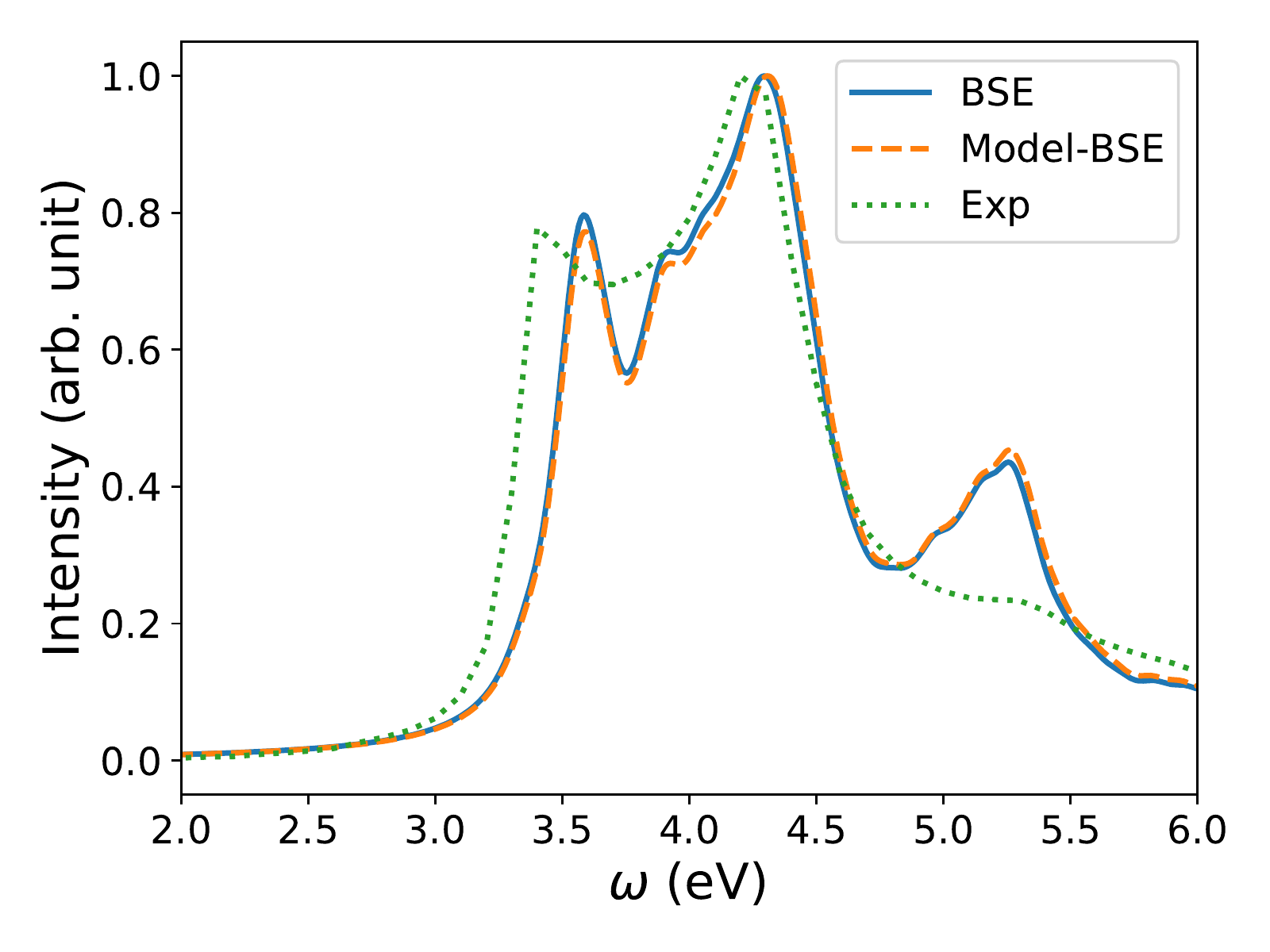}

    \caption{Absorption spectrum of crystalline Si computed by solving the Bethe-Salpeter equation (BSE) starting from PBE\cite{perdew_generalized_1996} wavefunctions, using a 2-atom cell and $12\times12\times12$ $k$-point sampling (blue line). The orange dashed line (Model-BSE) shows the same spectrum computed using a diagonal dielectric matrix with diagonal elements equal to $\epsilon_\infty=12.21$ (see text). Experimental results\cite{aspnes1983dielectric} are shown by the green dotted line.}
    \label{fig:si_12_yambo}
\end{figure}

We note that the FF-BSE algorithm uses the $\Gamma$ point and is efficient and appropriate for large systems. In order to verify that a diagonal dielectric matrix is an accurate approximation also when using unit cells and fine grids of $k$-points, we computed the absorption spectrum of Si with a 2-atom cell and a $12\times12\times12$ $k$-point grid, using the Yambo\cite{marini2009yambo,sangalli2019many} code. We then compared the results with those obtained using a diagonal approximation of the dielectric matrix, and elements derived from the long-wavelength dielectric constant computed with the same cell and $k$-point grid. Fig.~\ref{fig:si_12_yambo} shows that we found an excellent agreement between the two calculations, of the same quality as that obtained for water in the previous section.  

It is important to note that the method presented here to learn the filter between unscreened and screened integrals represents a way of obtaining a model dielectric function with ML techniques, and without the need of using ad hoc empirical parameters. Several model dielectric functions have been proposed to speed-up the solution of the BSE for solids over the years.\cite{penn1962wave,levine_new_1982,hybertsen_model_1988,baroni_ab_1986,cappellini_model_1993,djurisic_dielectric_2001,bokdam_role_2016,sun_low-cost_2020} Recently, Sun et al.\cite{sun_low-cost_2020} proposed a simplified BSE method that utilizes a model dielectric function (m-BSE). The authors used the model of Cappellini et al.\cite{cappellini_model_1993} with an empirical parameter, which they determined by averaging the values minimizing the RMSE between a model dielectric function and that obtained within the RPA for Si, Ge, GaAs, and ZnSe.\cite{walter1970wave} This simplified BSE method yields good agreement with the results of the full BSE solution. For example, in the case of LiF, the shift between the first peak obtained with m-BSE and BSE is 0.12 eV, to be compared to the shift of 0.04 eV found here, between ML-BSE and FF-BSE. A model dielectric function has been proposed also for 2D semiconductors\cite{trolle2017model} and silicon nanoparticles \cite{wang1994dielectric,tsu1997simple}. However, the important difference between our work and the models just described is that the latter requires empirical parameterization. One of the advantages of the ML approach adopted here is that it does not require the definition of empirical parameters and, importantly, it may also be applied to nanostructures and heterogeneous systems, such as solid/liquid interfaces, as discussed next.

\subsection*{Interfaces}
 
We have shown that for solids and liquids, the use of ML leads to the definition of a global scaling factor that, when utilized to model the screened Coulomb interaction, yields results for absorption spectra in very good agreement with those of the full FF-BSE calculations, at a much lower computational cost. We now discuss solid/liquid interfaces as prototypical heterogeneous systems.

We considered two silicon/water interfaces modeled by periodically repeated slabs. One is the H-Si/water interface, a hydrophobic interface with 420 atoms (72 Si atoms and 108 water molecules; Si surface capped by 24 H atoms); the other is a COOH-Si/water interface, a hydrophilic interface with 492 atoms (72 Si atoms and 108 water molecules; Si surface capped by 24 -COOH groups).\cite{pham_interfacial_2014} Not unexpectedly, we found that neither a global scaling factor nor a convolutional model is sufficiently accurate to reproduce the spectra obtained with FF-BSE, as shown in Figure~\ref{fig:sihwat_sf_cnn7} of the ESI. Therefore, we have developed a position-dependent ML model to describe the variation of the dielectric properties in the Si, water and interfacial regions. We divided the grid of $\tau_{vv^\prime}$ into slices, each spanning one $xy$ plane parallel to the interface; we then trained for a model on each slice. In this way we describe translationally invariant features along the $x$ and $y$ directions, and we obtain a $z$-dependent convolutional filter $K(z)$ or $z$-dependent scaling factors $f^{\text{ML}}(z)$.
We found that a position-dependent filter, $K(z)$, or a scaling factor for each slice, $f^{\text{ML}}(z)$, yield a comparable accuracy, and therefore we focus on the $f^{\text{ML}}(z)$ model, which is simpler.

\begin{figure*}
 \centering
 \includegraphics[width=17.1cm]{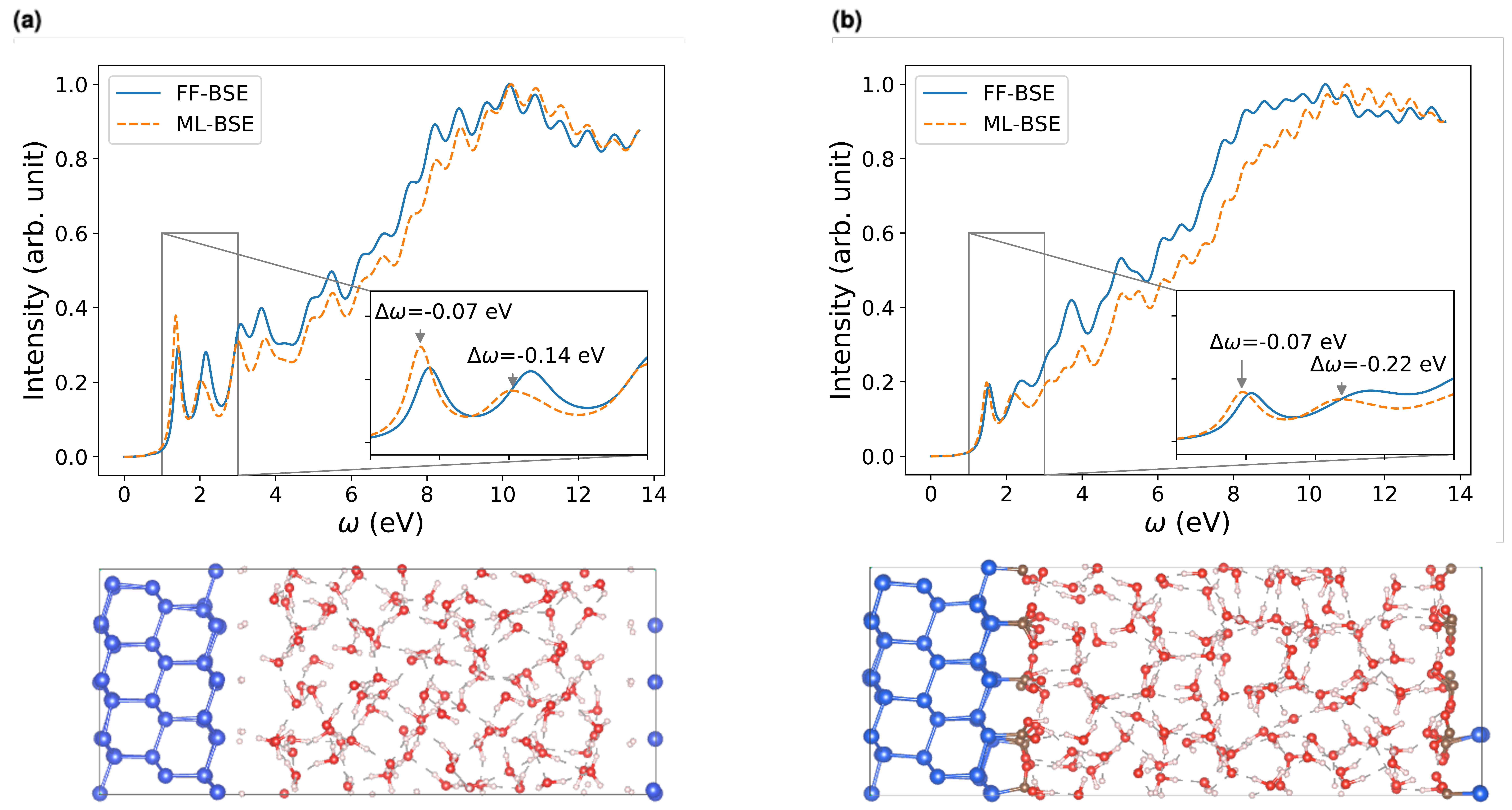}
 \caption{Comparison of absorption spectra obtained by solving the Bethe-Salpeter equation (BSE) in finite field (FF) and using machine learning (ML) techniques for (a) a H-Si/water interface shown in the lower left panel and (b) a COOH-Si/water interface (shown in the lower right panel). Blue, red and white spheres represent Si, oxygen and hydrogen respectively. C is represented by brown spheres. (See the ESI for results from using a kinetic energy cutoff of 60 Ry for wavefunctions.)}
 \label{fig:siwat_2interfaces_2par}
\end{figure*}

We found that the $z$-dependent ML model $f^{\text{ML}}(z)$ is accurate to represent the screening of the Si/water interfaces when computing absorption spectra (Figure~\ref{fig:siwat_2interfaces_2par}). Together with Figure~\ref{fig:sihwat_sf_cnn7} in the ESI, our finding show that a block diagonal dielectric matrix, where all the diagonal elements in the dielectric matrix have the same value, is not a good representation of the screening, unlike the case of water and ordered, periodic solids; instead taking into account the body of the dielectric matrix as in the $f^{\text{ML}}(z)$ model is critical in the case of an interface.

Depending on how the grid of $\tau_{vv^\prime}$ are divided, we obtain different $f^{\text{ML}}(z)$ profiles for Si/water interfaces. Figure~\ref{fig:siwat_2interfaces_2par} shows the spectra in the case of $f^{\text{ML}}(z)$ defined by two parameters (a constant value in the Si region, and a  different constant value in the water region); we name this profile $f^{\text{ML}}_{p2}(z)$. In Figure~\ref{fig:sihwat_sf_slice}(a) of the ESI, we present the spectra obtained using $f^{\text{ML}}(z)$ in the case of 108 slices evenly spaced in the $z$ direction, which we call $f^{\text{ML}}_{p108}(z)$. The function $\epsilon^{\text{ML}}_f(z)$ corresponding to $f^{\text{ML}}_{p108}(z)$ presents maxima at the interfaces, and  minima at the points furthest away from the interface, in the Si  and the water regions (Figure~\ref{fig:sihwat_sf_slice}(b) of the ESI). 

In order to interpret our findings, we express $\Delta\tau$ in terms of projective dielectric eigenpotentials, (PDEP)\cite{wilson2008efficient,wilson2009iterative} and we decompose $f^{\text{Avg}}(\mathbf{r})$ into contributions from each individual PDEP,\cite{zheng2019dielectric} i.e., $f^{\text{Avg}}=\sum_i f^{\text{Avg}}_i$, where 
\begin{equation}
    f^{\text{Avg}}_{i}(\mathbf{r})=\frac{1}{N_{v,v^\prime}}\sum_{v,v^\prime}\frac{\phi_{i}(\mathbf{r})(\lambda_{i}/(1-\lambda_{i}))\int\phi_{i}^{*}(\mathbf{r^{\prime\prime}})\tau^u_{vv^\prime}(\mathbf{r^{\prime\prime}})d\mathbf{r^{\prime\prime}}}{\tau^u_{vv^\prime}(\mathbf{r})}
    \label{eq:f_pdep_i}
\end{equation}
and $\phi_i$ is the $i$-th eigenpotential of the static dielectric matrix corresponding to the eigenvalue $\lambda_i$. We find that the largest contribution to $f^{\text{Avg}}(\mathbf{r})$ comes from the eigenvectors corresponding to the most negative PDEP eigenvalue. This PDEP component has its maximum near the interfaces, with the square modulus of the corresponding PDEP eigenpotential being localized at the interfaces (Figure~\ref{fig:sihwat_charge_f} of the ESI). This shows that the maximum of $\epsilon^{\text{ML}}_{f}(z)$ at the interfaces stem from the contribution of the PDEP eigenpotential with the most negative eigenvalue. 

Interestingly, $f^{\text{ML}}_{p2}(z)$ and $f^{\text{ML}}_{p108}(z)$ yield absorption spectra of similar quality. This suggests that the absorption spectrum is not sensitive to the details of the profile at the interface, at least in the case of the H-Si/water interface (Figure~\ref{fig:siwat_2interfaces_2par}(a) and Figure~\ref{fig:sihwat_sf_slice} of the ESI) and the COOH-Si/water interface (Figure~\ref{fig:siwat_2interfaces_2par}(b) and Figure~\ref{fig:sicoohwat_sf_slice} of the ESI) studied here. However, knowing the functional form of $f^{\text{ML}}_{p108}(z)$ is useful to determine the location of the interfaces, and it can be used to define where the discontinuities in $f^{\text{ML}}_{p2}(z)$ are located. 

We further developed a 3D grid model, $f^{\text{ML}}(\mathbf{r})$. This is a simple extension of the $z$-dependent model, where instead of slicing $\tau_{vv^\prime}$ in only one direction, we equally divided $\tau_{vv^\prime}$ into sub-domains in all three Cartesian directions. We tested cubic sub-domains of side lengths from 0.6 \AA~ to 2.6 \AA, and we found that the accuracy of the resulting spectrum is similar to that obtained with the $z$-dependent model, as shown in Figure~\ref{fig:sihwat_3D} of the ESI. 

In order to verify  the transferability of the position-dependent model derived for one snapshot extracted from FPMD to other snapshots, we computed absorption spectra by using the same $f^{\text{ML}}(z)$ for different snapshots generated at ambient conditions and we found that the screening is weakly dependent on the atomic positions, at these conditions, similar to the case of water discussed above (Figure~\ref{fig:sihwat_5_20} of the ESI).

In summary, by obtaining $\epsilon^{\text{ML}}_f(z)$ from machine learning, we have provided a way to define a position-dependent dielectric function for heterogeneous systems. For the Si/water interfaces, the acceleration to compute the net screening effect is $\alpha_d=86$ for H-Si/water if bisection techniques are used ($n_\text{int}=5574$), and $\alpha_d=224$ for COOH-Si/water, again if bisection techniques are used ($n_\text{int}=8919$).

\subsection*{Nanoparticles}

\begin{figure}[htbp]
 \centering
 \includegraphics[width=\linewidth]{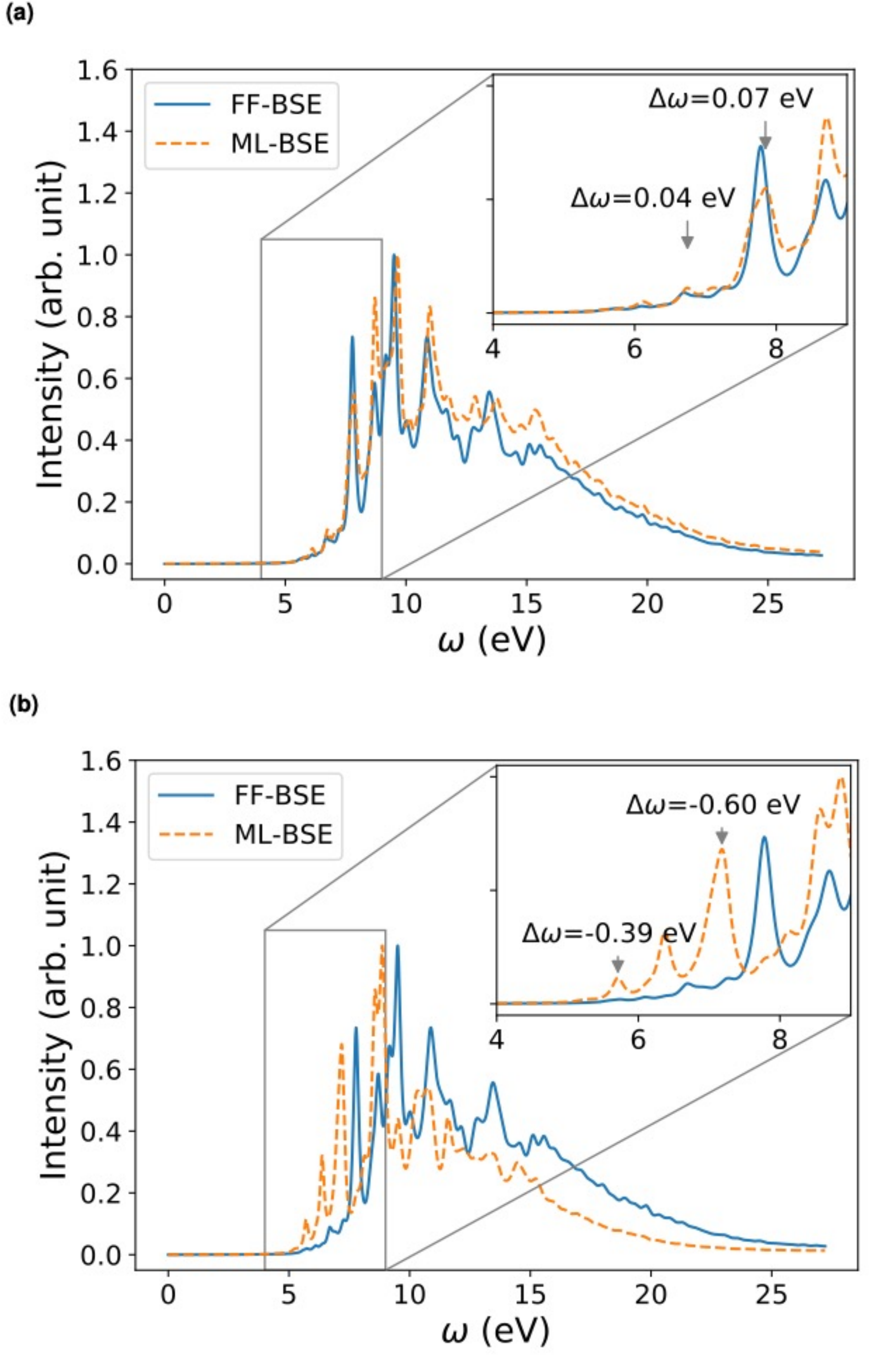}
 \caption{Comparison of absorption spectra of Si$_{10}$H$_{16}$ (40 \AA~cell) obtained by solving the Bethe-Salpeter equation (BSE) in finite field (FF) and using machine learning (ML) techniques for (a) convolutional layer with filter size $(7,7,7)$ from a cell of 30 \AA, and (b) a global scaling factor. The RMSE value between the FF-BSE and ML-BSE spectra is 0.067 for (a) and 0.141 for (b), respectively. The accuracy of using a convolutional layer with filter size $(7,7,7)$ from the 40 \AA~cell itself is similar to that of (a): RMSE=0.067.}
 \label{fig:si10h16_40ang_ml}
\end{figure}

As our last example we consider nanoparticles, i.e., 0D systems. We focus on silicon clusters Si$_{35}$H$_{36}$ and Si$_{87}$H$_{76}$\cite{govoni2012carrier,govoni_large_2015,brawand2016generalization} but we start from a small cluster Si$_{10}$H$_{16}$ first, to test the methodology. As shown in Figure~\ref{fig:si10h16_40ang_ml}(b), we found that a global scaling factor is not an appropriate approximation of the screening, e.g., for the spectrum of Si$_{10}$H$_{16}$ computed using PW basis set in a simulation cell with a large vacuum (cell length over 25 \AA). This finding points at an important qualitative difference with respect to the case of solids and liquids (condensed systems). Interestingly, we found that convolutional models are instead robust to different sizes of vacuum, and give absorption spectra in good agreement with FF-BSE calculations (Figure~\ref{fig:si10h16_40ang_ml}(a)). The inaccuracy of a global scaling factor stems from two reasons.  One is related to the fact that when the volume of the vacuum surrounding the cluster becomes large, the data of the training set is dominated by small matrix elements representing the vacuum region. Because the numerical noise is not translationally invariant, the use of Eq.~\ref{eq:CNN} overcomes this issue, as the noise from vacuum matrix elements is canceled out in the convolution process. We note that the presence of nonzero elements in the vacuum region is due to the choice of the PW basis set, which requires periodic boundary conditions. In the case of isolated clusters, the use of periodic boundary conditions could be avoided by choosing localized basis set. However, there are several systems of interest where using PW basis set is preferable and vacuum regions are present, such as nanoparticles deposited on surfaces. The second reason responsible for the inaccuracy of a global scaling factor, even if the noise arising from vacuum is eliminated, (see Figure~\ref{fig:si10h16_40ang_ml_rhocut100000} of the ESI) is that the  mapping between $\tau^u$ and $\Delta\tau$ being is simply more complex in nanoparticles than in homogeneous systems. Such a complexity can be accounted for when using Eq.~\ref{eq:CNN}.

\begin{figure}[htbp]
\centering
\includegraphics[width=\linewidth]{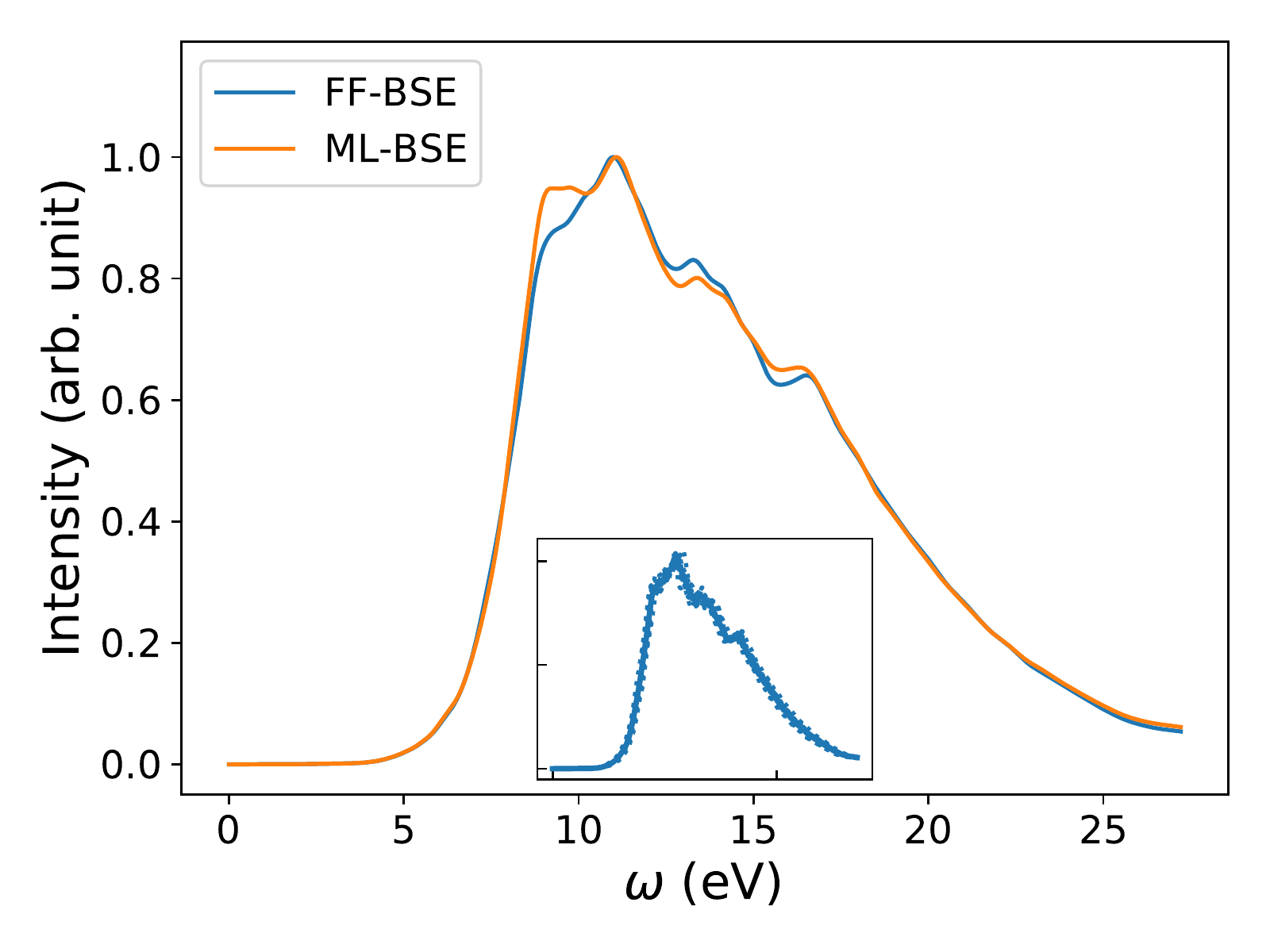}
\caption{Average spectra of Si$_{35}$H$_{36}$ obtained by solving the Bethe-Salpeter equation (BSE) in finite field (FF) and using machine learning techniques (ML). Results have been averaged over 10 snapshots obtained from first principles simulations at 500~K. The variability of the FF-BSE spectra within the 10 snapshots is shown in the inset. See also Figure~\ref{fig:si35h36_ave_ind} of the ESI for the same variability when using ML-BSE.}
\label{fig:si35h36_ave}
\end{figure}

In order to investigate the dependence of  the screening of nanoparticles on temperature, we transferedthe ML model trained for one specific snapshot of the Si$_{35}$H$_{36}$ cluster, to different snapshots extracted from a FPMD simulation, in order to predict absorption spectra at finite temperature. We applied the convolutional model with filter size $(7,7,7)$ obtained from the 0~K Si$_{35}$H$_{36}$ cluster to 10 snapshots of Si$_{35}$H$_{36}$ from an FPMD trajectory equilibrated at 500~K. As shown in Figure~\ref{fig:si35h36_ave}, the average ML-BSE spectrum can accurately reproduce the FF-BSE absorption spectrum at 500~K, with a small peak position shift of 0.08 eV. The ML-BSE spectra of individual snapshots  is also in good agreement with the corresponding spectra computed with  FF-BSE, shown in Figure~\ref{fig:si35h36_10} of the ESI. These results show that for nanoclusters, as for water, the screening is weakly dependent on atomic positions over a 500~K FPMD trajectory; note however that the 0~K spectrum (Figure~\ref{fig:si35h36_25ang_0K_ml} of the ESI) has different spectral features than the one collected at 500~K (Figure~\ref{fig:si35h36_ave}). 

\begin{figure}[htbp]
\centering
\includegraphics[width=\linewidth]{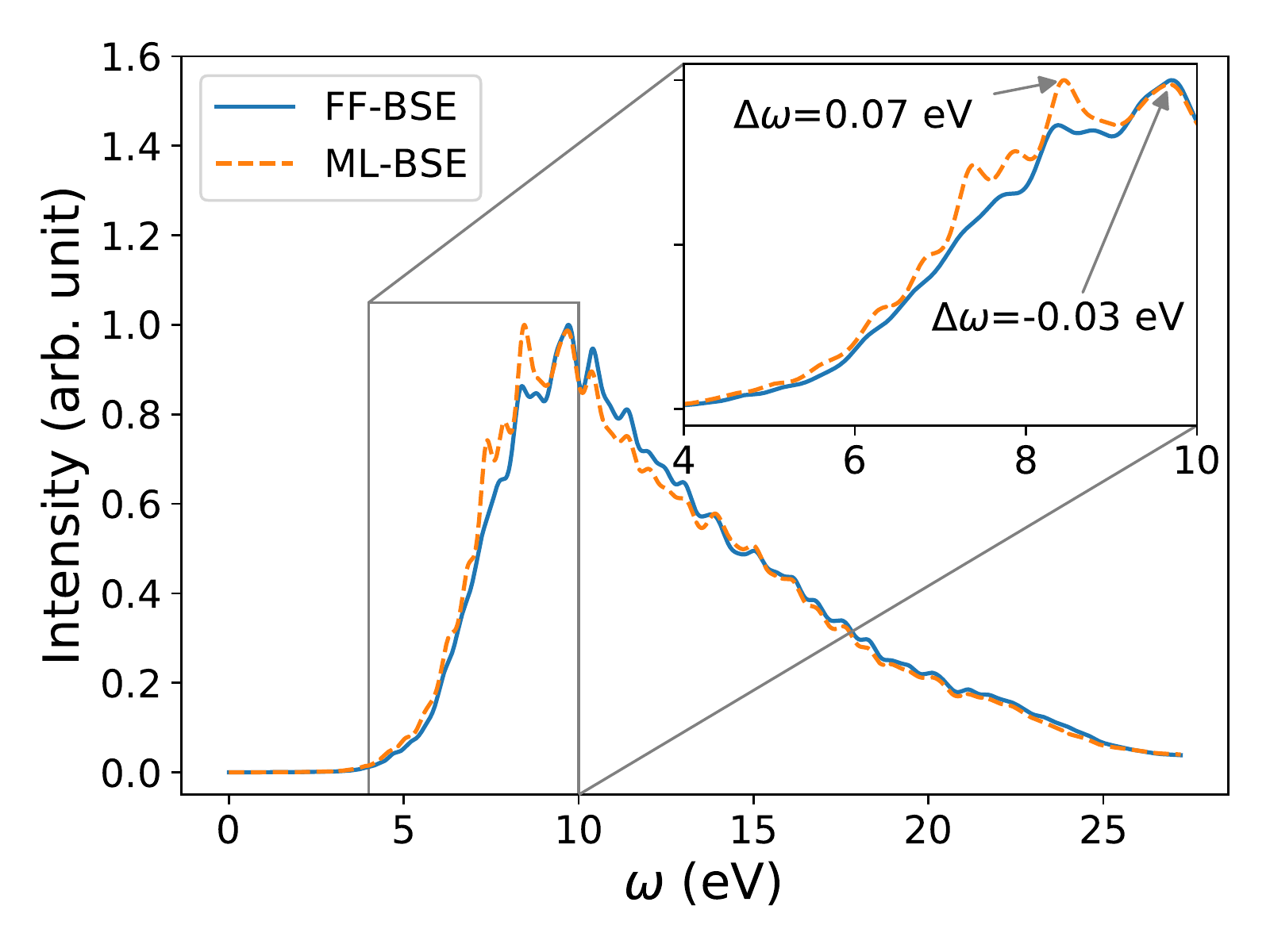}
\caption{Accuracy of the Si$_{87}$H$_{76}$ spectrum obtained from ML-BSE by applying a convolutional model with filter size $(7,7,7)$, trained from Si$_{35}$H$_{36}$. The RMSE value between the FF-BSE and ML-BSE spectra is 0.033.}
\label{fig:si87h76_si35model}
\end{figure}

We also found that the convolutional model trained for Si$_{35}$H$_{36}$ can be applied to Si$_{87}$H$_{76}$ with an error within 0.07 eV for peak positions (Figure~\ref{fig:si87h76_si35model}). The accuracy is comparable to the convolutional model from Si$_{87}$H$_{76}$ itself, as shown in Figure~\ref{fig:si87h76_ml} of the ESI. This shows that the convolutional model captures the nonlocality of the dielectric screening common to Si clusters of different sizes and is transferable from a smaller to a larger nanocluster (Si$_{87}$H$_{76}$) within the size range considered here. The FF-BSE calculation of Si$_{87}$H$_{76}$ is about 6 times more expensive in terms of core hours than that of Si$_{35}$H$_{36}$; hence, being able to circumvent the FF-BSE calculation of Si$_{87}$H$_{76}$ by using the model $K$ computed for Si$_{35}$H$_{36}$ is certainly an advantage.  

Conceptually, the convolutional model yields  filters that capture the translational invariant features of the dataset, and in our case they capture the nonlocality of the screening. In other words, the convolutional filters represent features in the mapping from $\tau^u_{vv^\prime}$ to $\Delta\tau_{vv^\prime}$ that are invariant across the simulation cell. For Si clusters, we found that the RMSE values between ML-BSE and FF-BSE spectra converges as the size of the filter increases. For example, for Si$_{35}$H$_{36}$, convergence is achieved at the filter size $(7,7,7)$, which corresponds to a cube with side length (2.24 \AA), corresponding approximately to the Si-Si bond length in the cluster (2.35 \AA). This result suggests that the screening of the Si cluster has features of the length of a nearest-neighbor bond that are translationally invariant.

The timing acceleration $\alpha_d$ for calculations of the absorption spectra of the Si$_{35}$H$_{36}$ cluster in a cubic cell of 20, 25, or 30 \AA~ in length, is 24, 47, or 90 times, respectively, when using bisection techniques (threshold 0.03, 4 levels in each Cartesian direction), as shown in Figure~\ref{fig:si10_si35_timing} of the ESI. In the case of Si$_{87}$H$_{76}$ cluster, $\alpha_d \simeq 160$.

\section*{Conclusions}

We presented a method based on machine learning (ML) to determine a key quantity entering many body perturbation theory calculations, the dielectric screening; this quantity determines the strength of the electron-hole interaction entering the BSE. In our ML model, the screening is viewed as a convolutional (linear) filter that transforms the unscreened into the screened Coulomb interaction. Our results show that such a model can be obtained for a chosen atomic configuration and then re-used to represent the screening of multiple configurations sampled in a FPMD at finite temperature for several systems, including water, solid/water interfaces, and silicon clusters.

 In particular, we found that in the case of homogeneous systems, e.g. liquid water and several insulating and semiconducting solids, absorption spectra can be accurately predicted by using a diagonal dielectric matrix. When using such a diagonal form, we found excellent agreement with spectra computed by the full solution of the BSE in finite field. In addition, our results showed that for liquid water the same diagonal approximation can be used to accurately compute spectra for different configurations from FPMD at ambient conditions, thus easily obtaining a thermal average representing a finite temperature spectrum. 

In the case of nanostructures and heterogeneous systems, such as solid/liquid interfaces, we found that the use of diagonal matrices or  block-diagonal dielectric matrices to describe the two portions of the system (Si and water, in the example chosen here) does not yield accurate spectra; through machine learning of the screening we could define simple models yielding accurate absorption spectra and a simple way of computing thermal averages. For nanostructures, it is necessary to use a convolutional model to properly represent the nonlocality of the dielectric screening. Similar to water and the Si/water interfaces, we found that the function describing the screening for hydrogenated Si-clusters of about 1~nm does not depend in any substantial way on the atomic coordinates of the snapshots sampled during our FPMD simulations, up to the maximum temperature tested here, 500~K. 

The time savings in the calculations of the screening using ML are remarkable, ranging from a factor of 13 to 87 for the solids and liquids studied here, with cells varying from 64 to 192 atoms. For the clusters and the interface, we obtained time savings ranging from 30 to 224 times, with cells varying from 26 to 492 atoms.

Finally, we note that the ML-based procedure presented here, in addition to substantially speeding up the calculation of spectra, especially at finite T, represents a general approach to derive model dielectric functions, which are key quantities in electronic structure calculations, utilized not only in the solution of the BSE. For example, our approach provides a strategy to develop dielectric-dependent hybrid functionals (DDH)\cite{skone2014self,skone2016nonempirical} for TDDFT calculations, as well as an interpretation of the parameters entering model dielectric functions.\cite{penn1962wave,levine_new_1982,hybertsen_model_1988,cappellini_model_1993,tsu1997simple,wang1994dielectric,bokdam_role_2016,sun_low-cost_2020} In particular, for homogeneous systems, our findings points at TDDFT with DDH functionals as an accurate method to obtain absorption spectra, consistent with the results of Sun et al.\cite{sun_low-cost_2020}, which were however derived semi-empirically. Work is in progress to further develop a strategy to develop parameters entering hybrid DFT functionals using machine learning.\cite{dick2020machine}

\section*{Conflicts of interest}
There are no conflicts to declare.

\section*{Acknowledgements}
The authors thank Bethany Lusch, He Ma, Misha Salim, and Huihuo Zheng for helpful discussions. The work was supported by Advanced Materials for Energy-Water Systems (AMEWS) Center, an Energy Frontier Research Center funded by the U.S. Department of Energy, Office of Science, Basic Energy Sciences (DOE-BES), and Midwest Integrated Center for Computational Materials (MICCoM) as part of the Computational Materials Science Program funded by DOE-BES. This research used resources of the Argonne Leadership Computing Facility, which is a DOE Office of Science User Facility supported under Contract DE-AC02-06CH11357, and resources of the University of Chicago Research Computing Center (RCC). The GM4 cluster at RCC is supported by the National Science Foundation’s Division of Materials Research under the Major Research Instrumentation (MRI) program award no. 1828629.

\bibliography{mlbse} 
\bibliographystyle{rsc} 
\makeatletter\@input{suppaux.tex}\makeatother
\end{document}

\maketitle

\thispagestyle{empty}

\section{Computational details}

In the following sections we report the computational details used in this work. 

\subsection{Systems}

We considered the systems reported in Table~\ref{tab:systems}. The electronic structure of each system was computed at the density functional theory (DFT) level of theory using plane wave basis sets and the ONCV pseudopotentials,\cite{schlipf_optimization_2015} with the Perdew–Burke–Ernzerhof (PBE)\cite{perdew_generalized_1996} exchange and correlation functional. Quantum Espresso\cite{giannozzi_quantum_2009} (version 6.1.0) and the Qbox\cite{gygi_architecture_2008} (version 1.66.2) codes were used. For each system the macroscopic dielectric constant, $\epsilon_\infty$, was calculated by averaging the diagonal elements of the polarizability tensor computed using the Qbox code, and was labeled $\epsilon_\infty^{\text{PT}}$. The electronic dipole used to compute the polarizability tensor is defined from using the center of charge of maximally localized Wannier functions (MLWF) with the refinement correction by Stengel and Spaldin.\cite{stengel2006accurate}

\begin{table}
\centering
\caption{\label{tab:systems} Systems considered in this work.}
\begin{tabular}{|l|l|l|l|}
\hline 
System & Number of atoms & Size of the cell (\AA) & $\epsilon_{\infty}^{\text{PT}}$\tabularnewline
\hline 
\hline 
16-H$_{2}$O & 48 & $a=7.82$ & 1.83$\pm$0.01\tabularnewline
\hline 
64-H$_{2}$O & 192 & $a=12.41$ & 1.87$\pm$0.004\tabularnewline
\hline 
Si & 64 & $a=5.43$ & 10.01\tabularnewline
\hline 
SiC & 64 & $a=4.36$ & 6.12\tabularnewline
\hline 
C & 64 & $a=3.57$ & 5.29\tabularnewline
\hline 
MgO & 64 & $a=4.21$ & 3.16\tabularnewline
\hline 
LiF & 64 & $a=4.03$ & 2.04\tabularnewline
\hline 
Si$_{10}$H$_{16}$ & 26 & $a=20.00$ to $50.00$ & -\tabularnewline
\hline 
Si$_{35}$H$_{36}$ & 71 & $a=25.00$ & -\tabularnewline
\hline 
Si$_{87}$H$_{76}$ & 163 & $a=25.00$ & -\tabularnewline
\hline 
H-Si/water & 420 & $a=11.62,b=13.42,c=33.43$ & 3.34\tabularnewline
\hline 
COOH-Si/water & 492 & $a=11.62,b=13.42,c=35.73$ & 3.84\tabularnewline
\hline 
\end{tabular}
\end{table}

\subsection{First principles molecular dynamics simulations}
First principles molecular dynamics (FPMD) simulations were carried out using the Qbox\cite{gygi_architecture_2008} code (version 1.66.2).

For liquid water, we considered unit cells with 16 or 64 water molecules. For 64-water samples, we considered snapshots from each of 10 independent FPMD trajectories (samples s0022-s0031) of the PBE400 dataset.\cite{dawson_equilibration_2018} For 16-water samples, we generated 20 independent MD trajectories starting from 20 independent snapshots of 16 water molecules initiated randomly with the same atomic density as the PBE400 dataset (1.11 g/cm$^3$ D$_2$O). The initiation method is the same as the method used to generate the PBE400 dataset.\cite{dawson_equilibration_2018} FPMD simulations were carried out using the Bussi-Donadio-Parrinello (BDP) thermostat\cite{bussi_canonical_2007} at 400 K with a thermostat time constant of 10000 a.u. and a time step of 10 a.u. ($\sim$0.24 fs). 

We modeled a hydrophobic Si/water interface with a slab containing 72 Si atoms, 24 H atoms terminating Si, and 108 water molecules. We modeled a hydrophilic Si/water interface with a slab containing 72 Si atoms, 24 -COOH groups terminating Si, and 108 water molecules.
The geometrical configurations were taken from Pham et al.\cite{pham_interfacial_2014} 

The FPMD simulations of the Si$_{35}$H$_{36}$ cluster were carried out using the BDP thermostat at 500 K. The thermostat time constant was 5000 a.u., and the time step was 20 a.u. ($\sim$0.48 fs). Equilibration was reached within the first 20000 time steps ($\sim$9.7 ps). The finite temperature absorption spectrum was obtained by averaging the spectra obtained for ten snapshots extracted every 5000 time steps from the FPMD trajectory, and  starting after 10.16 ps.

\subsection{Calculations of absorption spectra}
Absorption spectra calculations were carried out with the WEST\cite{govoni_large_2015}-Qbox\cite{gygi_architecture_2008} coupled codes using the FF-BSE scheme reported by Nguyen et al.\cite{nguyen_finite-field_2019}. 

For exited-state energies, a scissor operator was applied to the ground-state PBE energy levels to obtain a band gap corresponding to the value obtained at the G$_0$W$_0$@PBE level. Band gaps at the G$_0$W$_0$@PBE level were taken from the literature\cite{shishkin_self-consistent_2007,pham_probing_2014,pham_interfacial_2014,govoni_large_2015} except for 16-H$_2$O systems and the Si$_{10}$H$_{16}$ clusters, which we computed in this work. The values used are summarized in Table~\ref{tab:BSE_parameters}. For 16-H$_2$O snapshots, the G$_0$W$_0$@PBE band gap was computed using 640 PDEP eigenpotentials. For Si$_{10}$H$_{16}$, the G$_0$W$_0$@PBE band gap was computed using 1024 PDEP eigenpotentials and a cubic unit cell with a side length of 50 \AA. 

FF-BSE calculations were done at the $\Gamma$ point. Parameters used in FF-BSE simulations are in Table~\ref{tab:BSE_parameters}. 

The screening in FF-BSE\cite{nguyen_finite-field_2019} in the reciprocal space is expressed as follows:
\begin{equation}
    \Delta\tau_{vv^\prime}(\mathbf{G})=\begin{cases}
(\epsilon_{\infty}^{-1}-1)\tau_{vv^\prime}^{u}(\mathbf{G=0}), & \mathbf{G}=\mathbf{0}\\
\frac{4\pi e^{2}}{\mid\mathbf{G}\mid^{2}}\frac{\rho^+_{vv^\prime}(\mathbf{G})-\rho^-_{vv^\prime}(\mathbf{G})}{2}, & \mathbf{G}\neq\mathbf{0}
\end{cases}
\label{eq:dtau_G}
\end{equation}
where $\mathbf{G}$ is the reciprocal space lattice vector, and $\rho^\pm_{vv^\prime}(\mathbf{G})$ is the Fourier component of $\rho^\pm_{vv^\prime}(\mathbf{r})$. As described in the main text, we focus on getting a surrogate model corresponding to the $\mathbf{G}\neq\mathbf{0}$ terms; the term corresponding to $\mathbf{G=0}$, the long-wavelength limit, is added separately.

\begin{table}
\centering
\caption{\label{tab:BSE_parameters} Parameters used to obtain the FF-BSE spectra.}

\begin{tabular}{|l|P{0.15\linewidth}|P{0.18\linewidth}|P{0.1\linewidth}|p{0.1\linewidth}|p{0.08\linewidth}|}
\hline 
System & G$_{0}$W$_{0}$@PBE band gap (eV) & Scissor operator (eV) & Bisection levels in each Cartesian direction & Bisection threshold & Kinetic energy cutoff (Ry)\tabularnewline
\hline 
\hline 
16-H$_{2}$O & 10.41 & - & 2 & 0.02 & 60\tabularnewline
\hline 
64-H$_{2}$O & 8.1$^{\text{Ref.}}$\cite{pham_probing_2014} & 3.99$\pm$0.16 & 2 & 0.05 & 60\tabularnewline
\hline 
Si & 1.37 ($X_{1c}$-point)$^{\text{Ref.}}$\cite{govoni_large_2015} & 0.70 & 3 & 0.02 & 40\tabularnewline
\hline 
SiC & 2.28 ($X_{1c}$-point)$^{\text{Ref.}}$\cite{govoni_large_2015} & 0.95 & 2 & 0.04 & 70\tabularnewline
\hline 
C & 5.50$^{\text{Ref.}}$\cite{shishkin_self-consistent_2007} & 1.04 & 2 & 0.02 & 80\tabularnewline
\hline 
MgO & 7.25$^{\text{Ref.}}$\cite{shishkin_self-consistent_2007} & 2.48 & 2 & 0.01 & 60\tabularnewline
\hline 
LiF & 13.27$^{\text{Ref.}}$\cite{shishkin_self-consistent_2007} & 4.13 & 2 & 0.02 & 60\tabularnewline
\hline 
Si$_{10}$H$_{16}$ & 8.50 & 3.80 & 4 & 0.03 & 20\tabularnewline
\hline 
Si$_{35}$H$_{36}$ & 6.29$^{\text{Ref.}}$\cite{govoni_large_2015} & 2.80 (0~K); 3.58$\pm$0.17 (500~K) & 4 & 0.03 & 25\tabularnewline
\hline 
Si$_{87}$H$_{76}$ & 4.77$^{\text{Ref.}}$\cite{govoni_large_2015} & 2.21 & 4 & 0.03 & 25\tabularnewline
\hline 
H-Si/water & 1.34$^{\text{Ref.}}$\cite{pham_interfacial_2014} & 0.33 & 5 & 0.05 & 25$^a$\tabularnewline
\hline 
COOH-Si/water & 1.34$^{\text{Ref.}}$\cite{pham_interfacial_2014} & 0.27 & 5 & 0.05 & 25$^a$\tabularnewline
\hline 
\end{tabular}
%\begin{tablenotes}
\small
 \raggedright $^a$Comments on this choice is discussed in Section~\ref{Siwat}.
%\end{tablenotes}
\end{table}

For bulk Si, we used the Yambo\cite{marini2009yambo,sangalli2019many} code (version 4.4.0) to compute BSE spectra at the $\Gamma$ point and with $k$-point samplings. The Lanczos-Haydock solver was used, with scissor operators reported in Table~\ref{tab:si_yambo_par}. Details of the calculations with Yambo are reported in Table~\ref{tab:si_yambo_par}. 
For Model-BSE, the screened Coulomb potential $W=\epsilon^{-1}V_c$ was computed using $\epsilon$ defined as: 

\begin{equation}
    \epsilon^{-1}_{\mathbf{G},\mathbf{G^\prime}}=\begin{cases}
\epsilon_{\infty}^{-1}, & \mathbf{G}=\mathbf{G^\prime}=\mathbf{0}\\
1+f, & \mathbf{G}=\mathbf{G^\prime},\mathbf{G}\neq\mathbf{0} \\
0, & \mathbf{G}\neq\mathbf{G^\prime}
\end{cases}
\label{eq:epsilon_yambo}
\end{equation}

where $\epsilon_\infty$ is the dielectric constant computed by Yambo, and $f$ is the scaling factor defined in the main text. $f$ is an input value and is specified in the caption of each Model-BSE spectrum reported in this article.

\begin{table}[H]
\centering
\caption{\label{tab:si_yambo_par} Parameters of bulk Si for BSE calculations with the Yambo code.}
\begin{tabular}{|l|p{0.08\linewidth}|p{0.13\linewidth}|p{0.13\linewidth}|p{0.13\linewidth}|p{0.08\linewidth}|p{0.08\linewidth}|}
\hline
System & Number of atoms & $k$-point & Size of dielectric matrix (in no. PWs) & Size of exchange term in the BSE kernel (in no. PWs) & Number of bands & Scissor operator (eV) \\
\hline
\hline
Si & 64 & $\Gamma$ & 1000 & 295667 & 320 & 0.70 \\
\hline
Si & 2 & $12\times12\times12$ & 80 & 9185 & 20 & 0.85\\
\hline
\end{tabular}
\end{table}

\subsection{Machine learning}

We carried out ML-BSE calculations by implementing an interface between WEST\cite{govoni_large_2015} and Tensorflow\cite{tensorflow2015-whitepaper} (version 1.13). The convolutional model used in Eq.~\ref{eq:CNN} of the main text is defined as:
\begin{equation}
\Delta\tau_{vv^\prime}(x,y,z) = \sum_{i=0}^{n_x-1}\sum_{j=0}^{n_y-1}\sum_{k=0}^{n_z-1} K_{i,j,k} \tau^u_{vv^\prime}[x +(il-m_x) \Delta x, y+(jl-m_y)\Delta y, z+(kl-m_z)\Delta z]
\label{eq:CNN_detail}
\end{equation}
where $K$ is the convolutional filter of size $(n_x,n_y,n_z)$, and $\Delta x$, $\Delta y$, $\Delta z$ are the spacings of the uniform 3D-mesh used to represent periodic functions in real-space. $m_i,i=x,y,z$ is the size of padding in each of the $x,y,z$ directions, and $m_i=\lfloor \frac{p_i}{2} \rfloor$ where $p_i=n_i+(n_i-1)(l-1)-1$, and $l$ is the dilation rate. 
We performed a  hyperparameter search to determine the parameters in the training procedure. In the optimization procedure, we used the Adam optimizer\cite{kingma2014adam} with a learning rate of 0.001. The loss was evaluated by mean squared error (MSE). Early stopping was used to stop training based on validation loss after 25 epochs without an improvement. Periodic boundary conditions are satisfied using padded arrays; we considered the same size for input and output arrays in Eq.~\ref{eq:CNN}. In order to save memory and training time, we have trained parameters for convolutional models skipping every other element of the  $\tau^u_{vv^\prime}$ and $\Delta\tau_{vv^\prime}$ arrays in each Cartesian direction (except for 16-H$_2$O samples, where the size of the system allows us to consider all elements). In this way the arrays entering the training procedure have an 8 fold smaller memory footprint when the coarse grid is used. When the trained models are applied, we reconcile the fact that training was done on coarse grids, by applying a dilation rate of 2 in Eq.~\ref{eq:CNN}. In this way, the convolutional filter can be applied, after training, to $\tau^u_{vv^\prime}$ arrays that are defined on the original FFT grid. Therefore, we used $l=2$ in Eq.~\ref{eq:CNN_detail} except for the 16-H$_2$O system, for which we chose $l=1$.

For each system considered here, the training and validation data come from one snapshot, and we use data from snapshots different from the training snapshot as the test set. We have trained ML models with either a global scaling factor or a convolutional model with filter size $(n, n, n)$ with $n=3,5,7,9,12,15,20$. To study the effect of the training/validation split, we considered a given snapshot (which we call s00001) of the 16-H$_2$O system as an example, where there are 735 pairs of $\tau^u$ and $\Delta\tau$ arrays. Three different training/validation splits of the data set were considered: (1) all pairs used for training and validation, (2) 80\% pairs used for training and 20\% pairs used for validation, and (3) 60\% pairs used for training and 40\% pairs used for validation. Note that case~(1) does not give the same training and validation loss because minibatches of size 35 were used in training but all data were used in each validation. To evaluate the accuracy of the models, we trained for the models using snapshot s00001, and we used two other snapshots (i.e. s00003 and s00007) as test sets. None of these models have significant differences in the accuracy of reproducing the peak positions of the FF-BSE spectra. For all three split schemes, the average $\Delta\omega$ of the lowest-energy peak over different model architectures is -0.03 eV for s00001, -0.01 eV for s00003, and 0.00 eV for s00007. For the spectrum RMSE from ML-BSE using each split scheme, s00003 is 0.009 greater than s00001, and s00007 is 0.019 greater than s00001. These results suggest that the training/validation split within the range tested here does not significantly impact the accuracy or transferability of the ML models for predicting the absorption spectra. A subset of the dataset ($\tau^u_{vv^\prime}$ and $\Delta\tau_{vv^\prime}$ pairs) can be randomly selected and used as the training set in ML. The size of the subset does have an effect on the accuracy of the prediction. At least 10\% of data are needed in this subset to have a converged $f^{\text{ML}}$ value.

When computing $f^{\text{Avg}}$, we note that numerical errors may cause the absolute value of some elements in $\Delta\tau_{vv^\prime}(\mathbf{r})/\tau_{vv^\prime}^{u}(\mathbf{r})$ to be extremely large, and these outliers need to be discarded before computing $f^{\text{Avg}}$. No outliers in data need to be eliminated to carry out ML to obtain $f^{\text{ML}}$.

\subsection{Protocol to compute absorption spectra at finite temperature}
Below we present the protocol to obtain absorption spectra at finite temperature:

\begin{enumerate}
    \item Obtain representative snapshots of the target system at a finite temperature $T$ using FPMD. 
    \item For one snapshot among those selected in Step 1, perform a FF-BSE calculation to obtain $\tau^u_{vv^\prime}$ and $\Delta\tau_{vv^\prime}$. Note that the selected snapshot may come, in some cases, from a smaller system of similar nature as the target system (see, e.g., the example of water presented in the main text). 
    \item Use the $\tau^u_{vv^\prime}$ and $\Delta\tau_{vv^\prime}$ saved in Step 2 as the training/validation sets to machine learn the mapping between unscreened and screened Coulomb integrals. 
    \item For all snapshots determined in Step 1, except for the one already evaluated in Step 2-3, compute the ML-BSE spectrum using the ML model trained in the previous step as a surrogate model for the calculation of screened Coulomb integrals.
    \item Obtain the optical absorption spectrum at the finite temperature $T$ by computing the average of the absorption spectra obtained in Step 4. 
\end{enumerate}

\section{Comparison between FF-BSE and ML-BSE absorption spectra}

\subsection{Liquid water}

\begin{figure}[H]
    \centering
    \begin{subfigure}{0.45\textwidth}
    \caption{}
    \includegraphics[width=\linewidth]{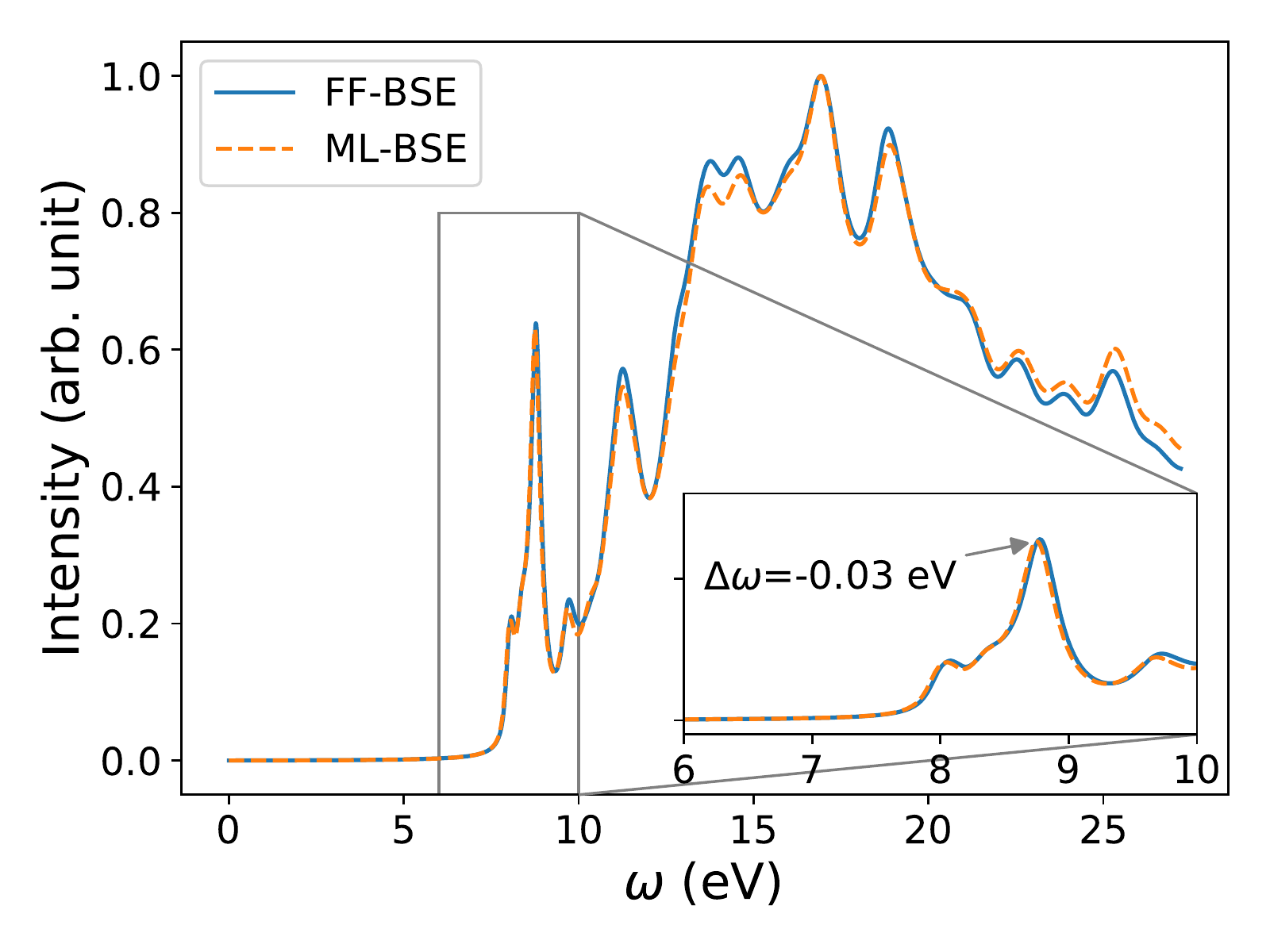}
    \end{subfigure}
    \hfill
    \begin{subfigure}{0.45\textwidth}
    \caption{}
    \includegraphics[width=\linewidth]{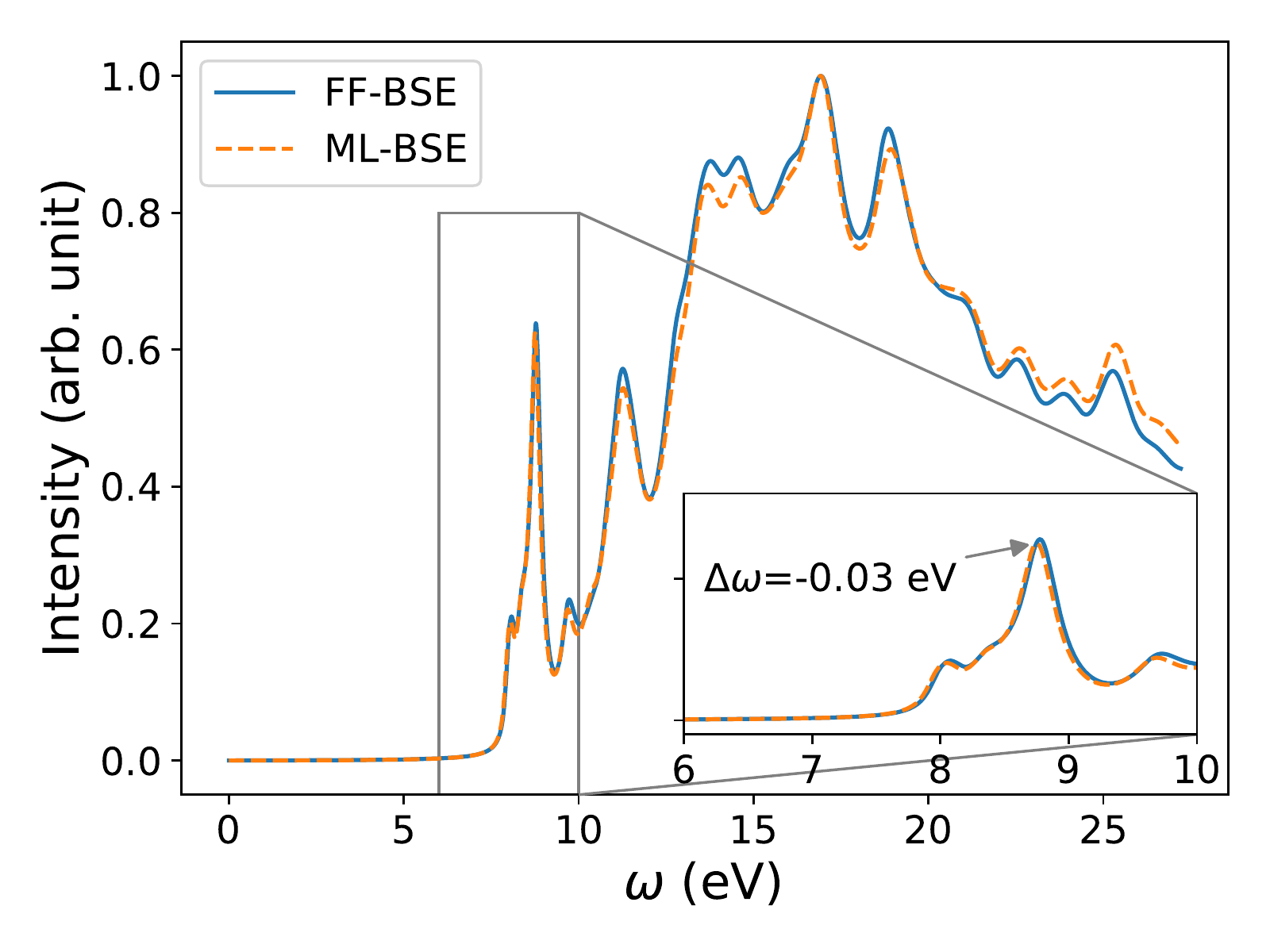}
    \end{subfigure}
\caption{Accuracy of ML-BSE spectra of liquid water (16-H$2$O) obtained using (a) a convolutional model with filter size $(7,7,7)$, (b) a global scaling factor model. RMSE of the spectra is 0.018 for (a) and 0.021 for (b).}
\label{fig:wat16_cnn7_sf}
\end{figure}

To quantify the accuracy of the ML-BSE spectrum, we compare it with the FF-BSE spectrum and compute the change of the energy of individual peaks ($\Delta\omega=\omega^{\text{ML-BSE}}-\omega^{\text{FF-BSE}}$, where $\omega$ is the position of the peak in energy) and the root mean square error (RMSE) of the whole spectrum in a given energy range (the range is 0.0-27.2 eV for all systems except for the interfaces, which is 0.0-13.6 eV).
For a representative snapshot of the 16-H$_2$O cell, when using a simple scaling factor model or a convolutional model of filter size $(7,7,7)$, we find that for the lowest-energy peak, $\Delta\omega=-0.03$ eV in both cases, and RMSE is 0.021 or 0.018 when the scaling factor or the convolutional model is used, respectively (Figure~\ref{fig:wat16_cnn7_sf}).This shows that, for the 16-H$_2$O system, the difference introduced by using different ML models (a convolutional model versus a global scaling factor) is negligible.

\begin{figure}[H]
    \centering

    \includegraphics[width=0.6\linewidth]{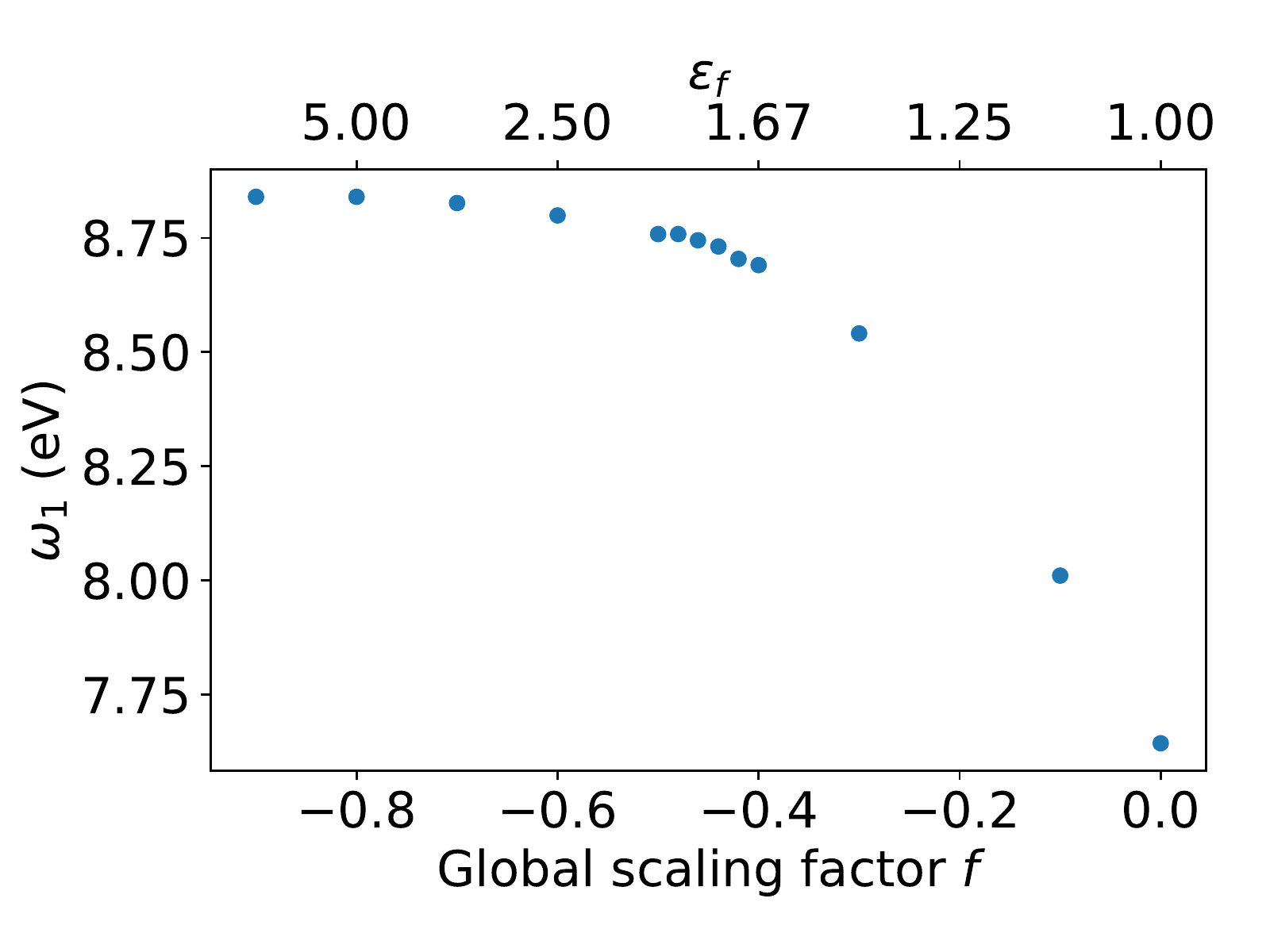}
    
    \caption{Sensitivity of the position of the lowest-energy peak ($\omega_1$) of the computed absorption spectrum of water, obtained for a snapshot with 16 H$_2$O molecules), to the global scaling factor ($f$). $\epsilon_f=(1+f)^{-1}$.}
    \label{fig:wat16_sf_sensitivity}
\end{figure}

Figure~\ref{fig:wat16_sf_sensitivity} shows the sensitivity of the position of the first peak of the spectrum to the value of the global scaling factor. We used the 16-H$_2$O case as an example.

\begin{figure}[H]
    \centering
    \begin{subfigure}{0.45\textwidth}
    \caption{}
    \includegraphics[width=\linewidth]{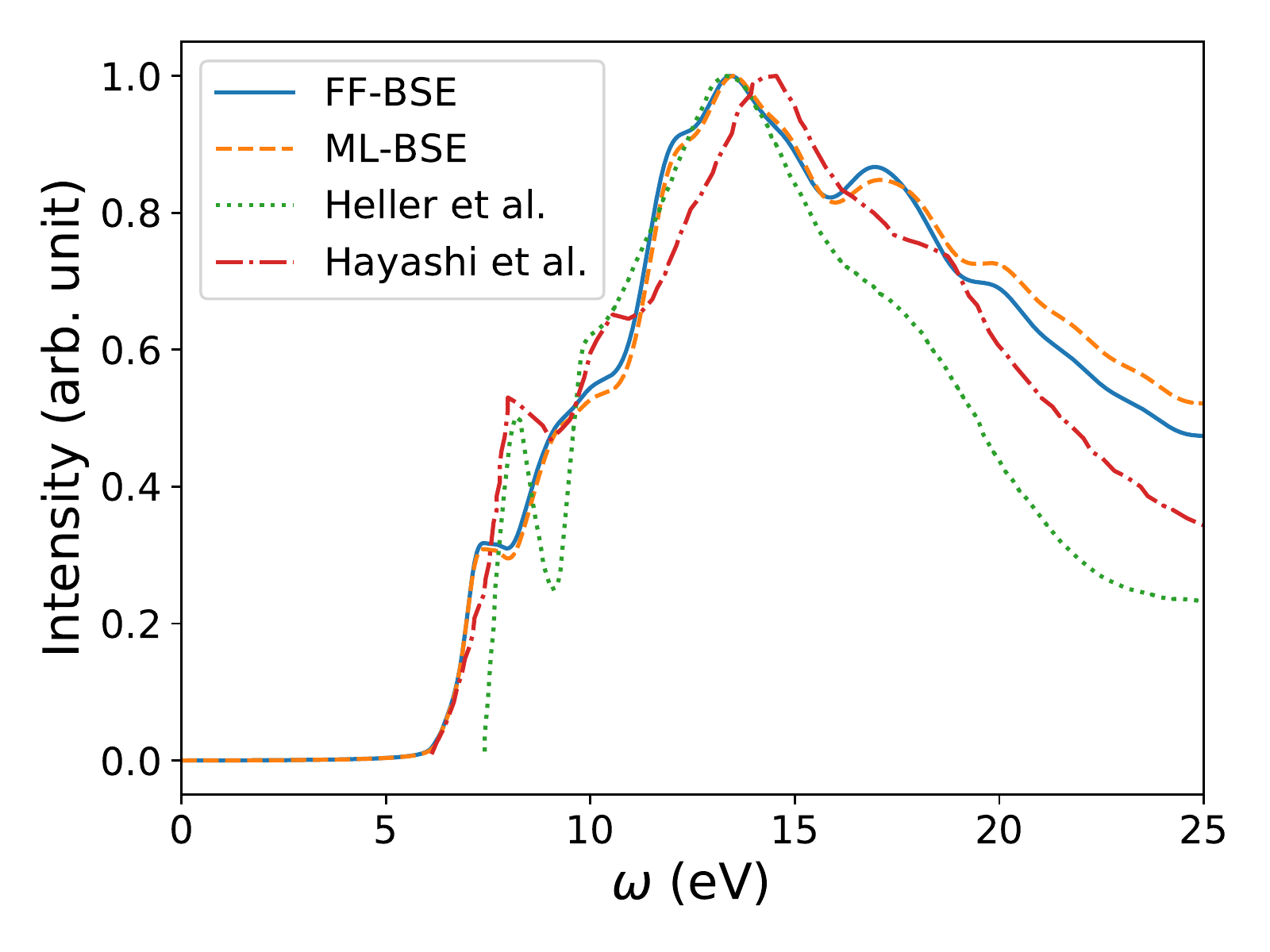}
    \label{fig:wat64_ave_exp}
    \end{subfigure}
    \hfill
    \begin{subfigure}{0.45\textwidth}
    \caption{}
    \includegraphics[width=\linewidth]{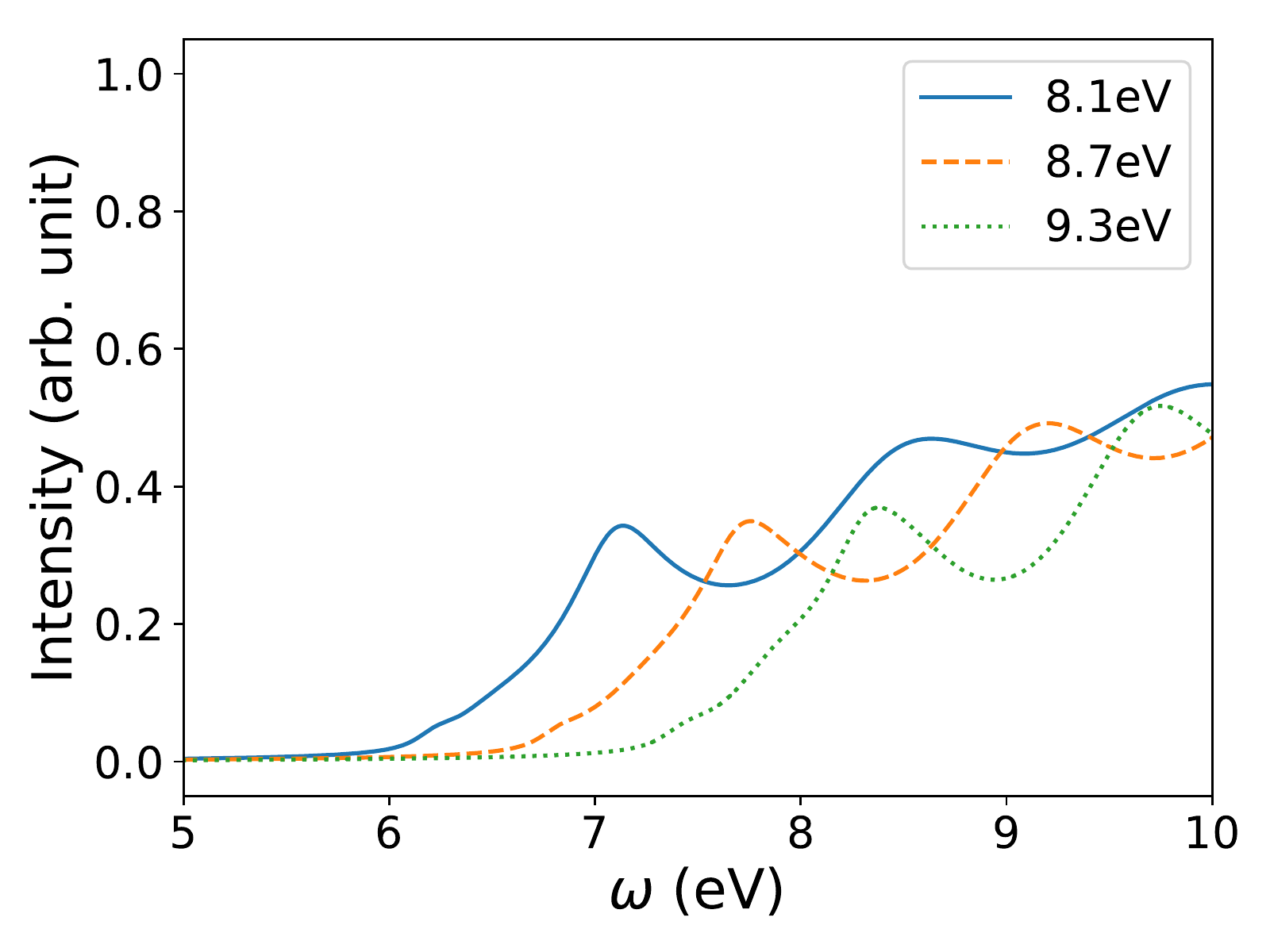}
    \label{fig:wat64_s0031_scissor}
    \end{subfigure}
    \caption{Absorption spectrum of liquid water (64-H$_2$O). (a) FF-BSE and ML-BSE are obtained computing and averaging the spectra of 10 snapshots. The position of the first peak of the experimental spectra from Heller et al.\cite{heller_collective_1974} and from  Hayashi et al.\cite{hayashi_complete_2000} is located at 8.18 eV and 7.98 eV, respectively. The position of the first peak of the ML-BSE and FF-BSE spectra is located at 7.40 eV. (b) Sensitivity of the FF-BSE absorption spectrum of a chosen snapshot to the fundamental gap of the system. In (b), the values of the fundamental gap, obtained applying a scissor operator to the computed PBE band gap, were chosen based on the values of the experimental gap of water of 8.7$\pm$0.6 eV,\cite{bernas1997electronic} and the G$_0$W$_0$@PBE band gap of 8.1 eV.\cite{pham_probing_2014} For this particular snapshot, the energy of the first peak is 7.14 eV, 7.85 eV, and 8.47 eV, when the band gaps of the system is equal to 8.1 eV, 8.7 eV, and 9.3 eV, respectively.}
    \label{fig:wat64_ave_exp_scissor}
\end{figure}

Figure~\ref{fig:wat64_ave_exp_scissor}(a) shows a comparison of the experimentally measured absorption spectrum of liquid water with the one computed for the 64-H$_2$O system using either FF-BSE or ML-BSE. 
Figure~\ref{fig:wat64_ave_exp_scissor}(b) shows the sensitivity of the spectrum to different choices of the scissor operator.

\begin{figure}[H]
    \centering
    
    \begin{subfigure}{0.45\textwidth}
    \caption{}
    \includegraphics[width=\linewidth]{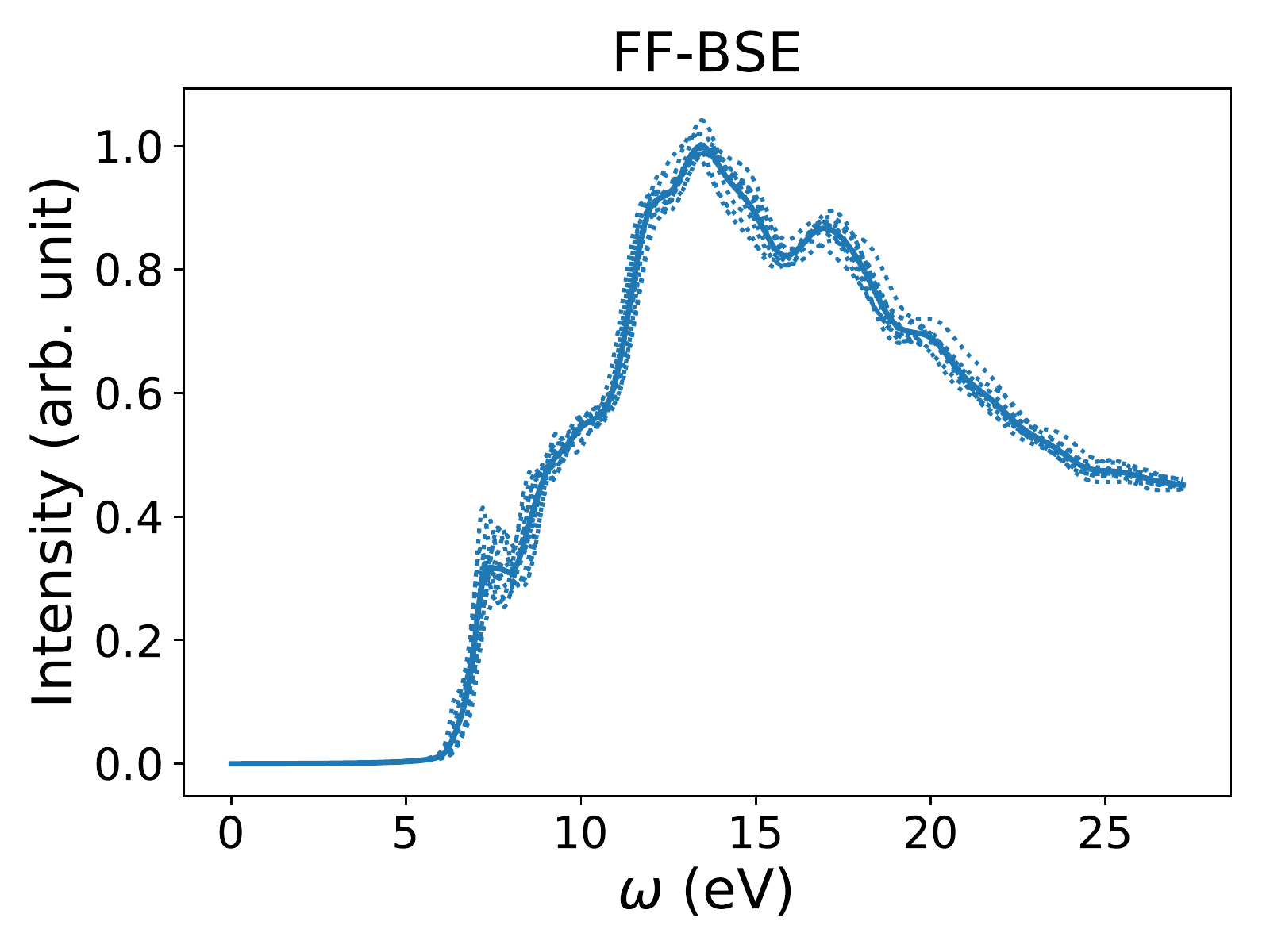}
    
    \end{subfigure}
    \hfill
    \begin{subfigure}{0.45\textwidth}
    \caption{}
    \includegraphics[width=\linewidth]{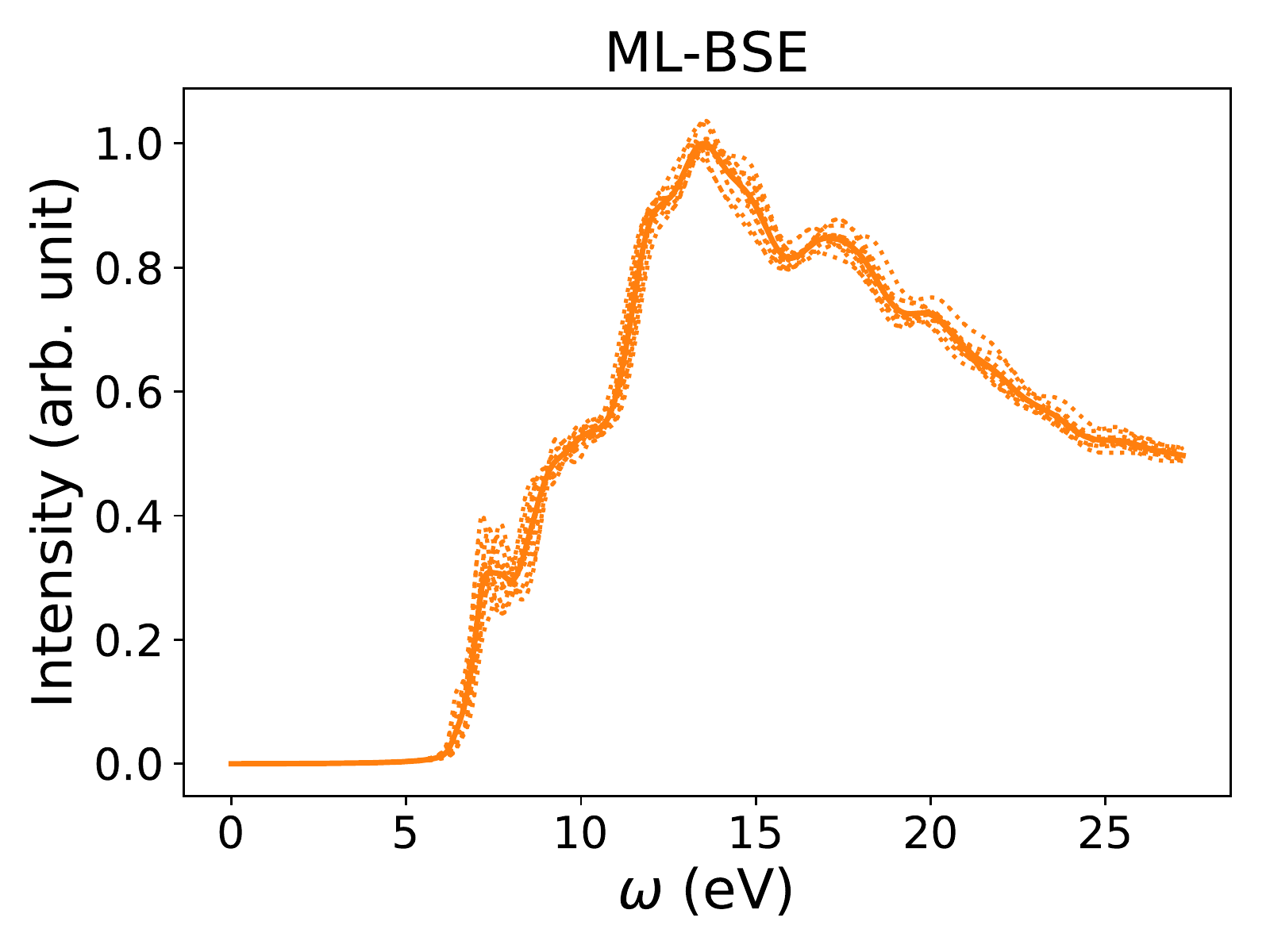}
    
    \end{subfigure}
    
    \caption{The absorption spectra of 10 individual snapshots of 64-H$2$O systems (dotted lines) and their averaged spectrum (solid line) from (a) FF-BSE and (b) ML-BSE calculations. The model used in ML-BSE is a global scaling factor obtained from simulations of a 16-water-molecule cell.}
    \label{fig:wat64_ave_ind}
\end{figure}

Figure~\ref{fig:wat64_ave_ind} shows the   variations of individual snapshots used to calculate the averaged spectrum of liquid water (64-H$_2$O system) in Figure~\ref{fig:wat64_ave}. The same comparison for each individual snapshot is reported in Figure~\ref{fig:wat64_10}. The model used in ML-BSE is the global scaling factor from 16-H$2$O.

\begin{figure}[H]
\centering

\begin{subfigure}{0.45\textwidth}
\includegraphics[width=\linewidth]{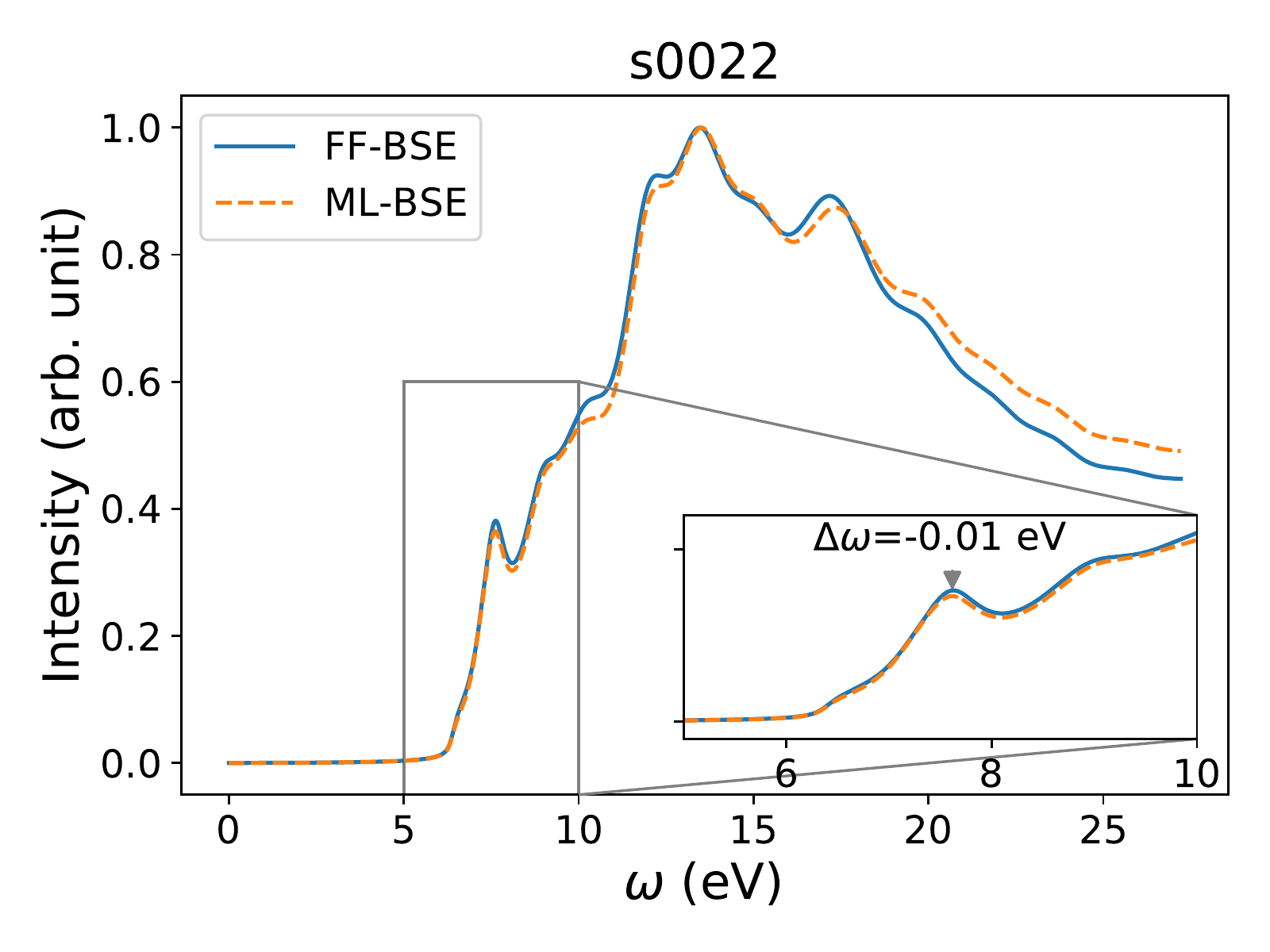}
\end{subfigure}
\hfill
\begin{subfigure}{0.45\textwidth}
\includegraphics[width=\linewidth]{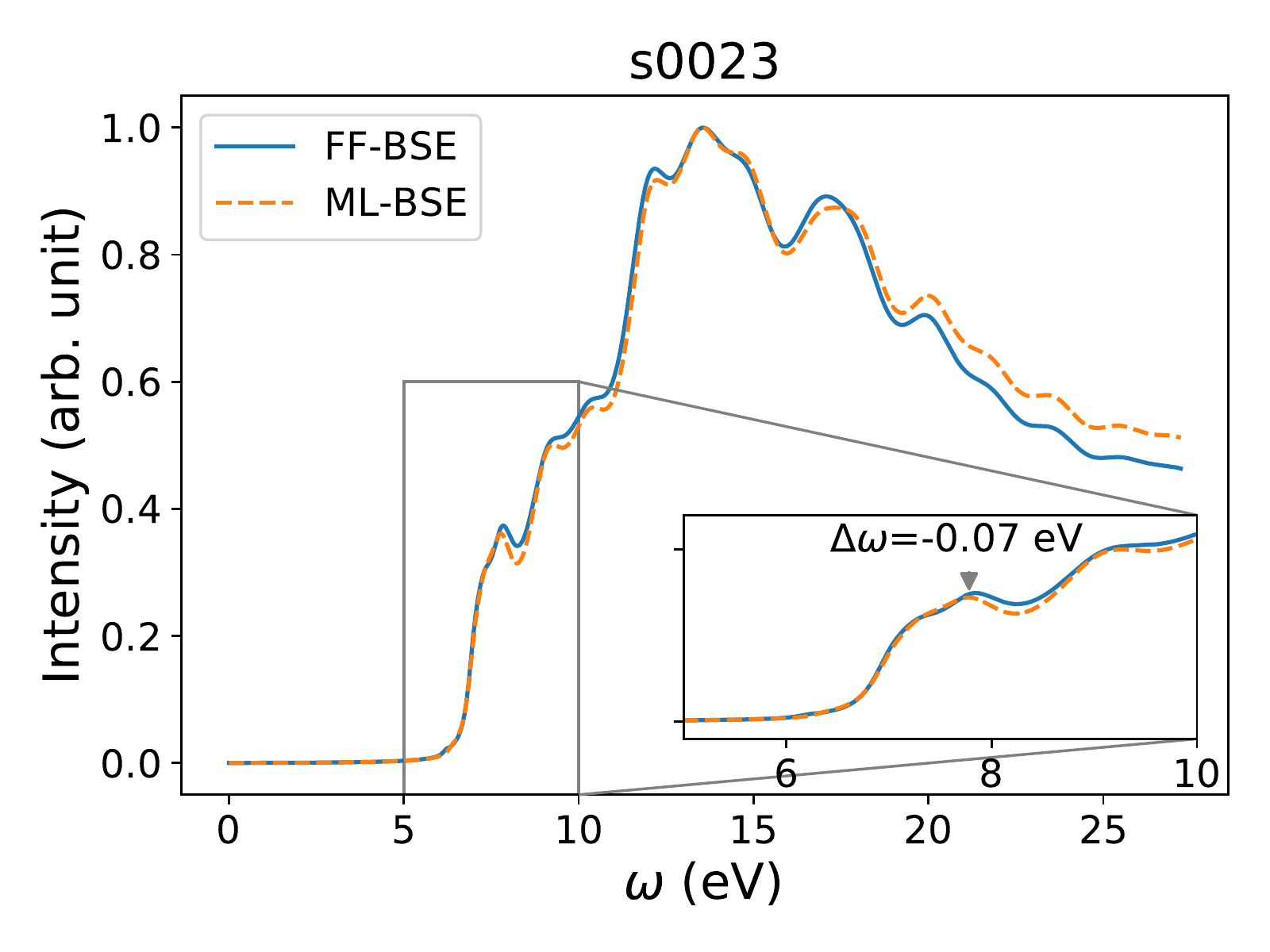}
\end{subfigure}
\hfill
\begin{subfigure}{0.45\textwidth}
\includegraphics[width=\linewidth]{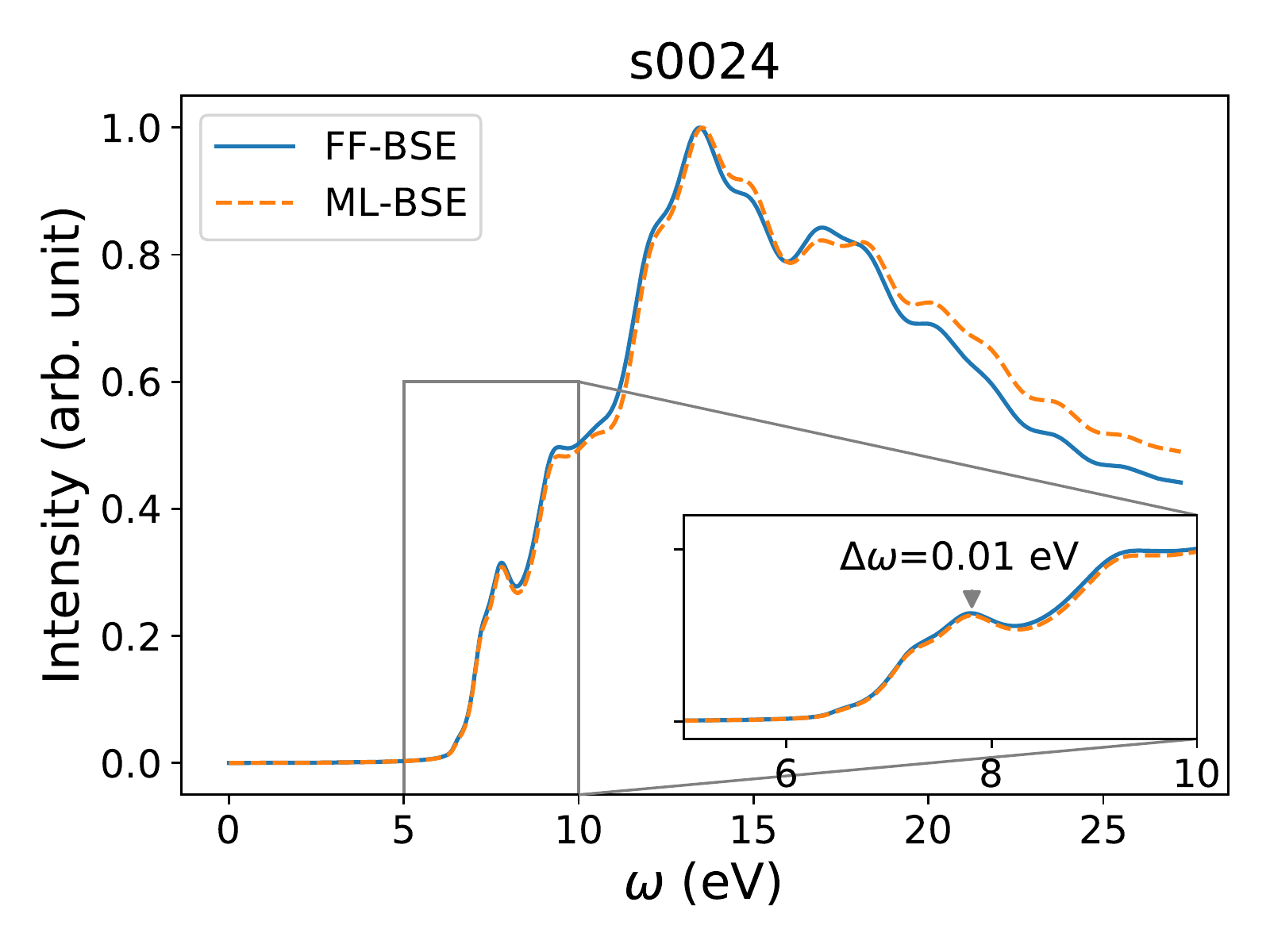}
\end{subfigure}
\hfill
\begin{subfigure}{0.45\textwidth}
\includegraphics[width=\linewidth]{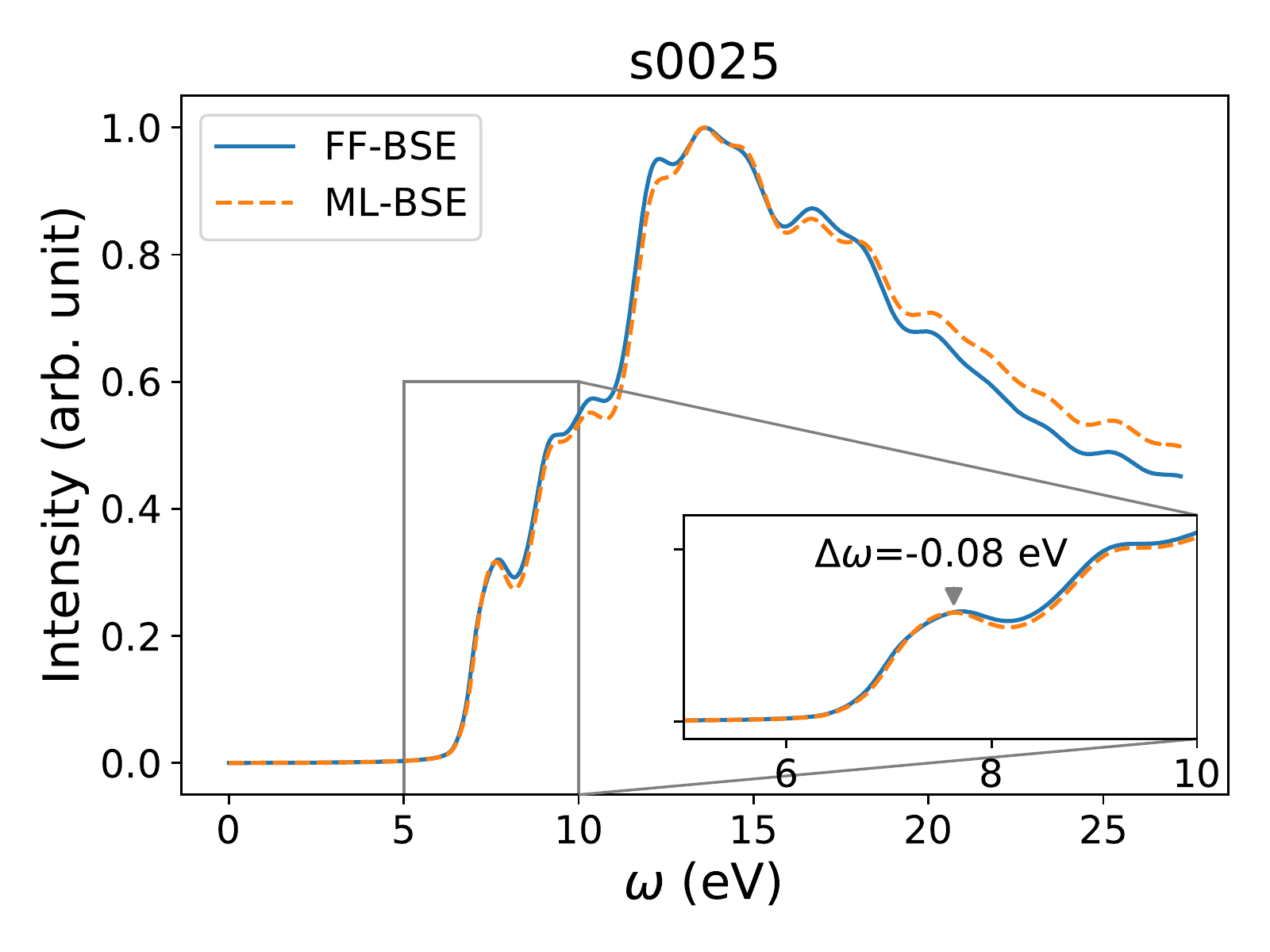}
\end{subfigure}
\hfill
\begin{subfigure}{0.45\textwidth}
\includegraphics[width=\linewidth]{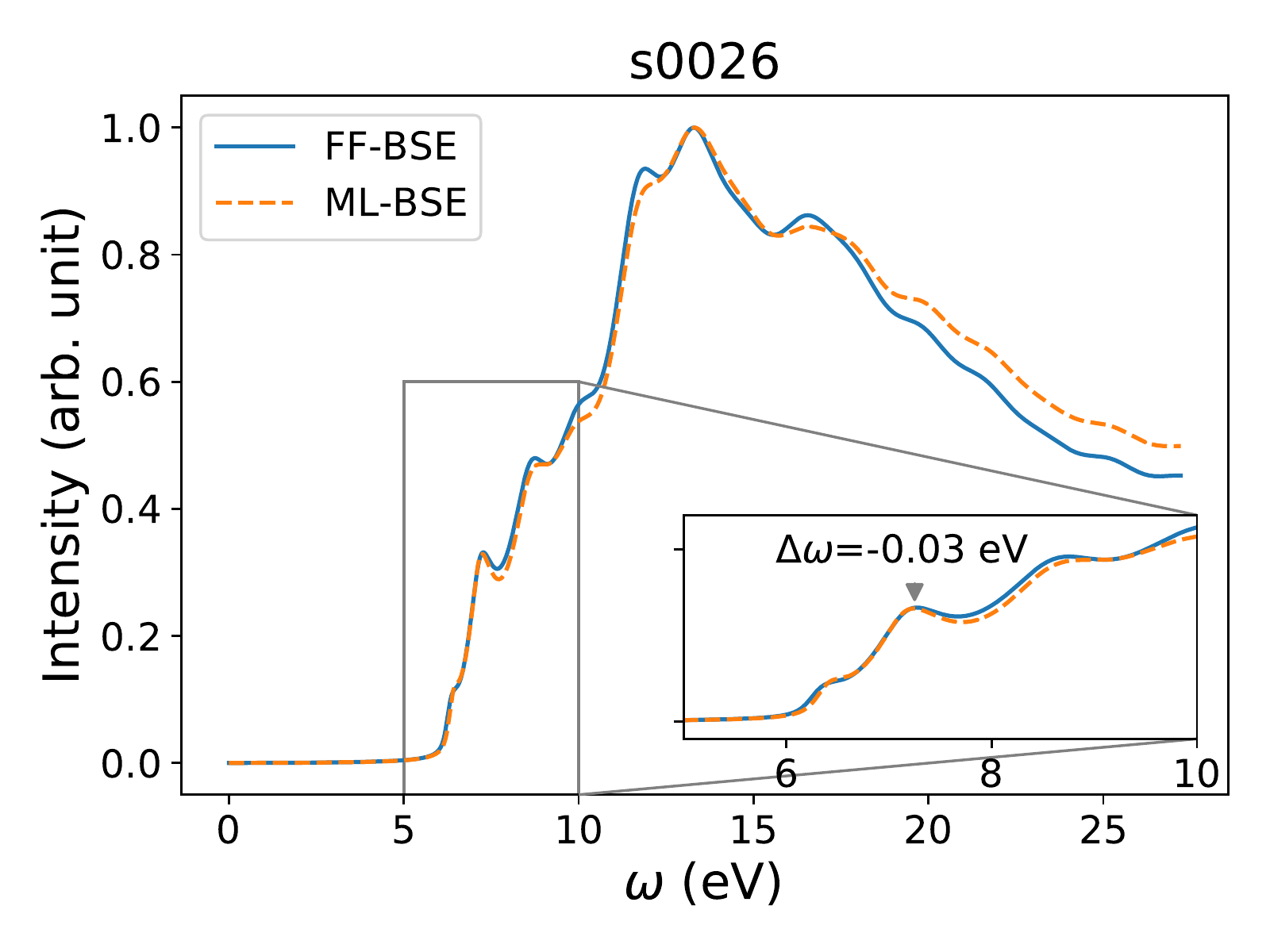}
\end{subfigure}
\hfill
\begin{subfigure}{0.45\textwidth}
\includegraphics[width=\linewidth]{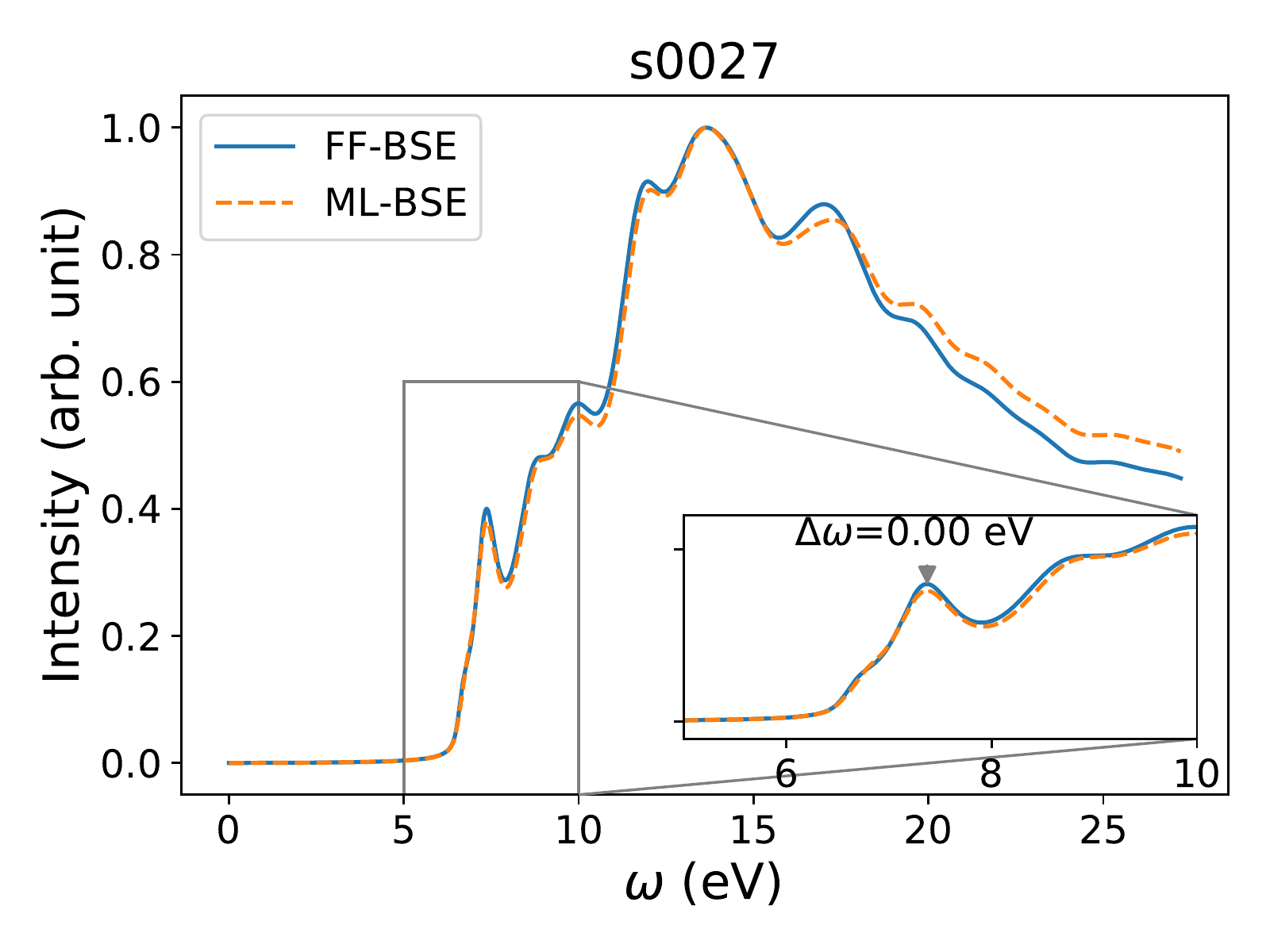}
\end{subfigure}

\caption{ML-BSE and FF-BSE spectra of 10 snapshots of the 64-H$_2$O system from FPMD trajectories at 400 K. The model used in ML-BSE is a global scaling factor obtained from simulations of a 16-water-molecule cell. The labels on the snapshots on top of each panel follows the labeling of snapshots of the PBE400 set (http://quantum-simulation.org/reference/h2o/pbe400/s32/index.htm).\cite{dawson_equilibration_2018}}
\label{fig:wat64_10}

\end{figure}

\begin{figure}[H]
\ContinuedFloat
\centering

\begin{subfigure}{0.45\textwidth}
\includegraphics[width=\linewidth]{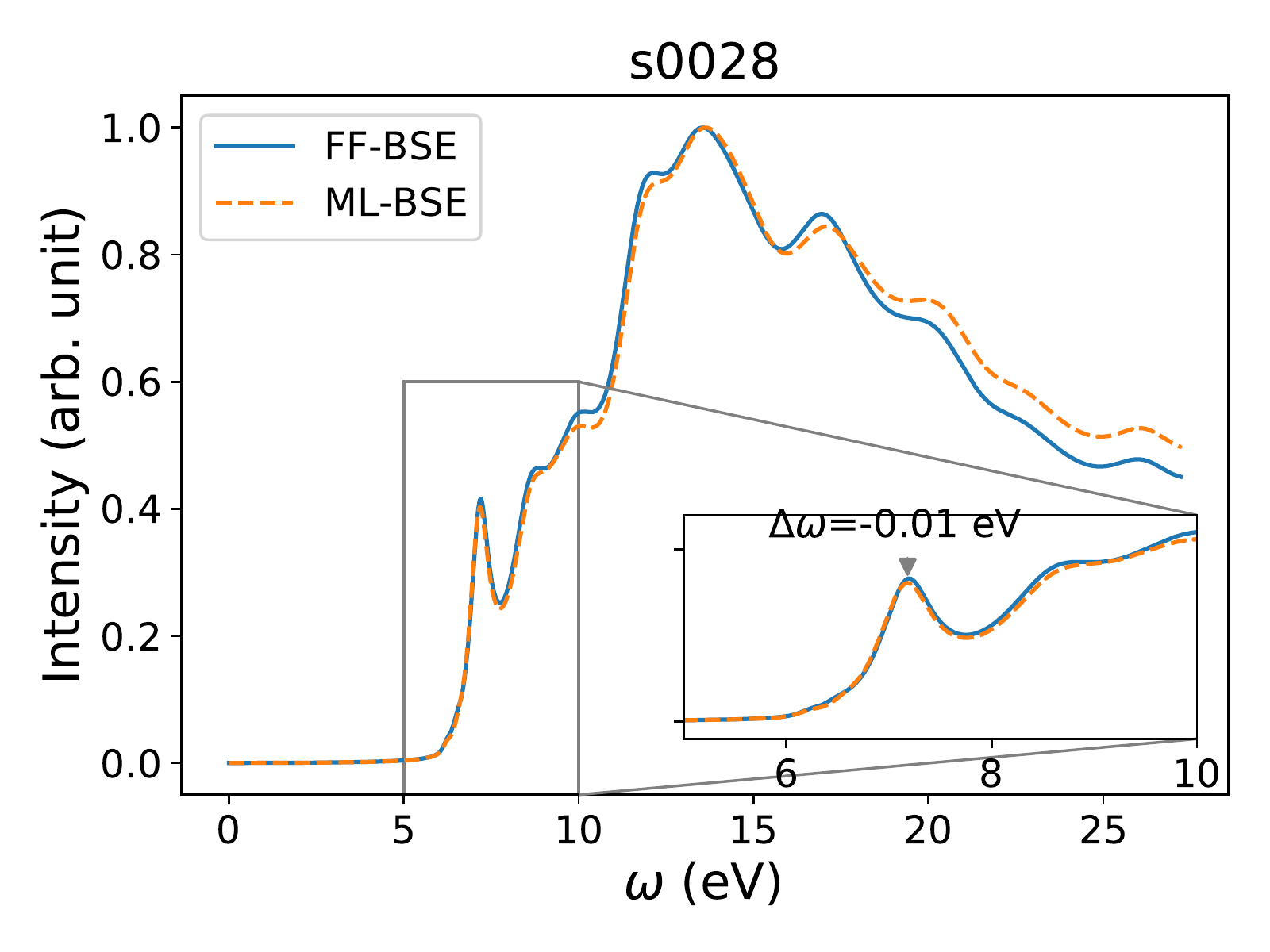}
\end{subfigure}
\hfill
\begin{subfigure}{0.45\textwidth}
\includegraphics[width=\linewidth]{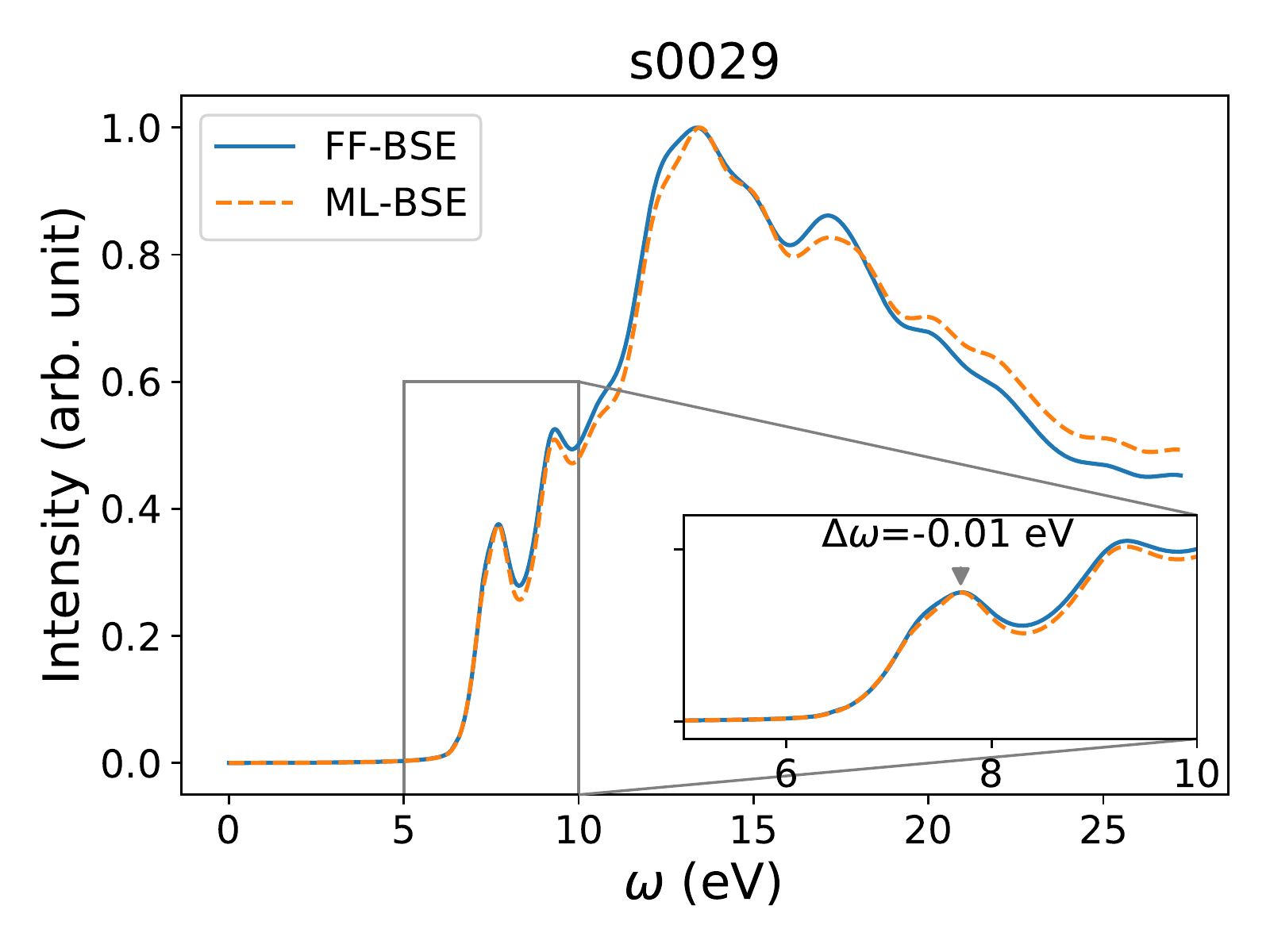}
\end{subfigure}
\hfill
\begin{subfigure}{0.45\textwidth}
\includegraphics[width=\linewidth]{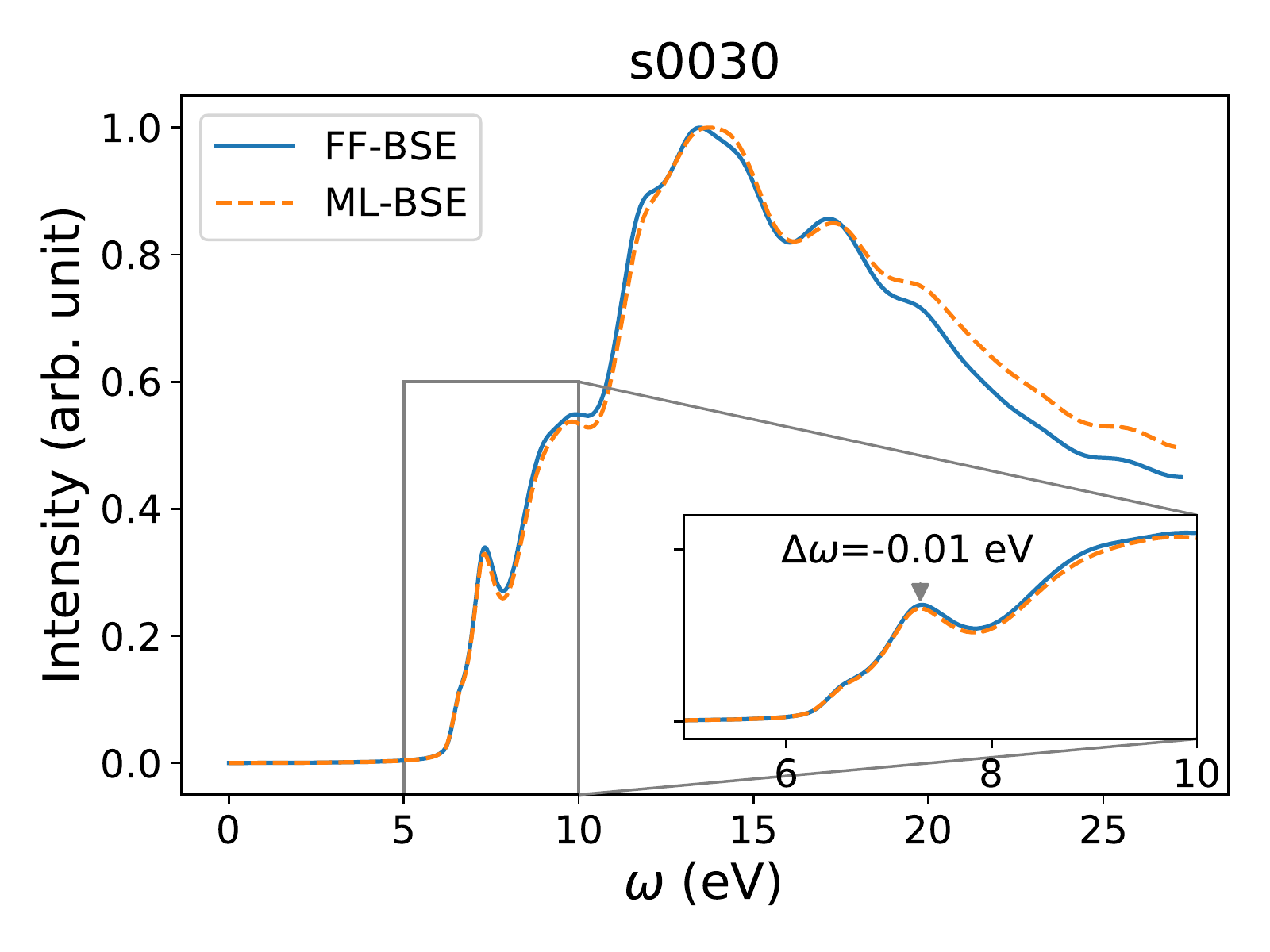}
\end{subfigure}
\hfill
\begin{subfigure}{0.45\textwidth}
\includegraphics[width=\linewidth]{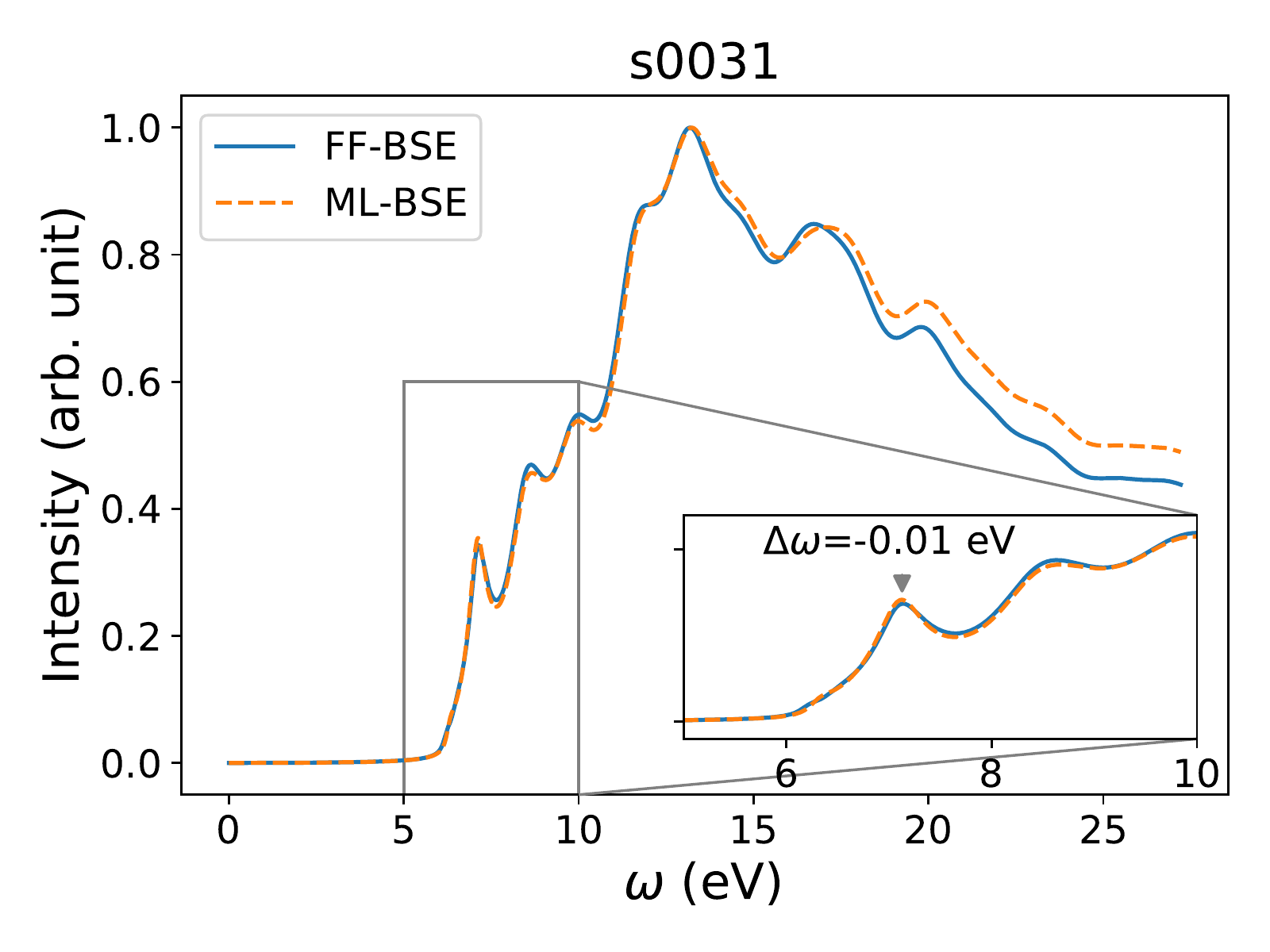}
\end{subfigure}

\caption{(Continued from the previous page) ML-BSE and FF-BSE spectra of 10 snapshots of the 64-H$_2$O system. The model used in ML-BSE is a global scaling factor obtained from simulations of a 16-water-molecule cell. The labels on the snapshots on top of each panel follows the labeling of snapshots of the PBE400 set (http://quantum-simulation.org/reference/h2o/pbe400/s32/index.htm).\cite{dawson_equilibration_2018}}
\label{fig:wat64_10}
\end{figure}

\subsection{Solids}

\begin{figure}[H]
    \centering
    \begin{subfigure}{0.45\textwidth}
    \caption{}
    \includegraphics[width=\linewidth]{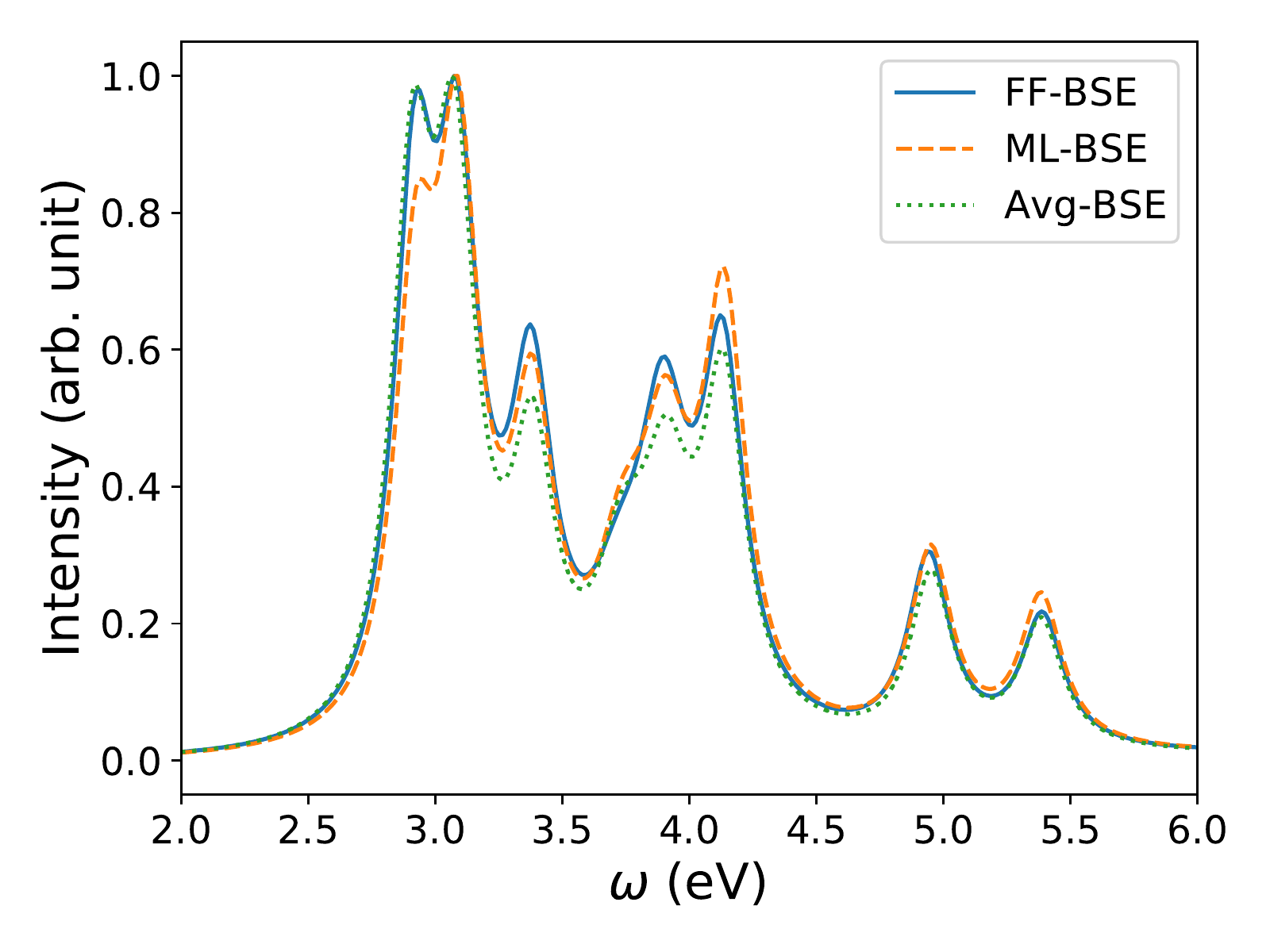}
    \end{subfigure}
    \hfill
    \begin{subfigure}{0.45\textwidth}
    \caption{}
    \includegraphics[width=\linewidth]{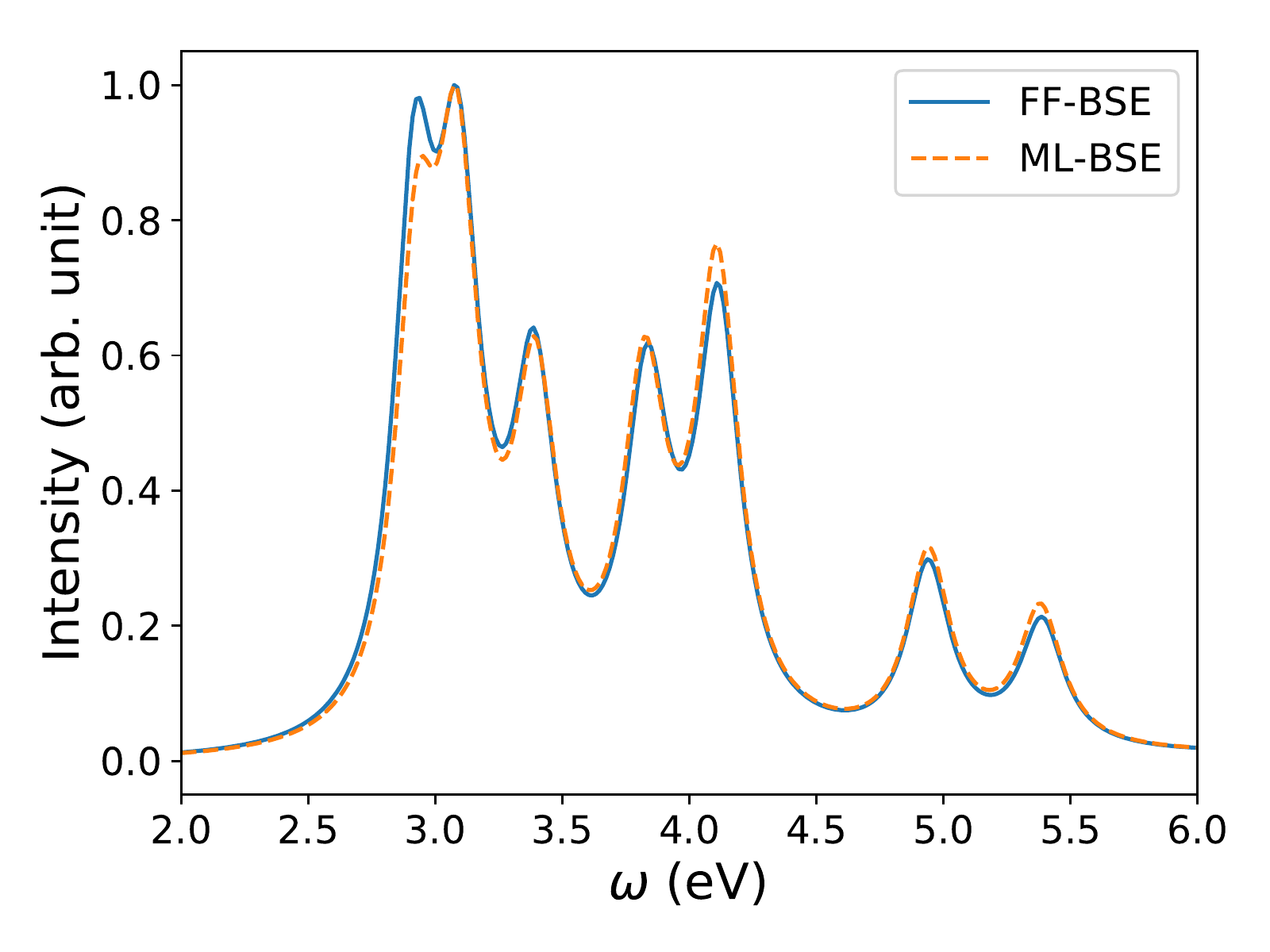}
    \end{subfigure}
    \caption{Accuracy of ML-BSE spectra of Si obtained using (a) a global scaling factor model and (b) a convolutional model (filter size $(7,7,7)$). Panel (a) also shows the BSE spectrum obtained by using a scaling factor computed from averaging $\Delta\tau/\tau^{u}$, labeled "Avg-BSE". The peak shifts of the ML-BSE spectrum from the FF-BSE spectrum are within 0.01 eV for all cases. The RMSE values between the ML-BSE and FF-BSE spectra are 0.019 for (a) and 0.015 for (b).}
    \label{fig:si_sf_cnn}
\end{figure}

\begin{figure}[H]
    \begin{subfigure}{0.45\textwidth}
    \caption{}
    \includegraphics[width=\linewidth]{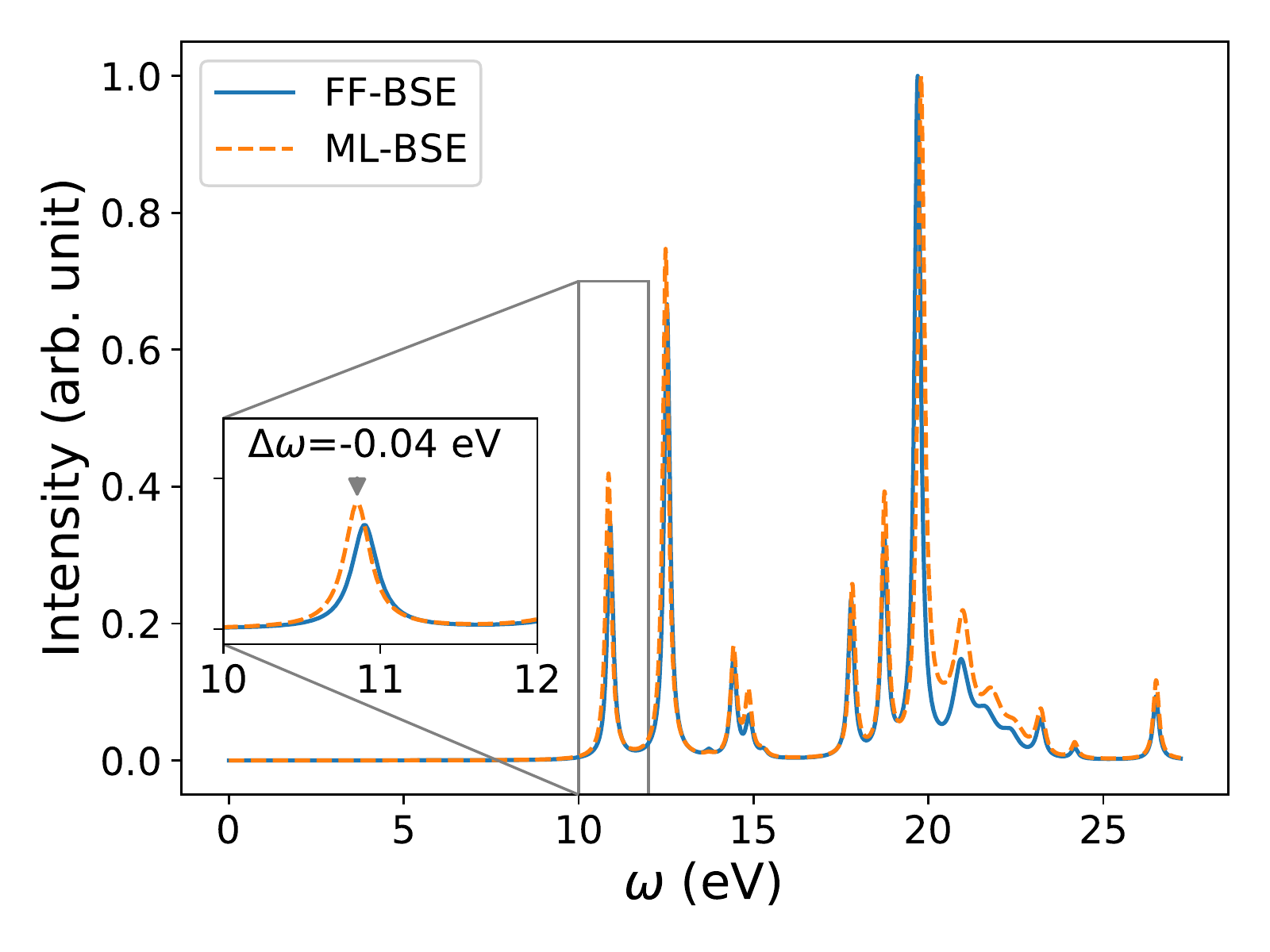}
    \end{subfigure}
    \hfill
    \begin{subfigure}{0.45\textwidth}
    \caption{}
    \includegraphics[width=\linewidth]{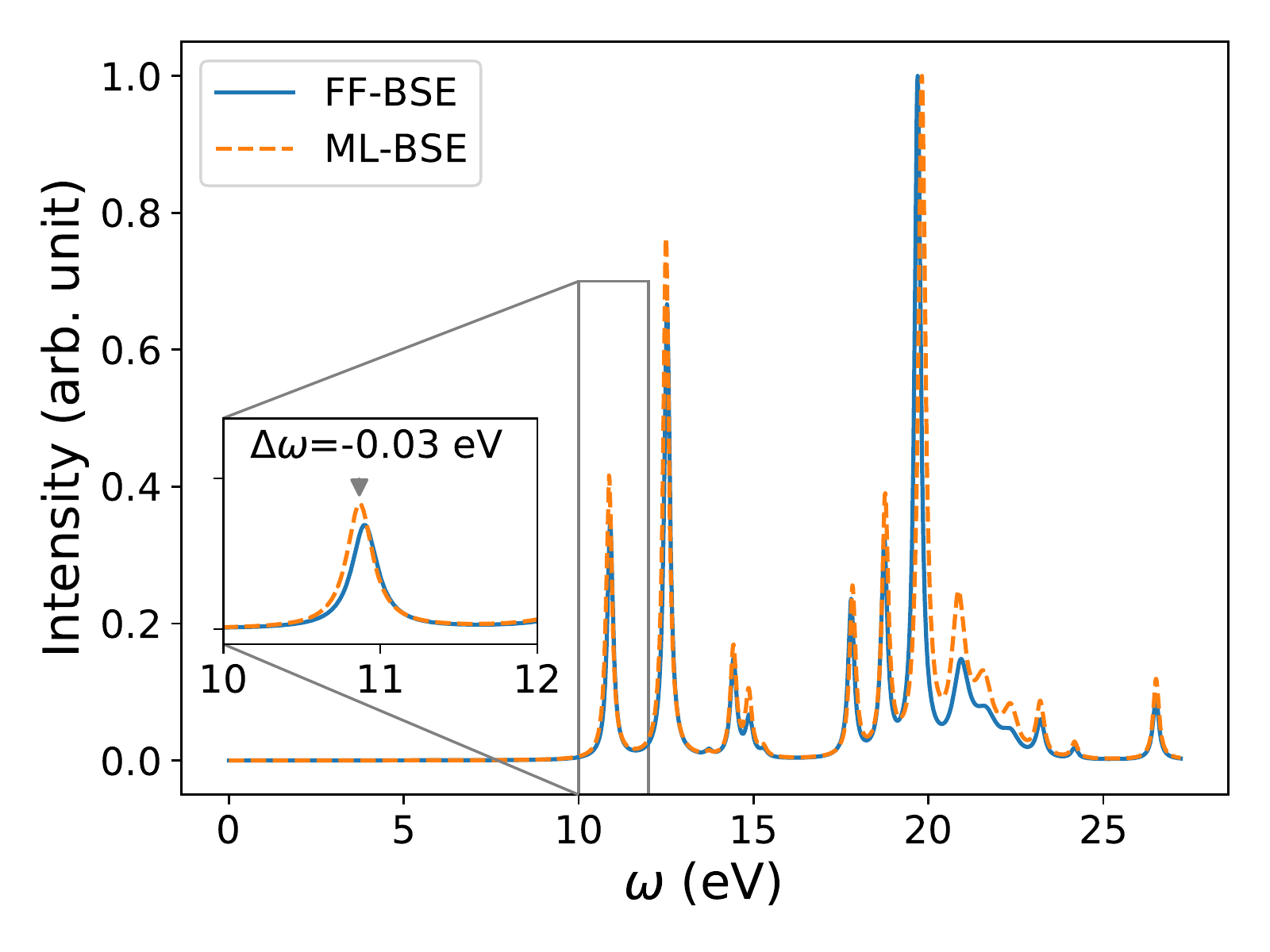}
    \end{subfigure}
    \caption{Accuracy of ML-BSE spectra of LiF obtained using (a) a global scaling factor model and (b) a convolutional model (filter size $(7,7,7)$). The RMSE values between the ML-BSE and FF-BSE spectra are 0.052 for (a) and 0.058 for (b).}
    \label{fig:lif_sf_cnn}
\end{figure}

\begin{figure}[H]
    \centering
        \begin{subfigure}{0.45\textwidth}
    \caption{}
    \includegraphics[width=\linewidth]{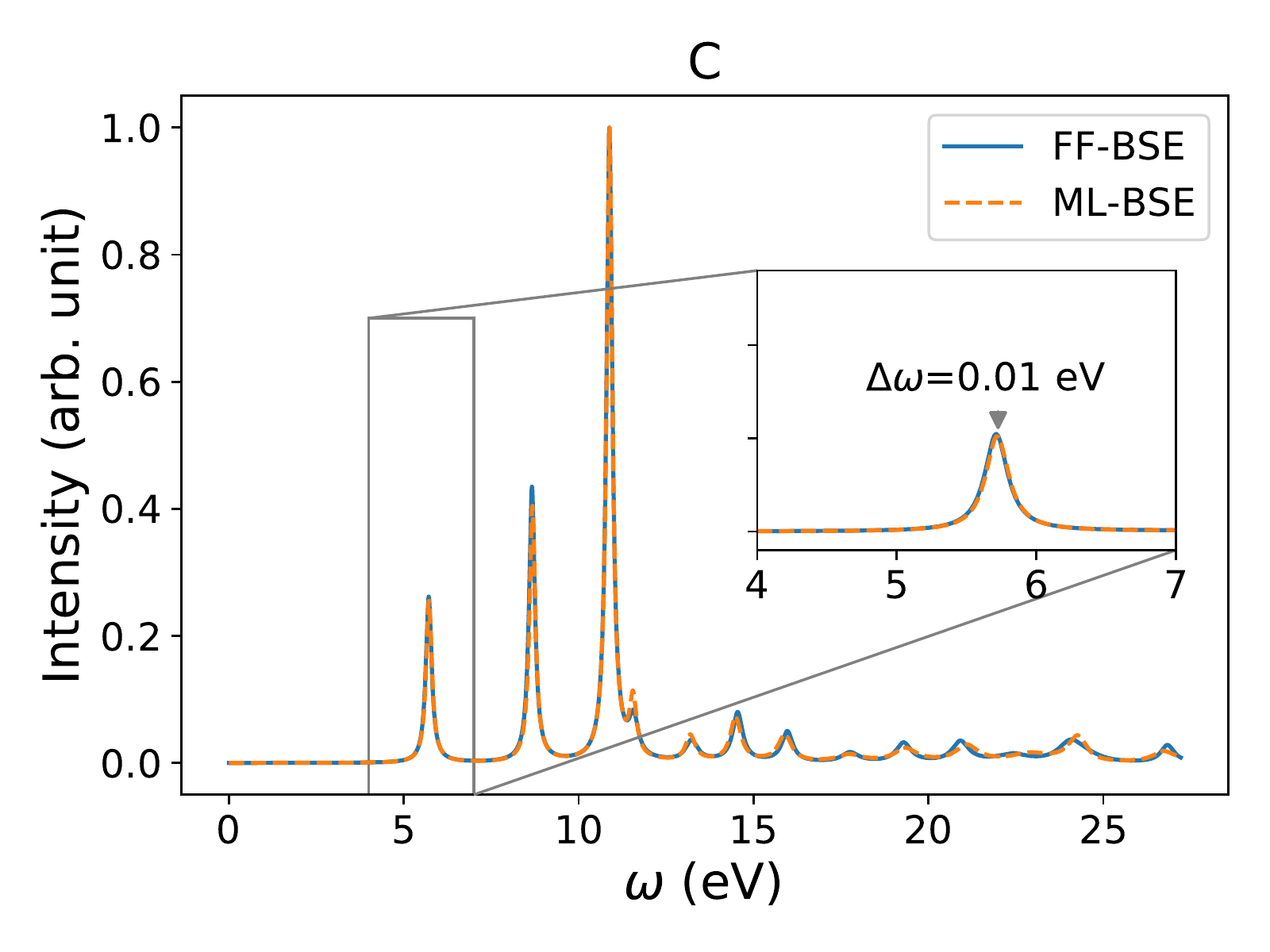}
    \end{subfigure}
    \hfill
    \begin{subfigure}{0.45\textwidth}
    \caption{}
    \includegraphics[width=\linewidth]{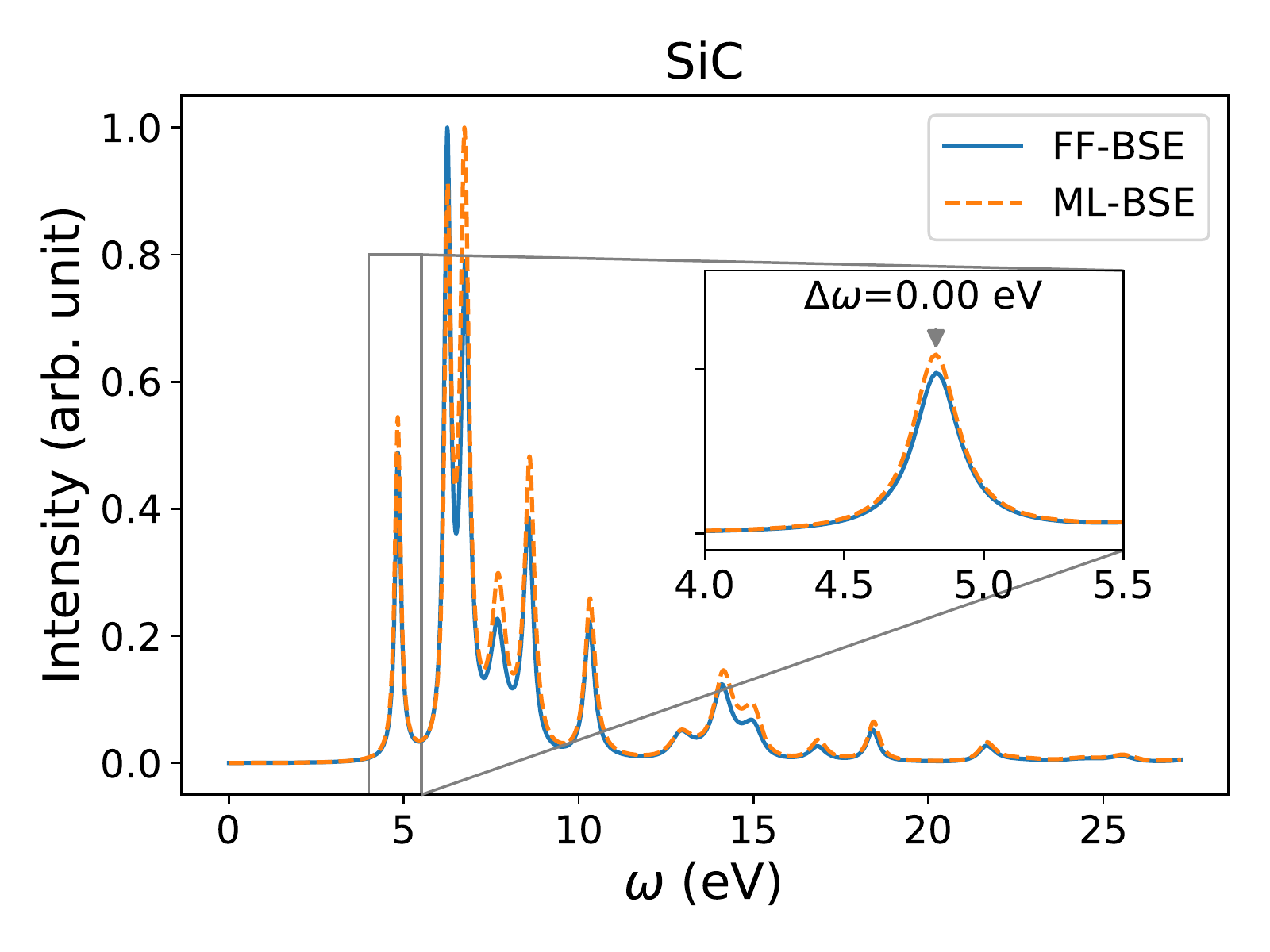}
    \end{subfigure}
    \hfill
    \begin{subfigure}{0.45\textwidth}
    \caption{}
    \includegraphics[width=\linewidth]{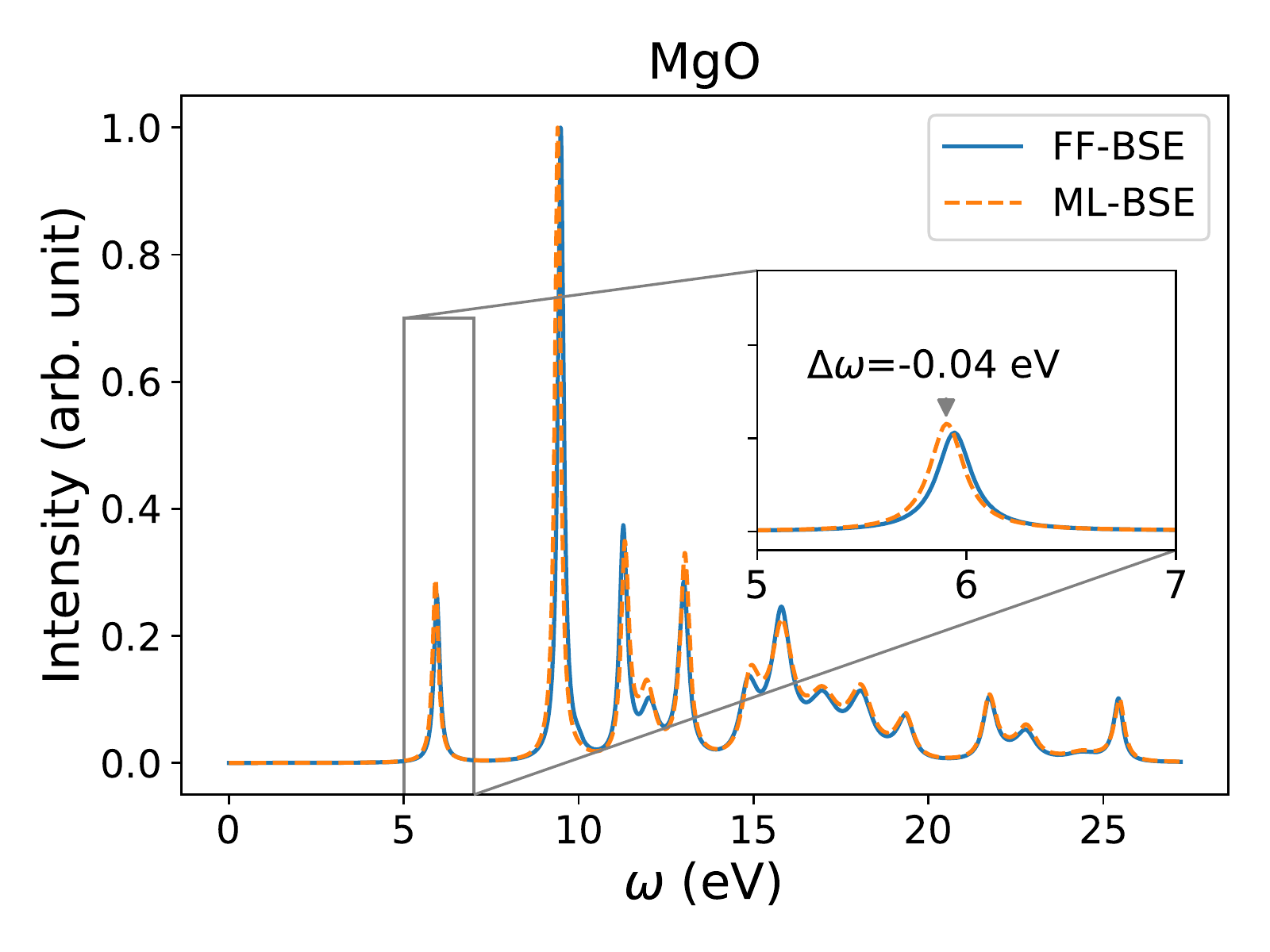}
    \end{subfigure}
    \caption{Comparison between FF-BSE and ML-BSE spectrum of (a) diamond, (b) SiC, and (c)MgO. ML-BSE results are obtained from a global scaling factor of the respective system. The RMSE values between the ML-BSE and FF-BSE spectra are 0.005 for (a), 0.027 for (b), and 0.044 for (c).}
    \label{fig:solids_spec}
\end{figure}

Figures~\ref{fig:si_sf_cnn} and \ref{fig:lif_sf_cnn} show that a convolutional model has similar accuracy as a global scaling factor for Si and LiF.
The comparisons between FF-BSE and ML-BSE spectrum for C (diamond), SiC, and MgO are reported in Figure~\ref{fig:solids_spec}. 

\begin{figure}[H]
    \centering
 
    \includegraphics[width=0.5\linewidth]{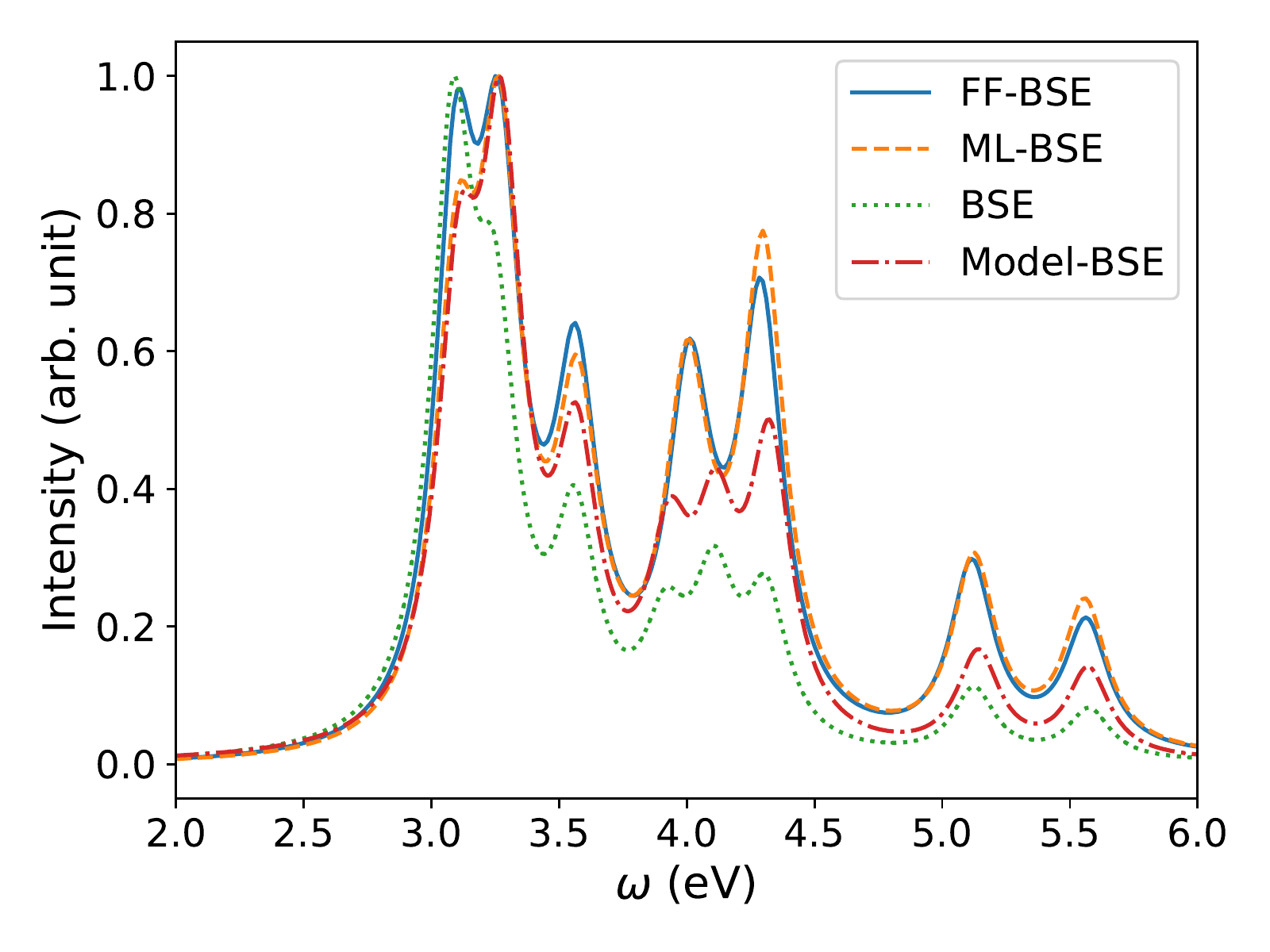}

    \caption{FF-BSE (WEST), ML-BSE (WEST) with a global scaling factor $f^{\text{ML}}=-0.81$, BSE (Yambo), and Model-BSE (Yambo) with $f=-0.81$ for Si with a 64-atom supercell at the $\Gamma$ point. The head of the dielectric matrix used in this figure is $\epsilon_\infty=22.11$, and was obtained for the same cell at the $\Gamma$ point in Yambo.}
    \label{fig:si64_ff_yambo_gamma}
\end{figure}

Figure~\ref{fig:si64_ff_yambo_gamma} shows the comparison between the absorption spectrum of Si computed with Yambo and WEST at the $\Gamma$ point.

\subsection{H-Si/water interface (hydrophobic interface)} \label{Siwat}
\subsubsection*{Comments on the kinetic energy cutoff}

\begin{figure}[H]
    \centering
    \begin{subfigure}{0.45\textwidth}
    \caption{}
    \includegraphics[width=\linewidth]{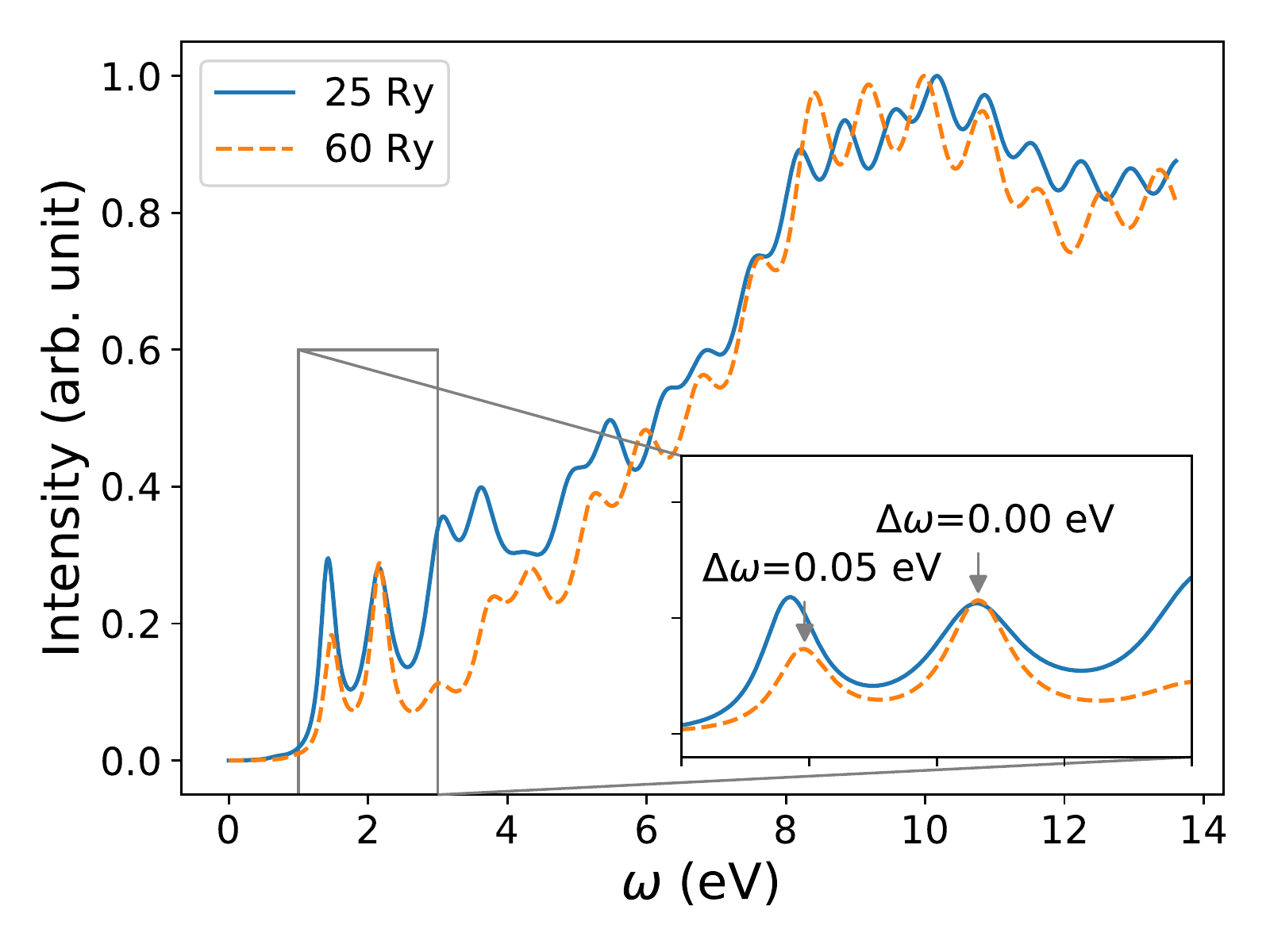}
    \end{subfigure}
    \hfill
    \begin{subfigure}{0.45\textwidth}
    \caption{}
    \includegraphics[width=\linewidth]{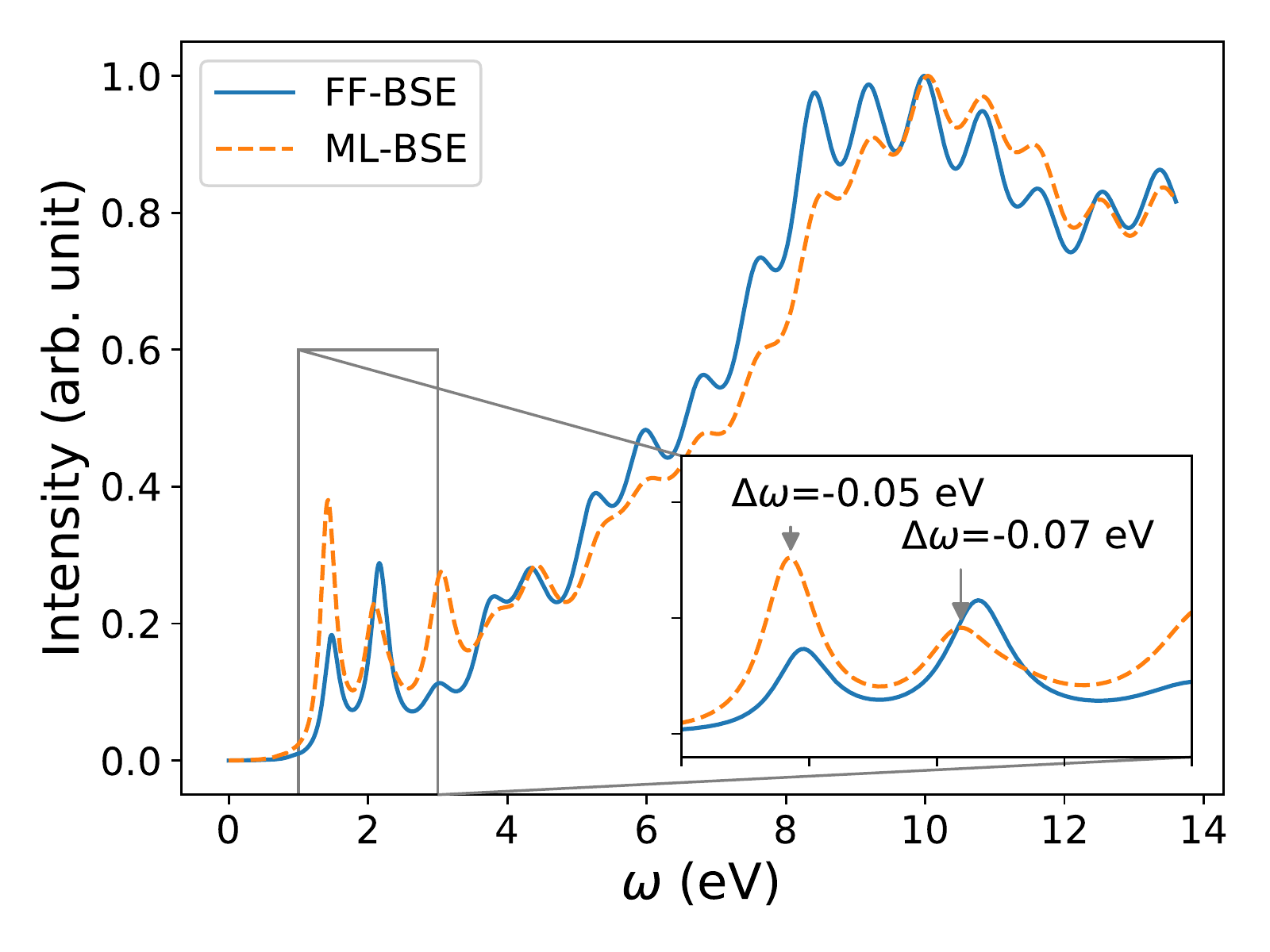}
    \end{subfigure}
    \hfill
    \caption{(a) FF-BSE spectrum from using a kinetic energy cutoff of 25 Ry or 60 Ry for wavefunctions. (b) Accuracy of ML-BSE using a $f^{\text{ML}}_{p2}(z)$ model and a kinetic energy cutoff of 60 Ry for wavefunctions.}
    \label{fig:sihwat_ecut}
\end{figure}

For the H-Si/water interface, we tested two different kinetic energy cutoff values for wavefunctions: 25 Ry and 60 Ry. From Figure~\ref{fig:sihwat_ecut}(a), we conclude that 25 Ry is sufficient to obtain the low-energy peaks of the Si/water interface FF-BSE spectrum. Comparing Figure~\ref{fig:siwat_2interfaces_2par}(a) and Figure~\ref{fig:sihwat_ecut}(b), we conclude that using 25 Ry or 60 Ry will not change our conclusions for the Si/water interface. The $f^{\text{ML}}_{p2}(z)$ models for the 25 Ry case and the 60 Ry case are very similar, with $\epsilon^{\text{ML}}_f(z)$ being 2.28 and 2.95 for $z$ in the respective regions of Si and water for the 25 Ry case, and 2.25 and 2.99 for $z$ in the respective regions of Si and water for the 60 Ry case.  Therefore, we used 25 Ry for all Si/water interfaces in this work.

\subsubsection*{A position-dependent model $f(\mathbf{r})$ is necessary}
\begin{figure}[H]

\centering
    \begin{subfigure}{0.45\textwidth}
    \caption{}
    \includegraphics[width=\linewidth]{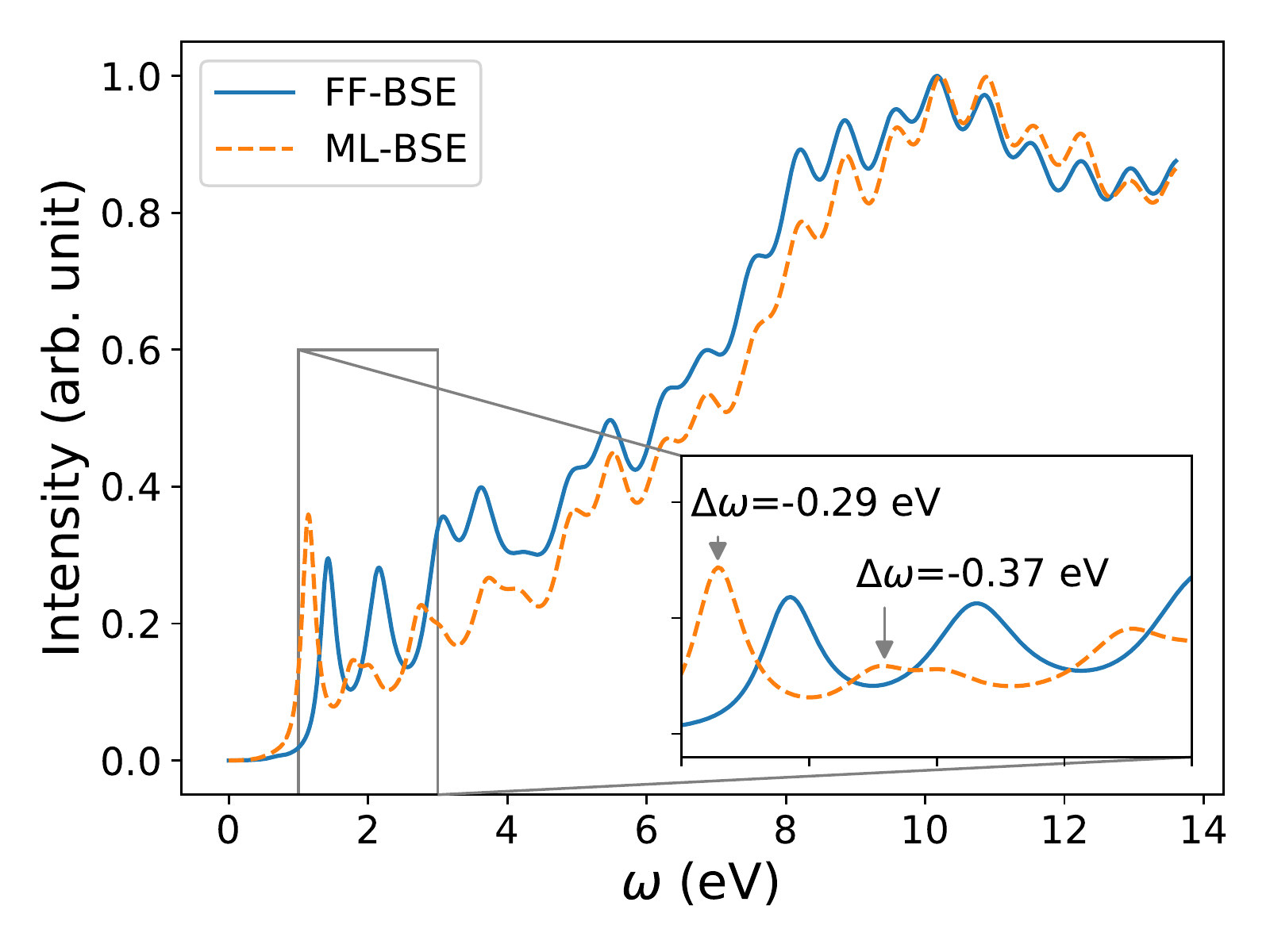}
    \end{subfigure}
    \hfill
    \begin{subfigure}{0.45\textwidth}
    \caption{}
    \includegraphics[width=\linewidth]{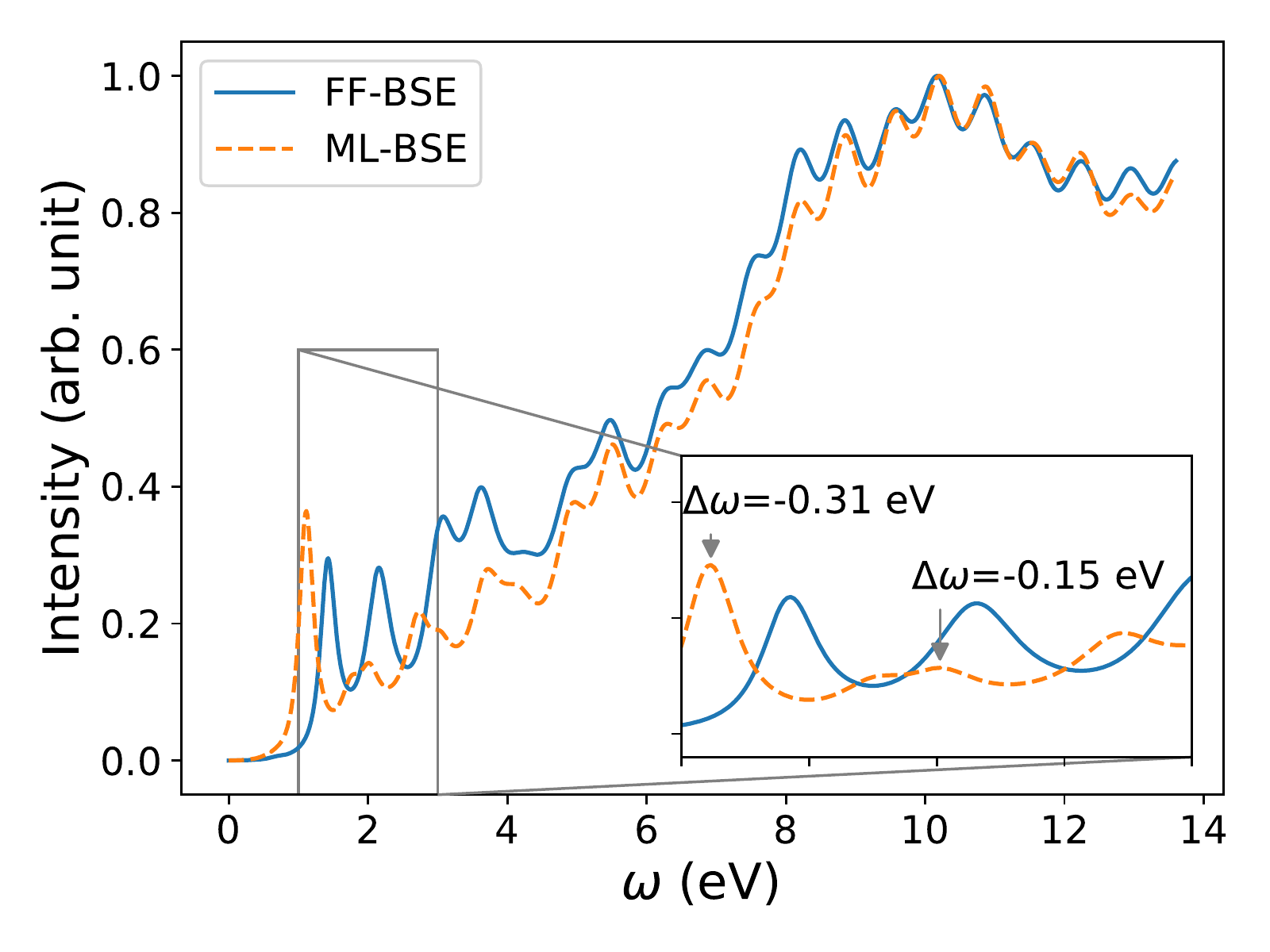}
    \end{subfigure}
    \hfill

\caption{Accuracy of ML-BSE using (a) a global scaling factor model, and (b) a convolutional model (filter size $(7,7,7)$)  for a snapshot representing a H-Si/water interface (see text). The RMSE values between the ML-BSE and FF-BSE spectra are 0.078 for (a) and 0.072 for (b).}
\label{fig:sihwat_sf_cnn7}
\end{figure}

From Figure~\ref{fig:sihwat_sf_cnn7}, we can see that when using a model which treats Si and water on the same footing, using average scaling factors, results for the absorption spectrum are not accurate.

\begin{figure}[H]

\centering
    \begin{subfigure}{0.45\textwidth}
    \caption{}
    \includegraphics[width=\linewidth]{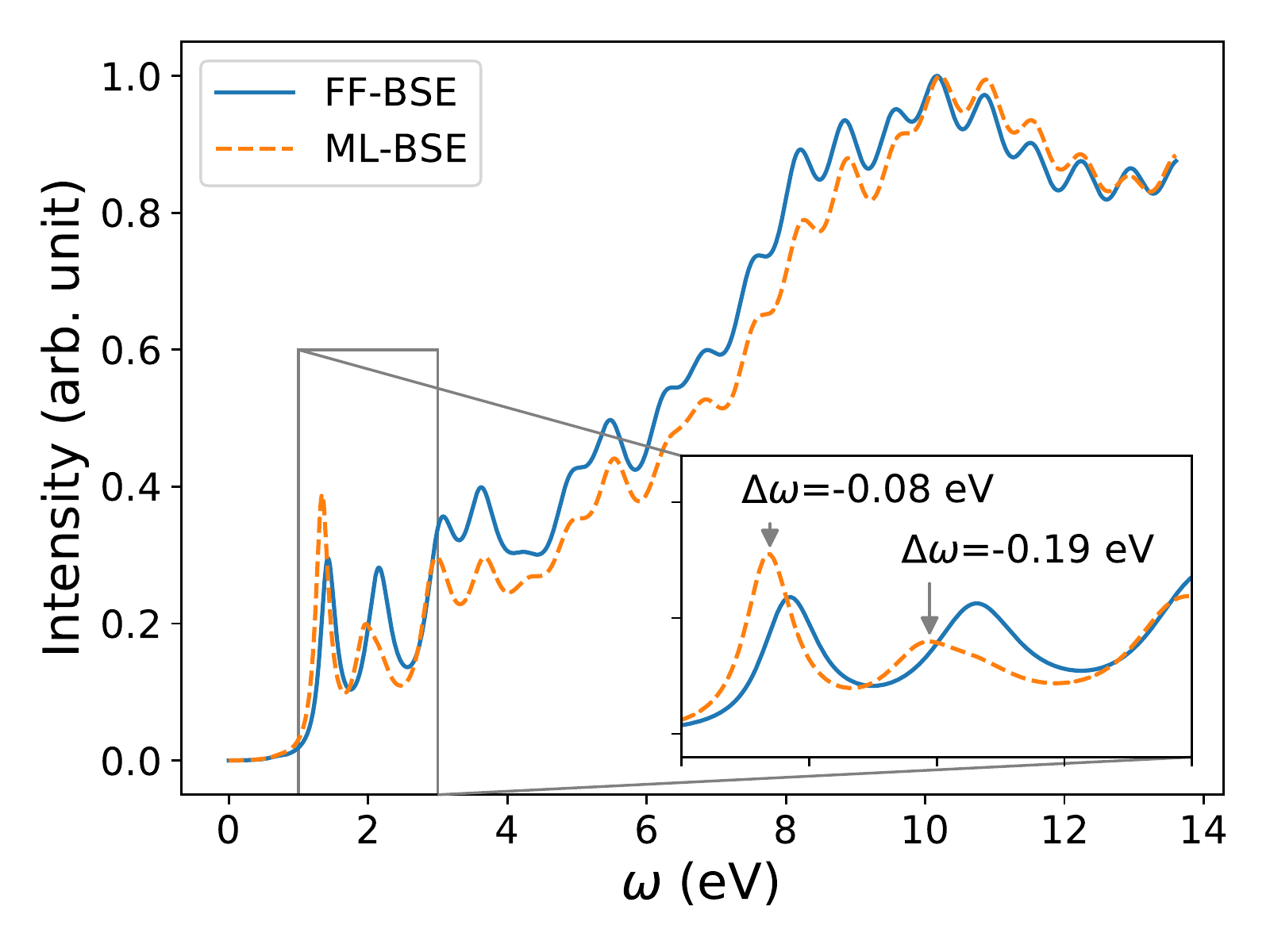}
    \end{subfigure}
    \hfill
    \begin{subfigure}{0.45\textwidth}
    \caption{}
    \includegraphics[width=\linewidth]{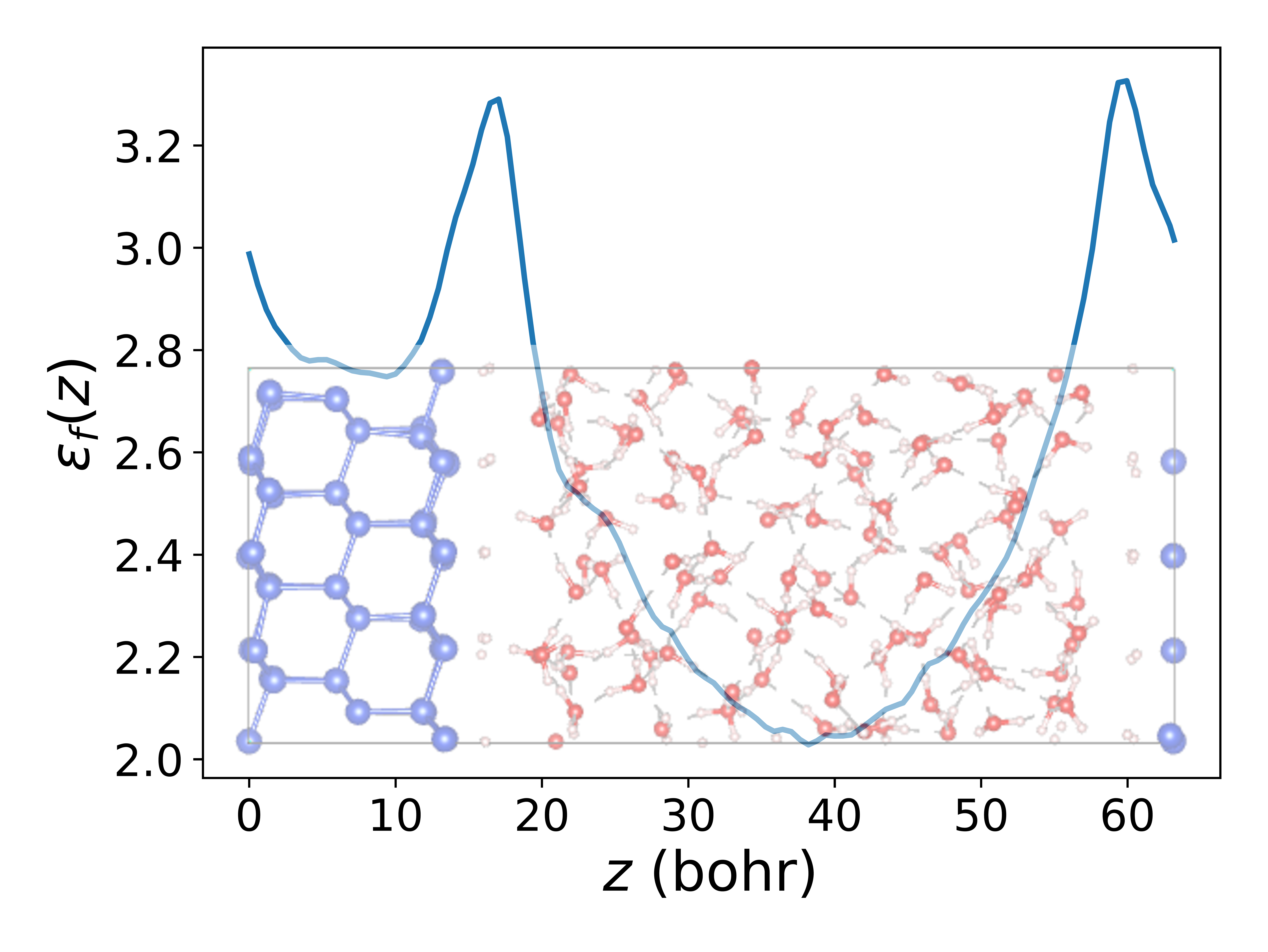}
    \end{subfigure}
\caption{Accuracy of ML-BSE using (a) a position-dependent model with 108 parameters ($f^{\text{ML}}_{p108}(z)$), and (b) $\epsilon^{\text{ML}}_f(z)$ corresponding to the $f^{\text{ML}}(z)$ profile used to compute (a), for a snapshot representing a H-Si/water interface (see text). The RMSE value between the ML-BSE and FF-BSE spectra is 0.059.}
\label{fig:sihwat_sf_slice}
\end{figure}

Comparing Figure~\ref{fig:sihwat_sf_cnn7} and Figure~\ref{fig:sihwat_sf_slice}, we find that a position-dependent model is necessary to obtain sufficiently accurate absorption spectra for the H-Si/water interface.

\subsubsection*{Interpretation of the $f(\mathbf{r})$ profile}
\begin{figure}[H]
\centering
\centering
    \begin{subfigure}{0.225\textwidth}
    \caption{}
    \includegraphics[width=\linewidth]{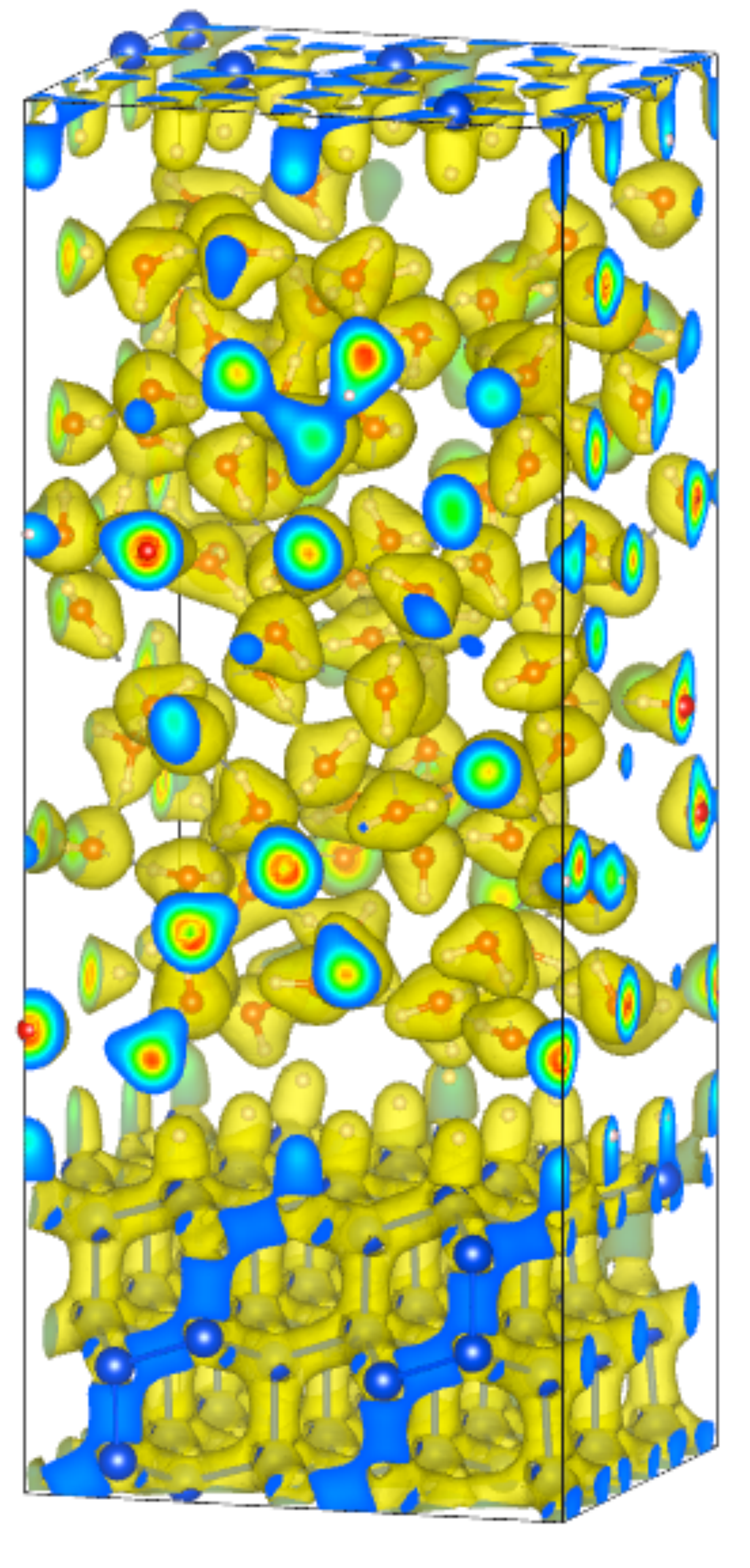}
    \end{subfigure}
    \hfill
    \begin{subfigure}{0.24\textwidth}
    \caption{}
    \includegraphics[width=\linewidth]{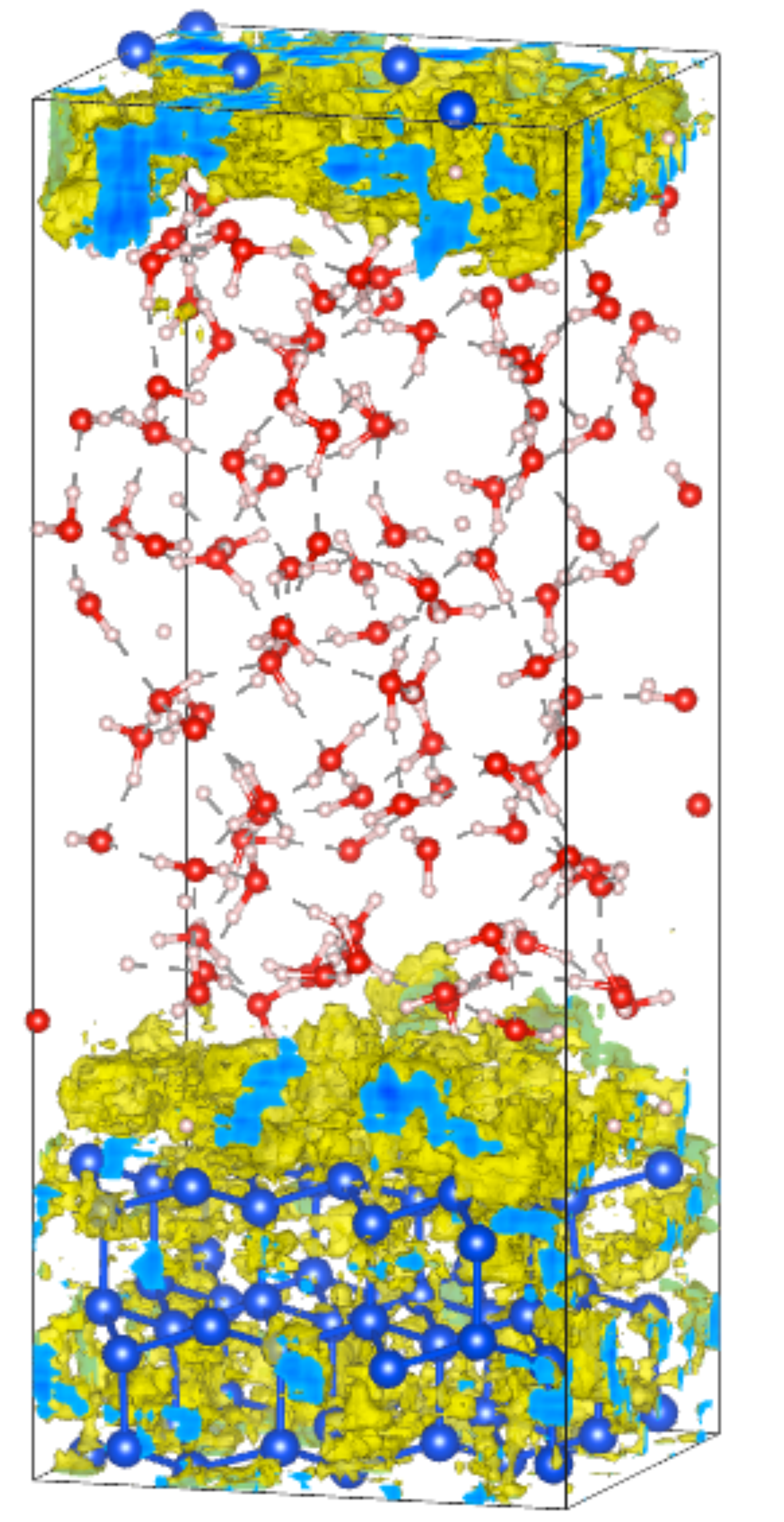}
    \end{subfigure}
        \hfill
    \begin{subfigure}{0.22\textwidth}
    \caption{}
    \includegraphics[width=\linewidth]{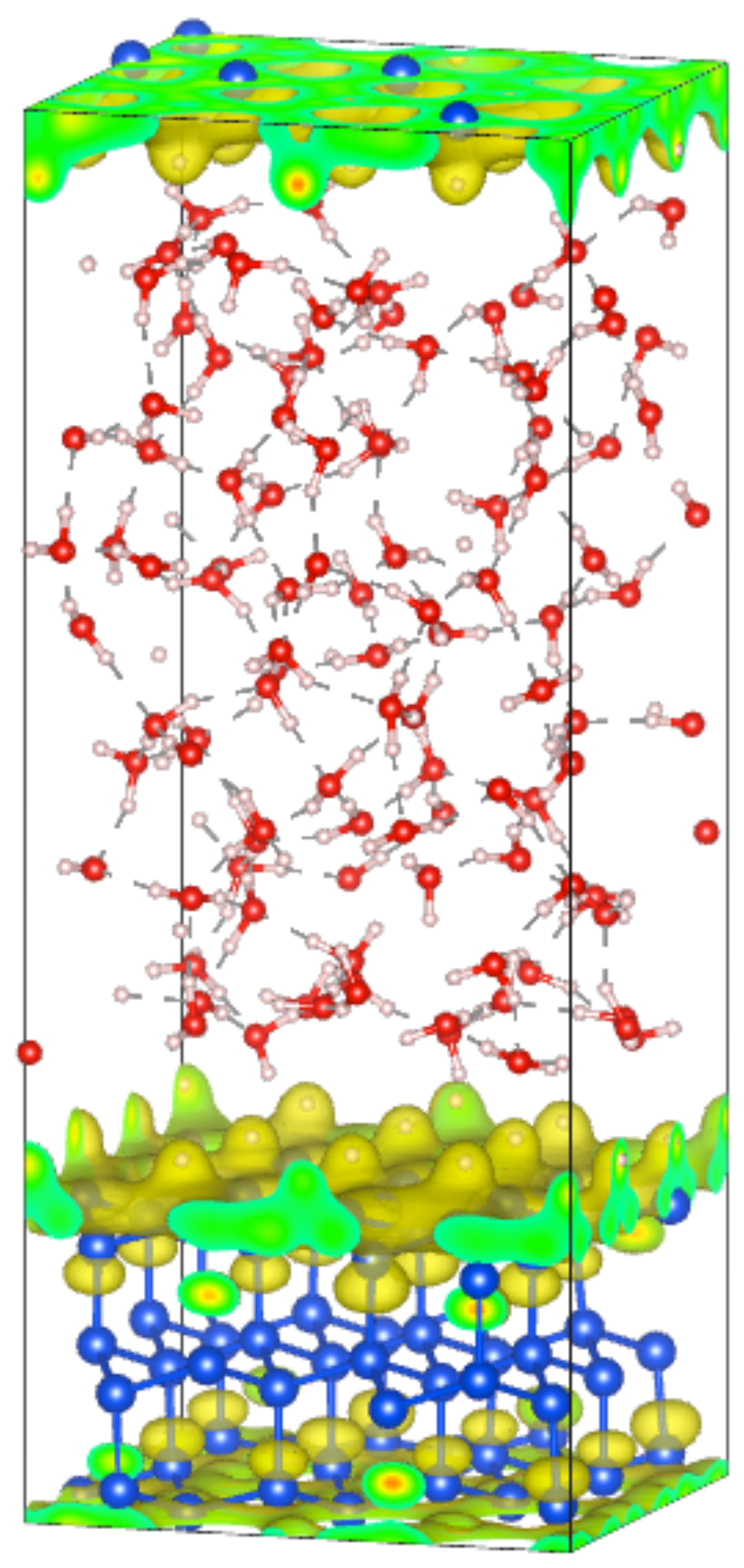}
    \end{subfigure}
        \hfill
    \begin{subfigure}{0.22\textwidth}
    \caption{}
    \includegraphics[width=\linewidth]{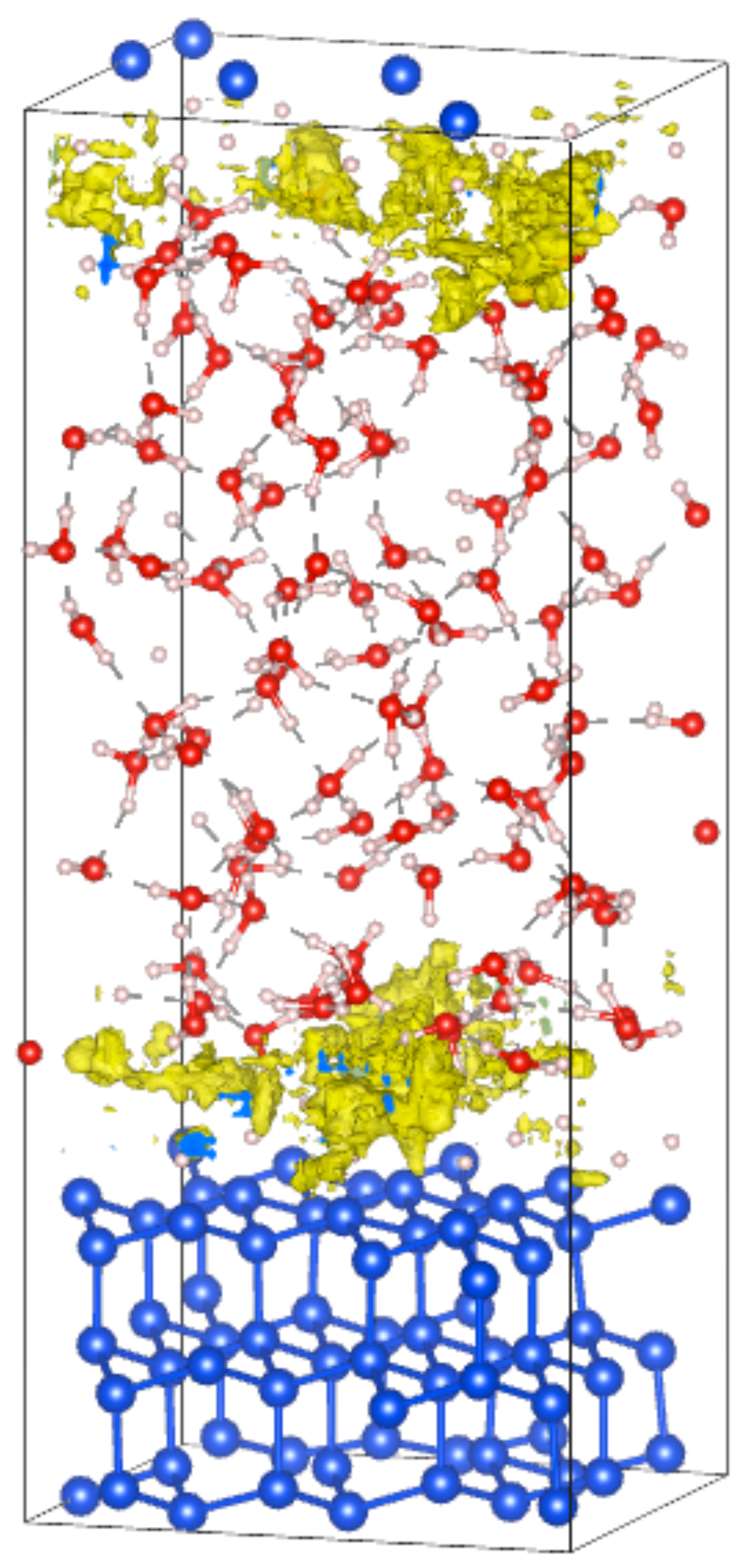}
    \end{subfigure}
\caption{(a) Charge density, isovalue 0.050; (b) $f^{\text{Avg}}(\mathbf{r})$ from averaging using a moving window of $8\times8\times8$, isovalue -0.644; (c) PDEP eigenpotential corresponding to the most negative eigenvalue, isovalue 3.64; (d) $f^{\text{Avg}}_1(\mathbf{r})$ from averaging using a moving window of $8\times8\times8$, isovalue -0.370. The isosurface at the isovalue (i.e. value that has the greatest absolute value) is in yellow. The 3D visualization was rendered in VESTA (version 3.4.0).\cite{momma2011vesta}} 
\label{fig:sihwat_charge_f}
\end{figure}

In Figure~\ref{fig:sihwat_charge_f} we show the charge density (a), $f^{\text{Avg}}(\mathbf{r})$ (defined in the main text) (b), PDEP eigenpotential corresponding to the most negative eigenvalue (c), and the component of $f^{\text{Avg}}(\mathbf{r})$ that corresponds to the most negative PDEP eigenvalue ($f_1^{\text{Avg}}(\mathbf{r})$, defined in Eq.~\ref{eq:f_pdep_i} of the main text) (d) for the H-Si/water interface. This shows that the maxima of 
$\epsilon^{\text{ML}}_{f}(z)$ (Figure~\ref{fig:sihwat_sf_slice}(b)) at the interfaces stem from the contribution of the PDEP eigenpotential with the most negative eigenvalue.

\subsubsection*{3D grid model from ML}
\begin{figure}[H]
\centering
\includegraphics[width=0.5\linewidth]{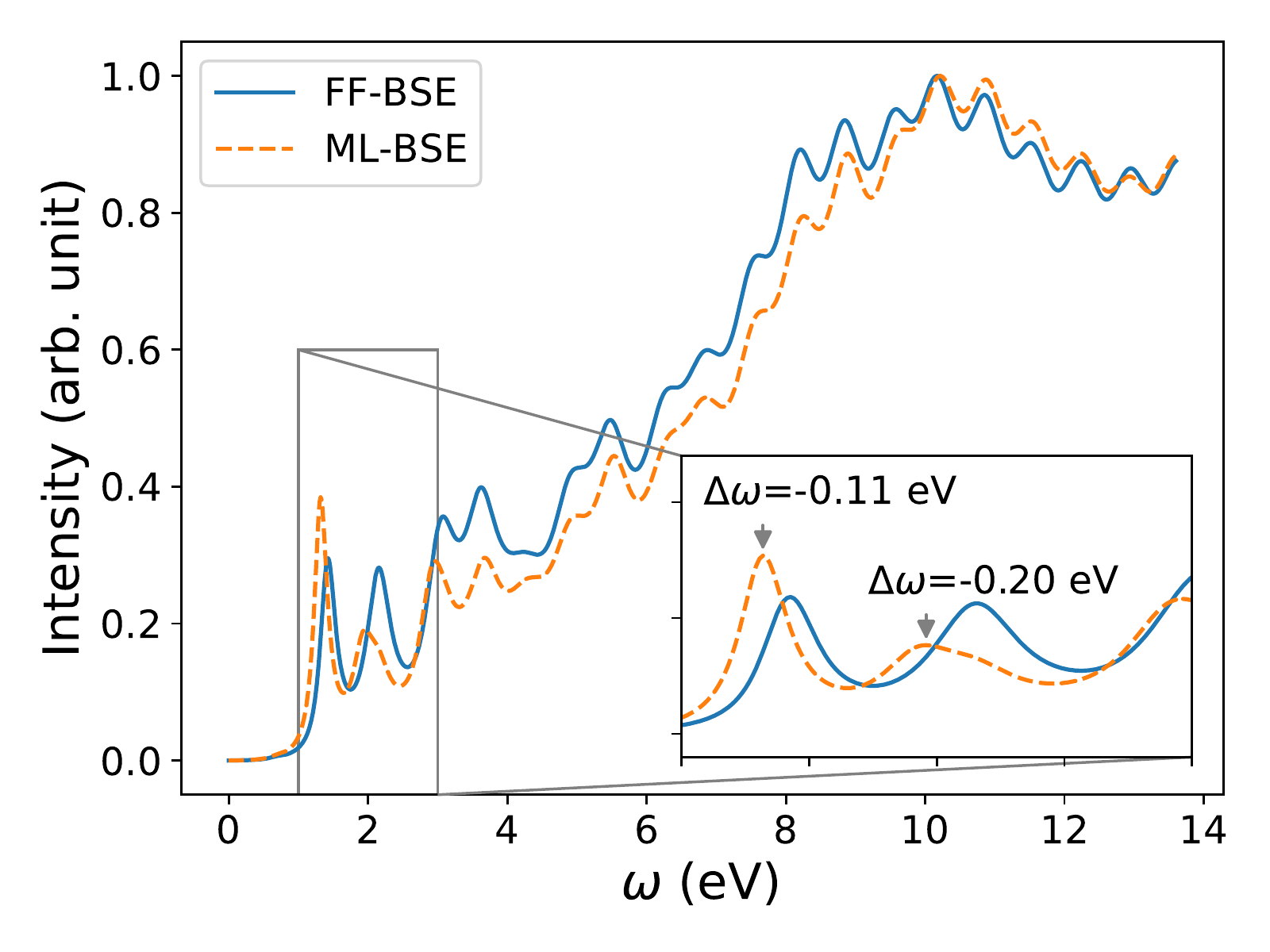}
\caption{Accuracy of ML-BSE using the 3D grid model of the H-Si/water interface. Each sub-domain is a cube, and the side length of each sub-domain is 2.6 \AA. The RMSE value between the ML-BSE and FF-BSE spectra is 0.059.}
\label{fig:sihwat_3D}
\end{figure}

In Figure~\ref{fig:sihwat_3D} we show that a 3D grid model $f^{\text{ML}}(\mathbf{r})$ yields an accurate ML-BSE spectrum for the H-Si/water interface. The 3D grid model has the same RMSE as the $z$-dependent model, and for the lowest-energy peak $\Delta\omega=-0.11$ eV, close to the $-0.08$ eV for the $z$-dependent model. The peak position from 3D grid model is slightly less accurate than the $z$-dependent model largely because the 3D grid model is more susceptible to the local variation in the training data.

\subsubsection*{Transferability of position-dependent models}
\begin{figure}[H]
\centering
    \begin{subfigure}{0.45\textwidth}
    \caption{}
    \includegraphics[width=\linewidth]{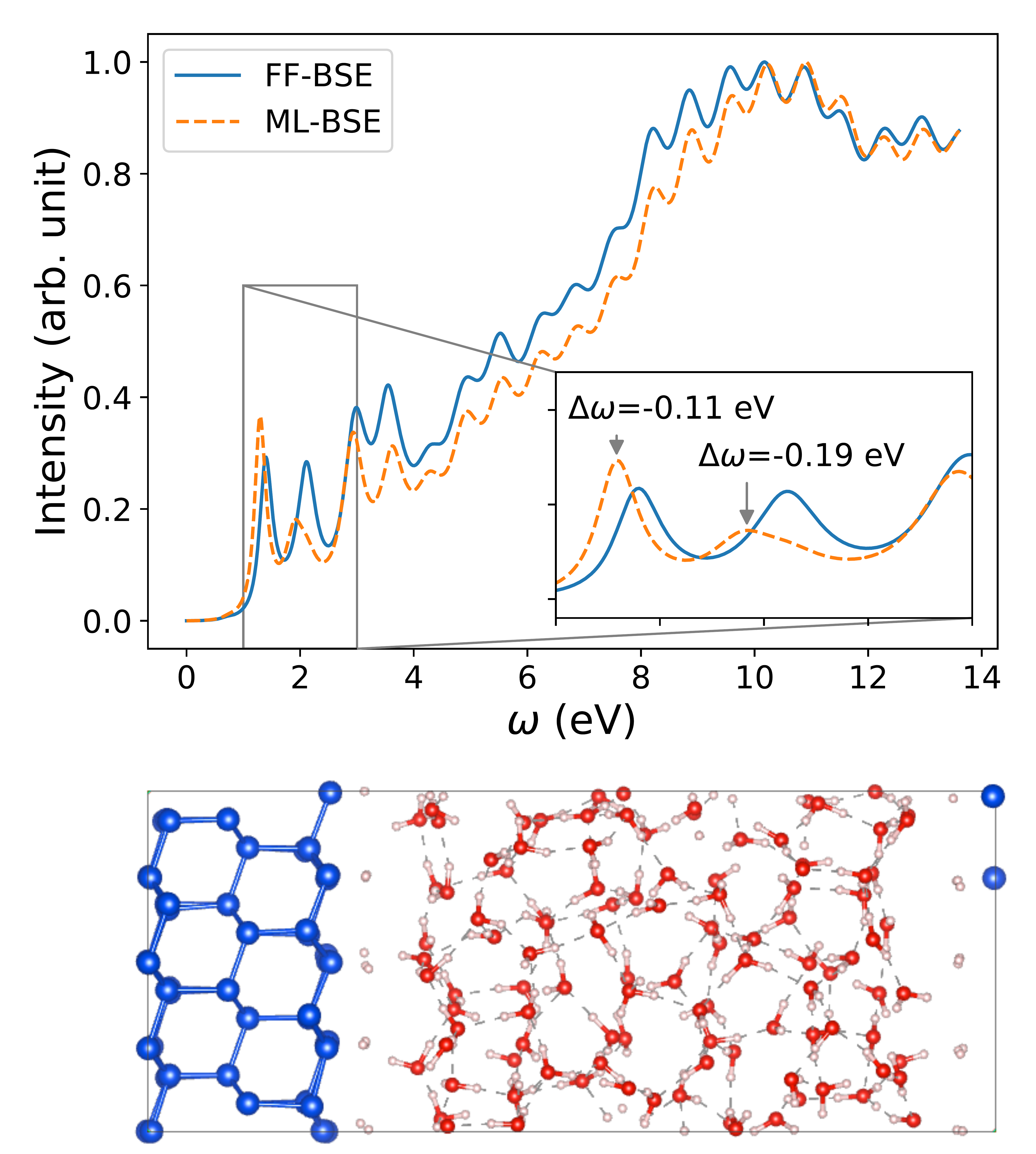}
    \end{subfigure}
    \hfill
    \begin{subfigure}{0.45\textwidth}
    \caption{}
    \includegraphics[width=\linewidth]{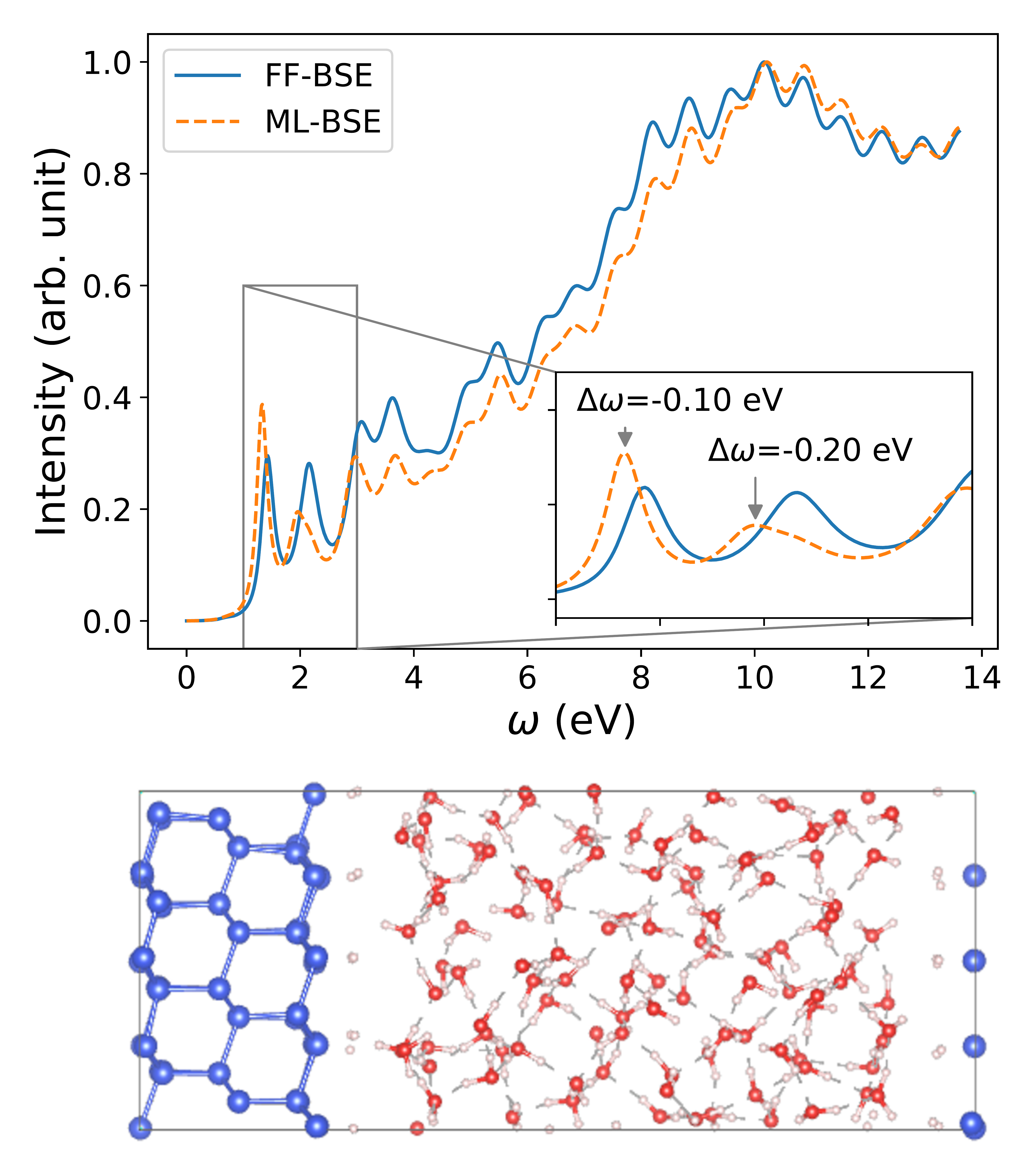}
    \end{subfigure}
\caption{Accuracy of ML-BSE using the $z$-dependent model $f^{\text{ML}}_{p216}(z)$ for the H-Si/water interface trained from a given snapshot (bottom of panel (b)) applied on a different snapshot (bottom of panel (a)). The spectrum corresponding to each snapshot is above its structural representation. The RMSE values is 0.064 for (a) and 0.059 for (b).} 
\label{fig:sihwat_5_20}
\end{figure}

To test the transferrability of the $f^{\text{ML}}(z)$ model across different snapshots, we consider a $f^{\text{ML}}(z)$ model, $f^{\text{ML}}_{p216}(z)$, more fine-grained than $f^{\text{ML}}_{p2}(z)$ and $f^{\text{ML}}_{p108}(z)$. The use of this more fine-grained model is to test whether overfitted $f^{\text{ML}}(z)$ models can still be transferrable across different snapshots. As shown in Figure~\ref{fig:sihwat_5_20}, the $z$-dependent model for the Si/water interface is transferable between different snapshots of the Si/water interface. 

\subsection{COOH-Si/water interface (hydrophilic interface)}
For the hydrophilic interface COOH-Si/water, we have found that the performance of the $f^{\text{ML}}_{p108}(z)$ model is also similar to $f^{\text{ML}}_{p2}(z)$ (Figure~\ref{fig:siwat_2interfaces_2par}(b) of the main text), consistent with our results for the H-Si/water interface.

\begin{figure}[H]

\centering
    \begin{subfigure}{0.45\textwidth}
    \caption{}
    \includegraphics[width=\linewidth]{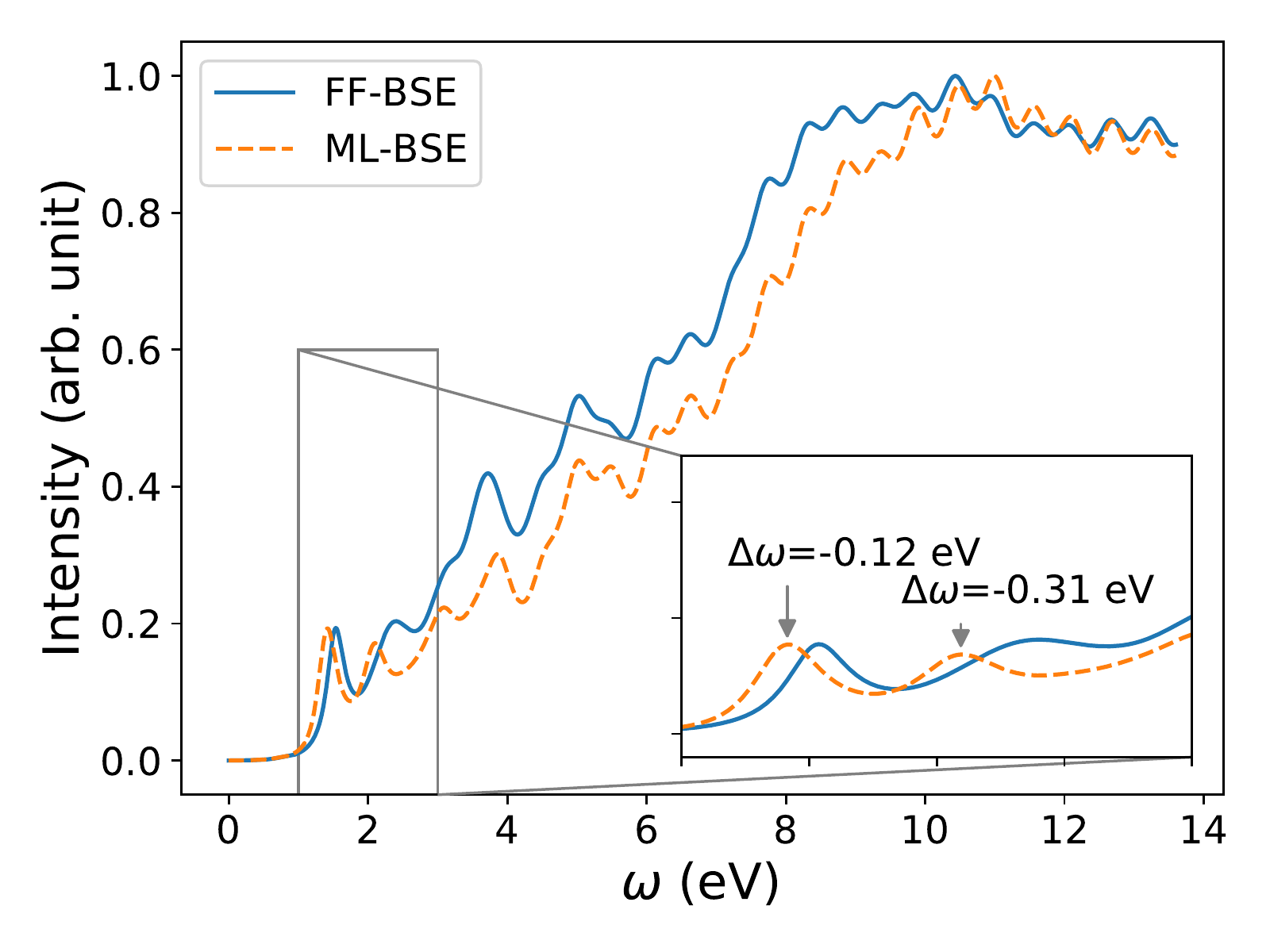}
    \end{subfigure}
    \hfill
    \begin{subfigure}{0.45\textwidth}
    \caption{}
    \includegraphics[width=\linewidth]{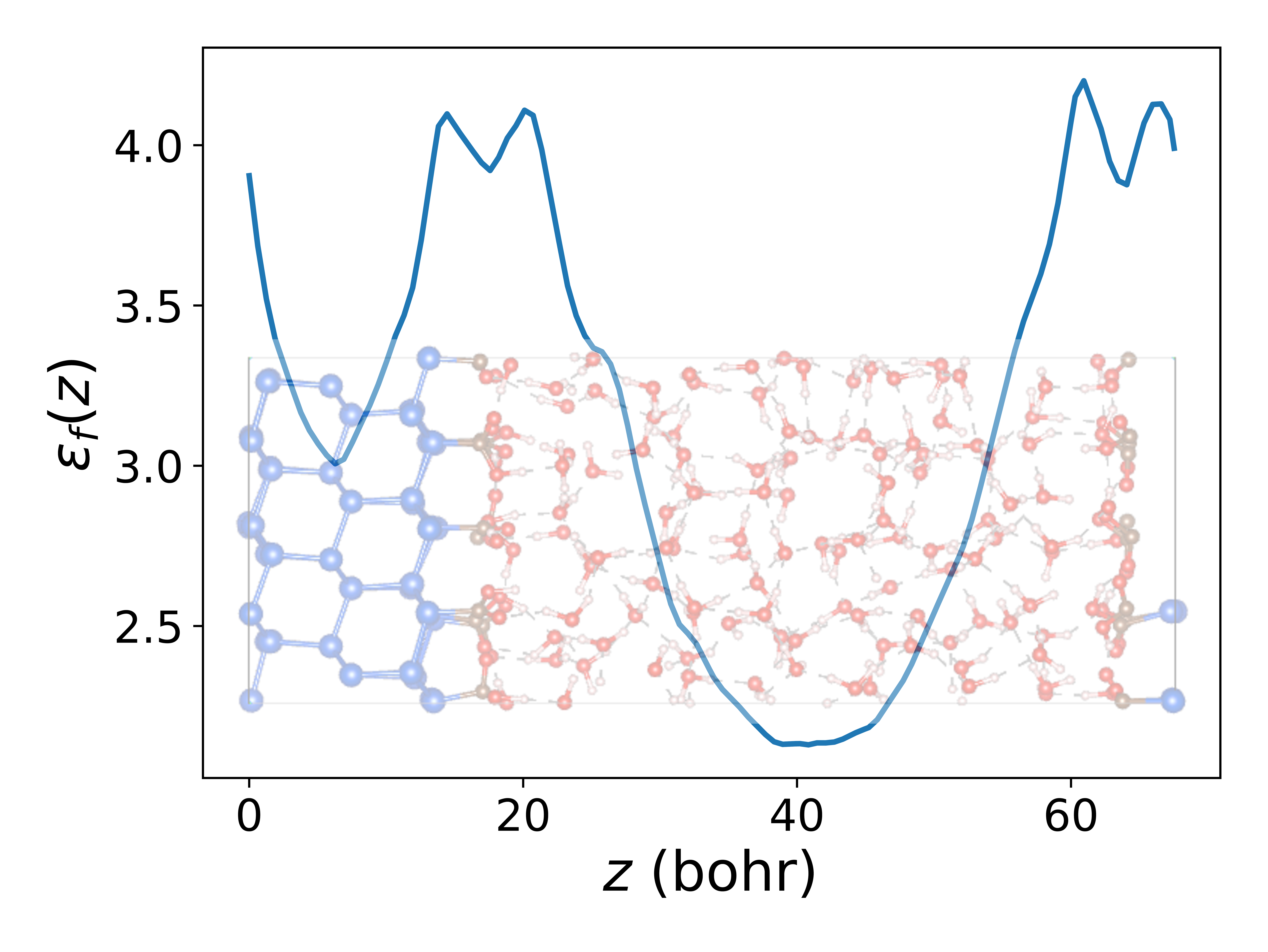}
    \end{subfigure}
\caption{Accuracy of ML-BSE using (a) a position-dependent model with 108 parameters ($f^{\text{ML}}_{p108}(z)$), and (b) $\epsilon^{\text{ML}}_f(z)$ corresponding to the $f^{\text{ML}}(z)$ profile used to compute (a), for a slab representing a COOH-Si terminated surface interfaced with water. The RMSE value between the ML-BSE and FF-BSE spectra is 0.078.}
\label{fig:sicoohwat_sf_slice}
\end{figure}

\subsection{Si clusters}

\begin{figure}[H]
\centering
    \begin{subfigure}{0.45\textwidth}
    \caption{}
    \includegraphics[width=\linewidth]{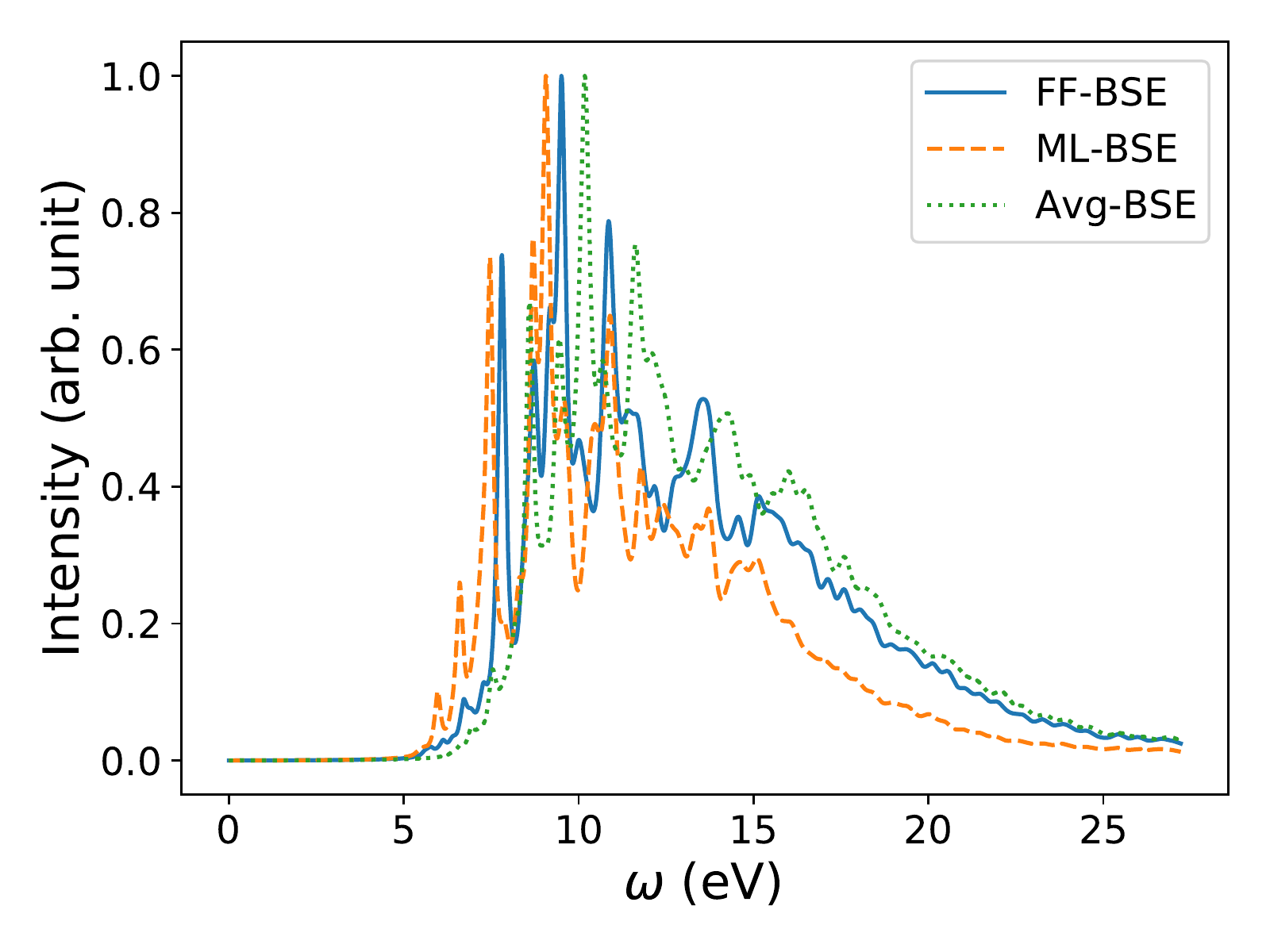}
    \end{subfigure}
    \hfill
    \begin{subfigure}{0.45\textwidth}
    \caption{}
    \includegraphics[width=\linewidth]{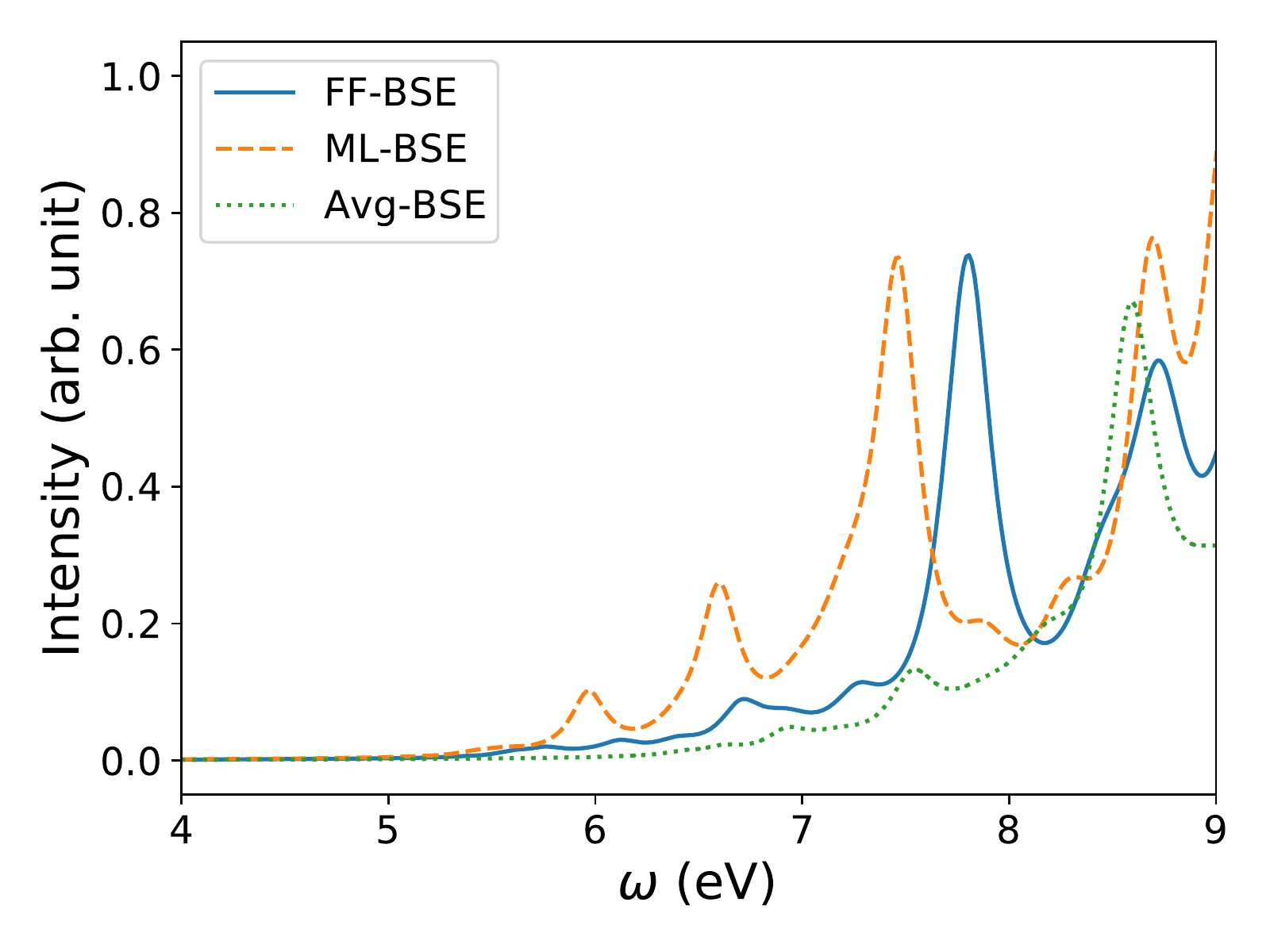}
    
    \end{subfigure}

\caption{Comparison of ML-BSE and Avg-BSE to FF-BSE for Si$_{10}$H$_{16}$ (30 \AA~ cell). ML-BSE spectra are obtained using $f^{\text{ML}}=-0.28$, and Avg-BSE is from using $f^{\text{Avg}}=-0.54$.  In (b) we show the same data as in (a) on a smaller energy interval (4 to 9 eV).}
\label{fig:si10h16_30ang_ML_avg}
\end{figure}

For Si clusters, we were able to obtain a linear regression model (that results in a global scaling factor $f^{\text{ML}}=-0.28$) and convolutional models in the same way as for homogeneous systems. However, the global scaling factor we obtained from linear regression does not yield accurate spectra, as shown in Figure~\ref{fig:si10h16_40ang_ml} of the main article. In addition, a global scaling factor derived from averaging $\Delta\tau/\tau^{u}$, $f^{\text{Avg}}=-0.54$, is different from from $f^{\text{ML}}$ and yields an even worse spectrum , as shown in Figure~\ref{fig:si10h16_30ang_ML_avg}. The value of the scaling factor may be different for simulation cells of different sizes , even if the absorption spectra have converged with respect to the size of the simulation cell. For a cubic simulation cell 30 (40) \AA~ in length, $f^{\text{ML}}=-0.28$ ($-0.22$). 

The reason behind this behavior is that, although the charge density of the Si cluster decays rapidly as a function of the distance from the cluster, 
the ratio $\Delta\tau/\tau^{u}$ does not. To have an ML model robust to the size of the simulation cell, we impose a threshold to the Si cluster data so that small elements (e.g. elements smaller than a chosen threshold) in $\tau^u_{vv^\prime}$ are set to 0. Then we use this data to obtain our ML models. 

\begin{figure}[H]
\centering
    \begin{subfigure}{0.45\textwidth}
    \caption{}
    \includegraphics[width=\linewidth]{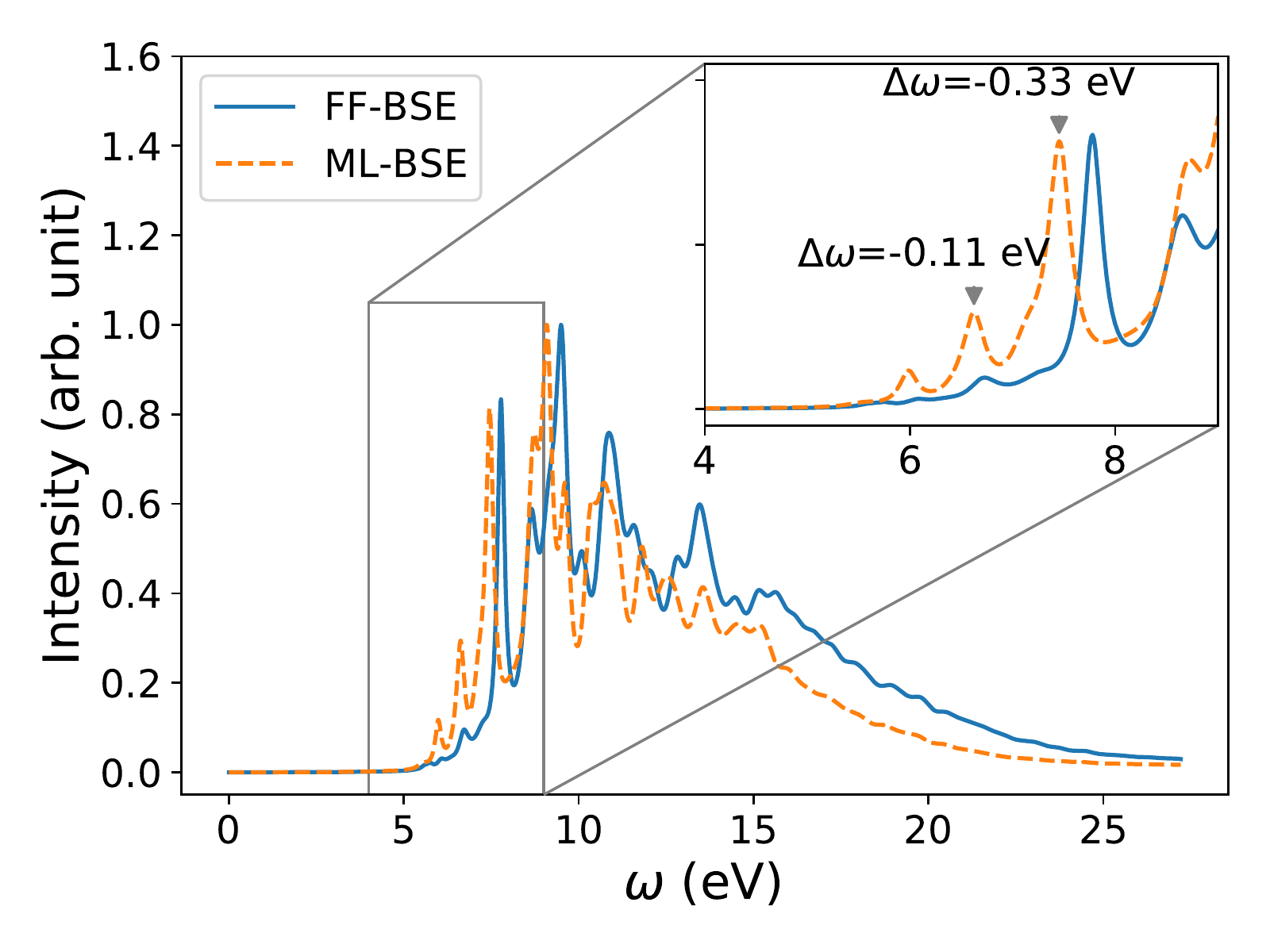}
    \end{subfigure}
    \hfill
    \begin{subfigure}{0.45\textwidth}
    \caption{}
    \includegraphics[width=\linewidth]{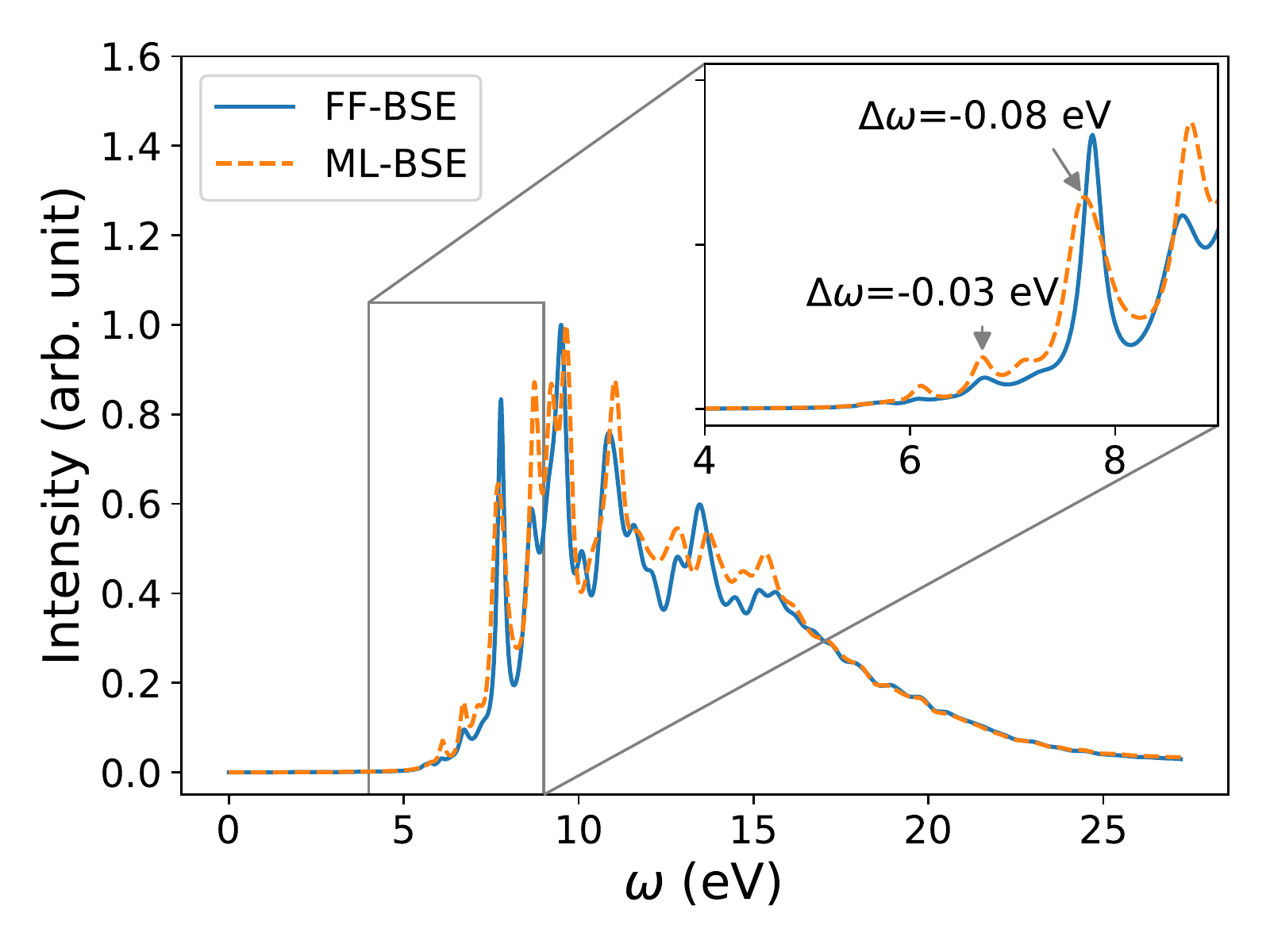}
    
    \end{subfigure}

\caption{Accuracy of ML-BSE spectra of Si$_{10}$H$_{16}$ (40 \AA~ cell) obtained using (a) a global scaling factor, and (b) a convolutional model (filter size $(7,7,7)$) from a smaller cell (30 \AA~ cell). The models are obtained by considering only regions of $\tau^u_{vv^\prime}$ above a charge density threshold ($10^{-5}$ times the largest charge density). The RMSE value between the FF-BSE and ML-BSE spectra are 0.119 for (a) and 0.061 for (b).}
\label{fig:si10h16_40ang_ml_rhocut100000}
\end{figure}

The spectra obtained from various models are shown in Figures~\ref{fig:si10h16_40ang_ml_rhocut100000}. Comparing Figures~\ref{fig:si10h16_40ang_ml_rhocut100000} and Figure~\ref{fig:si10h16_40ang_ml} of the main article, it is clear that imposing a threshold on the data improves the accuracy of the spectra. However, a global scaling factor model still performs poorly compared to convolutional models. The convolutional models give highly accurate absorption spectra regardless of whether the data are treated with a threshold. This suggests that convolutional models are robust to sparse data.

\begin{figure}[H]
\centering
    \begin{subfigure}{0.45\textwidth}
    \caption{}
    \includegraphics[width=\linewidth]{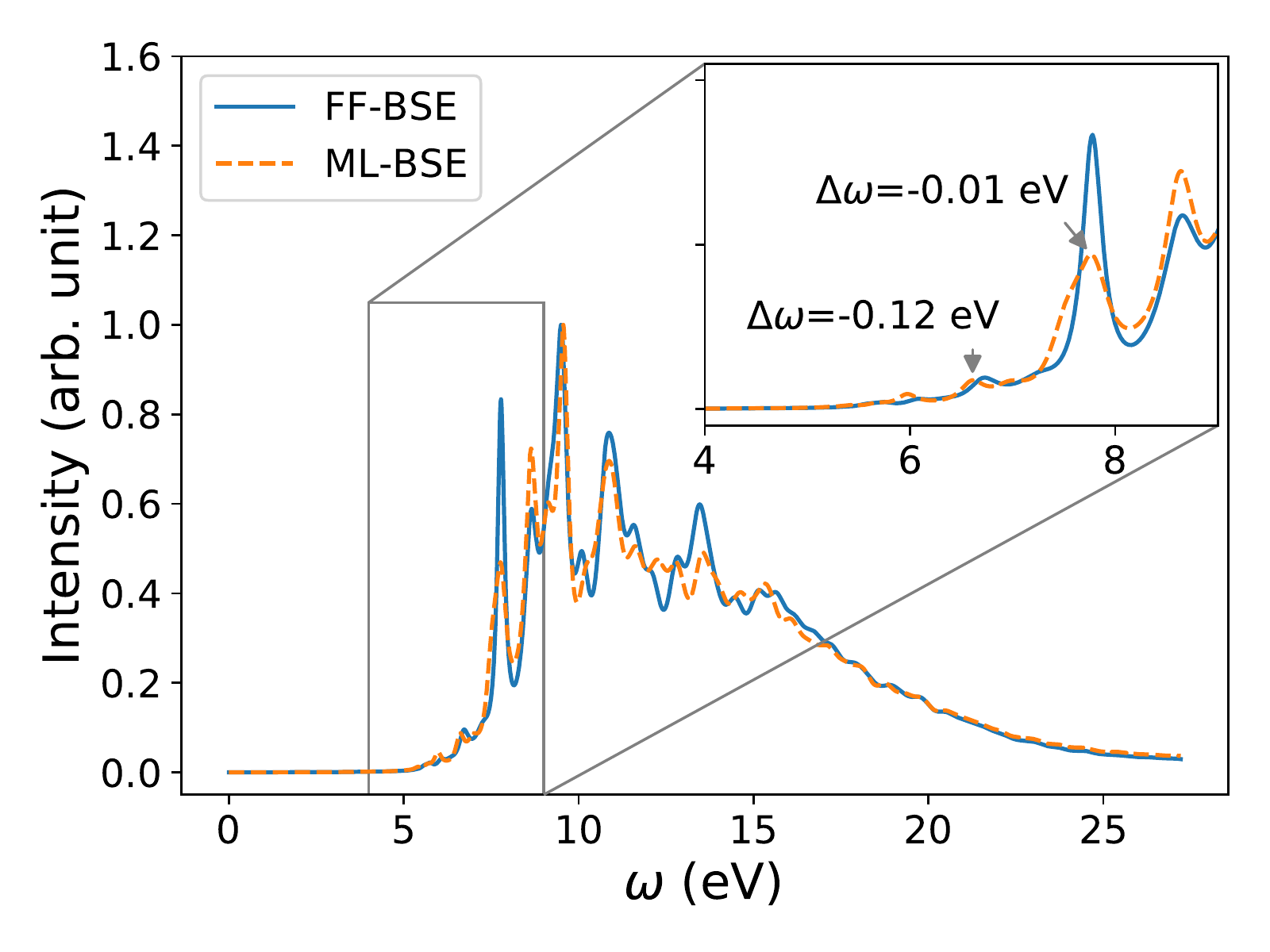}
    \end{subfigure}
    \begin{subfigure}{0.45\textwidth}
    \caption{}
    \includegraphics[width=\linewidth]{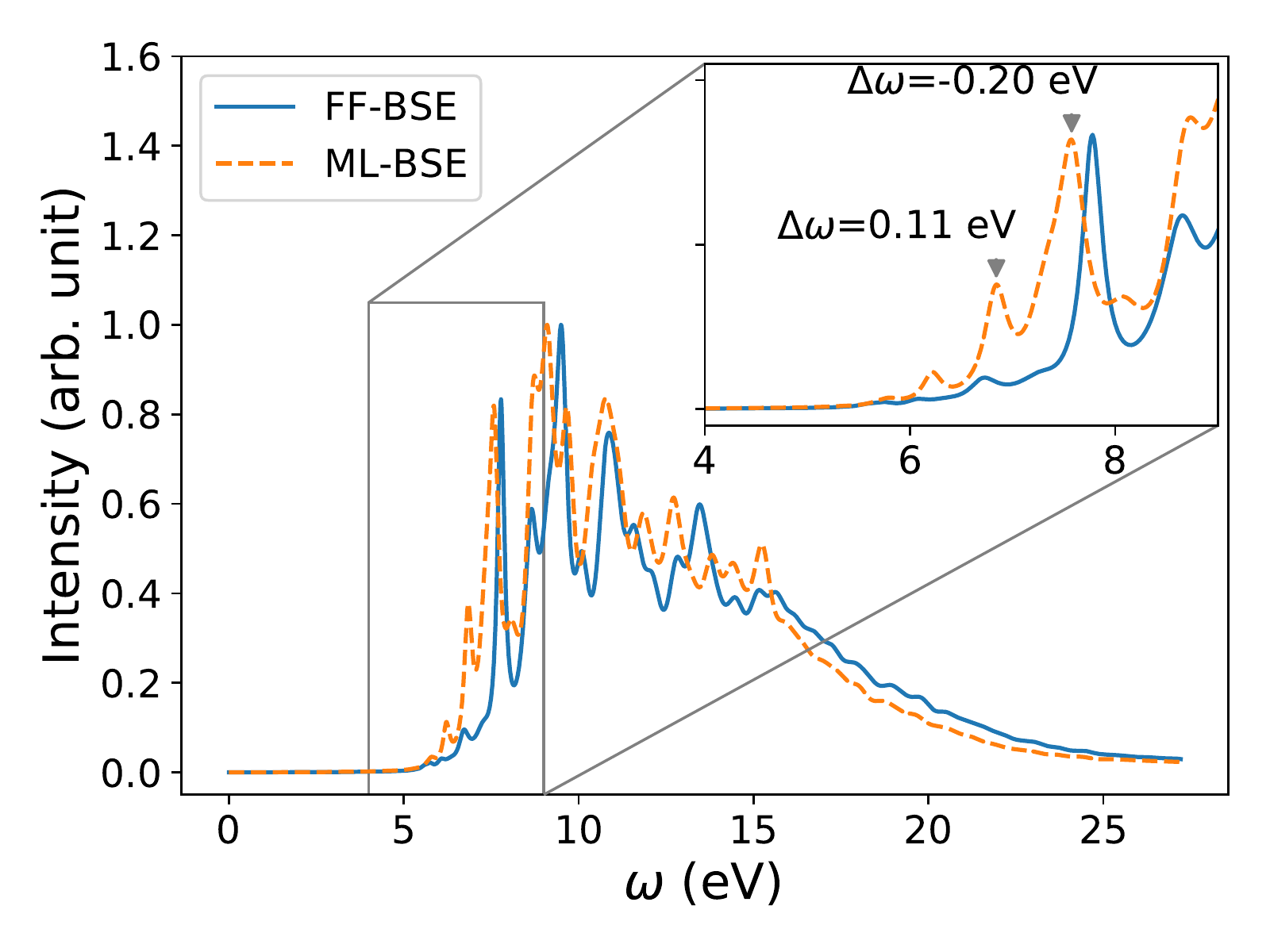}
    \end{subfigure}
    \begin{subfigure}{0.45\textwidth}
    \caption{}
    \includegraphics[width=\linewidth]{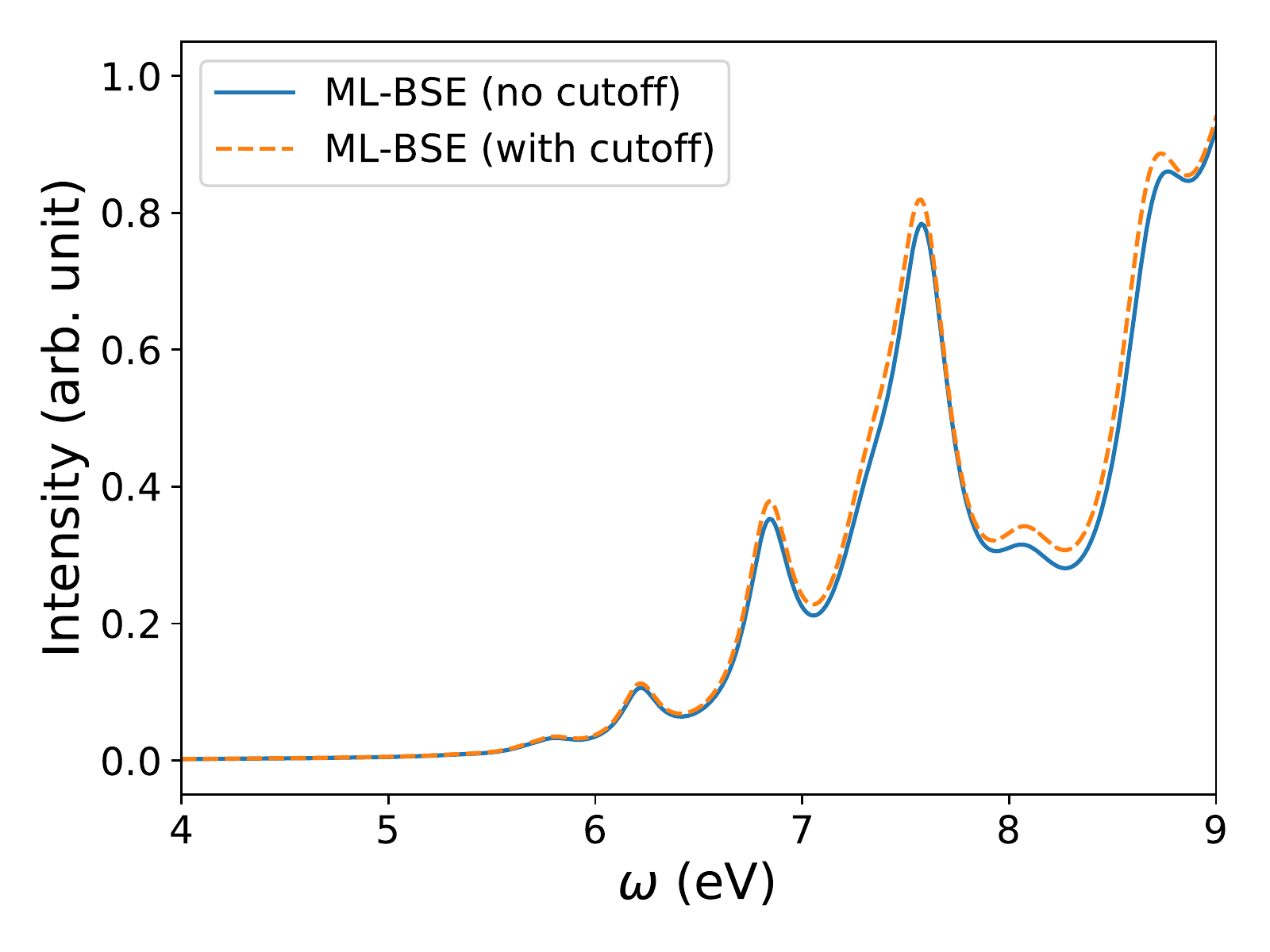}
    \end{subfigure}
    \hfill

\caption{Accuracy of ML-BSE spectra of Si$_{10}$H$_{16}$ (40 \AA~ cell) obtained from (a) a convolutional model (filter size $(7,7,7)$), and (b) a 3D grid model $f^{\text{ML}}(\mathbf{r})$. When the models are applied to obtain 
ML-BSE spectra in (a) and (b), only regions of $\tau^u_{vv^\prime}$ above a charge density threshold ($10^{-5}$ times the largest charge density) are considered. The RMSE value between the FF-BSE and ML-BSE spectra are 0.107 for (a) and 0.044 for (b). In (c) we compare ML-BSE spectra from the 3D grid model when the charge density threshold is applied (with cutoff) or not applied (no cutoff).}
\label{fig:si10h16_40ang_cg_cnn_rhocut100000}
\end{figure}

In the previous two paragraphs, we applied the threshold on the training data, and applied the resulting ML models without imposing a threshold. Another way to consider the effect of the vacuum is to use the original data in training, but use a threshold on $\tau^u_{vv^\prime}$ 
when the model is applied. Figure~\ref{fig:si10h16_40ang_cg_cnn_rhocut100000}(a) and Figure~\ref{fig:si10h16_40ang_ml}(a) show that, for convolutional models obtained using the original, un-thresholded data, imposing a threshold when applying the model can improve the accuracy of the spectrum. Figure~\ref{fig:si10h16_40ang_cg_cnn_rhocut100000}(b)(c) show that this will not improve much for the model based on scaling factors, even if a position-dependent model is used. This confirms that convolutional models are superior to models based on scaling factors.

\subsubsection*{Spectrum of Si$_{35}$H$_{36}$ at zero temperature}
\begin{figure}[H]
    \centering
     \includegraphics[width=0.5\linewidth]{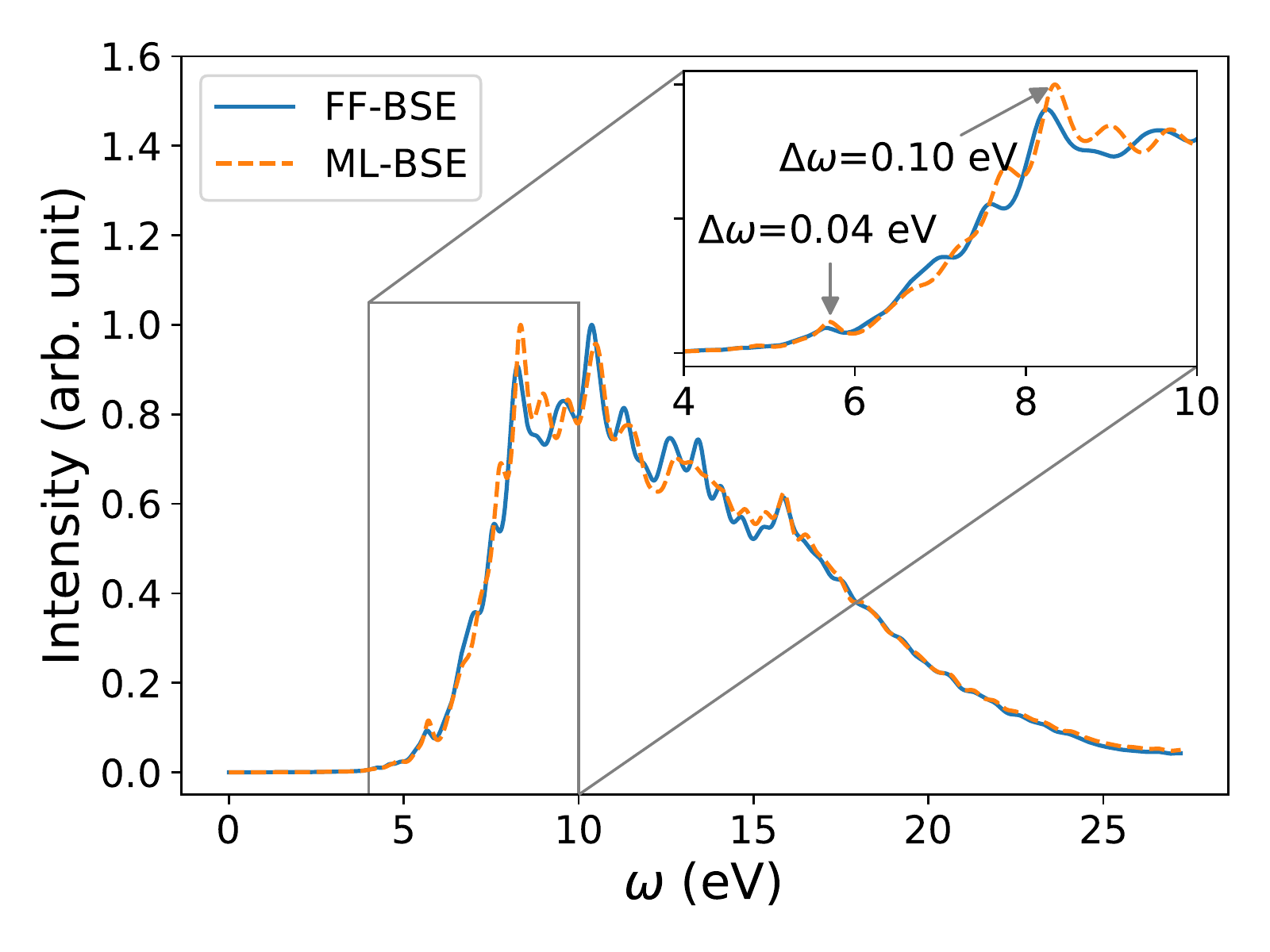}
    
    \caption{Accuracy of ML-BSE spectra obtained from a convolutional model (filter size $(7,7,7)$) for the 0 K geometrical configuration of Si$_{35}$H$_{36}$. The RMSE value between the ML-BSE and FF-BSE spectra is 0.029.}
    \label{fig:si35h36_25ang_0K_ml}
\end{figure}

The 0~K spectrum of Si$_{35}$H$_{36}$ differs from the 500~K spectra, as shown in Figure~\ref{fig:si35h36_25ang_0K_ml} and Figure~\ref{fig:si35h36_ave}.

\subsubsection*{Finite temperature spectra of Si$_{35}$H$_{36}$}
\begin{figure}[H]
    \centering
    
    \begin{subfigure}{0.45\textwidth}
    \caption{}
    \includegraphics[width=\linewidth]{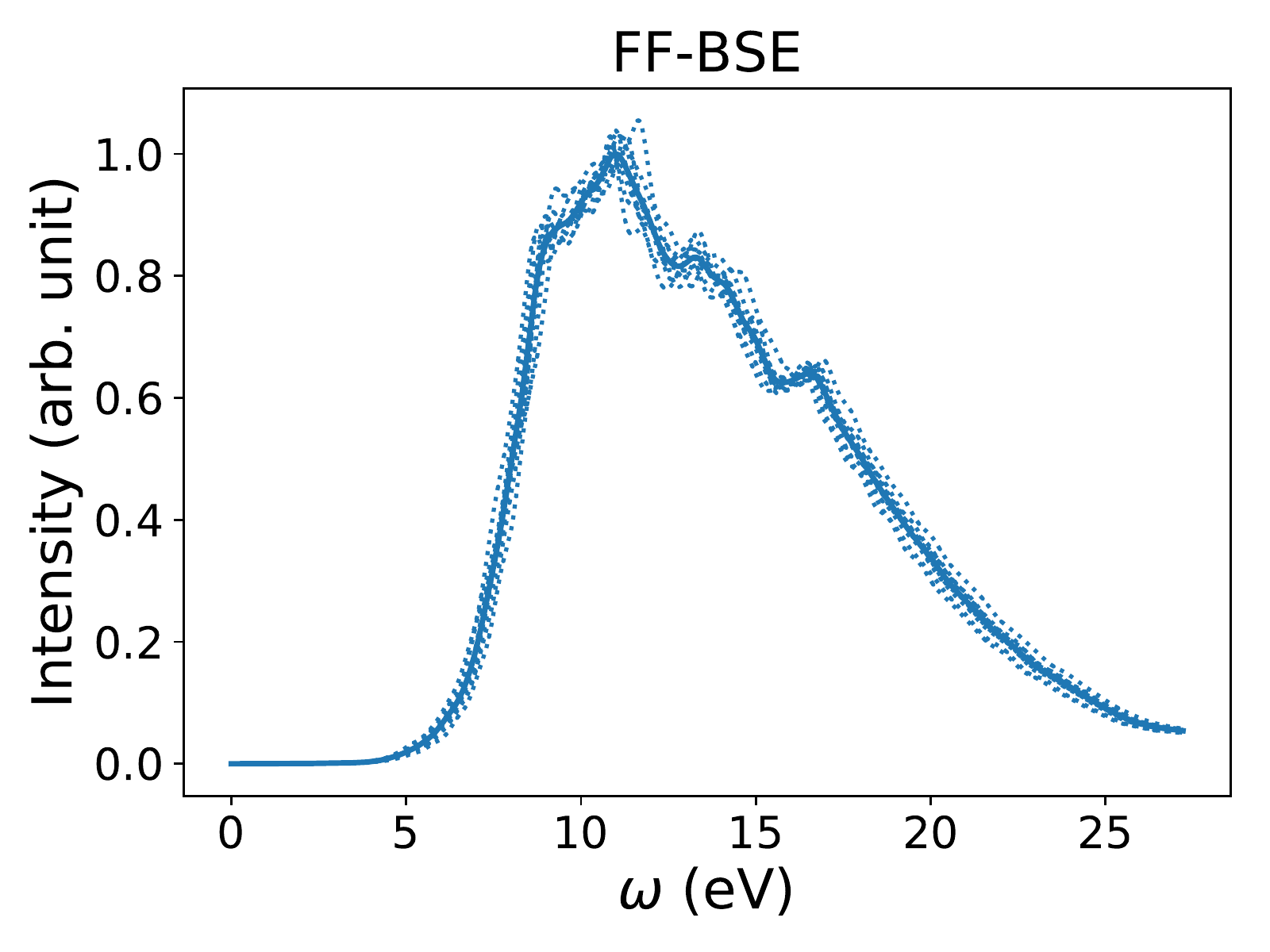}
    
    \end{subfigure}
    \hfill
    \begin{subfigure}{0.45\textwidth}
    \caption{}
    \includegraphics[width=\linewidth]{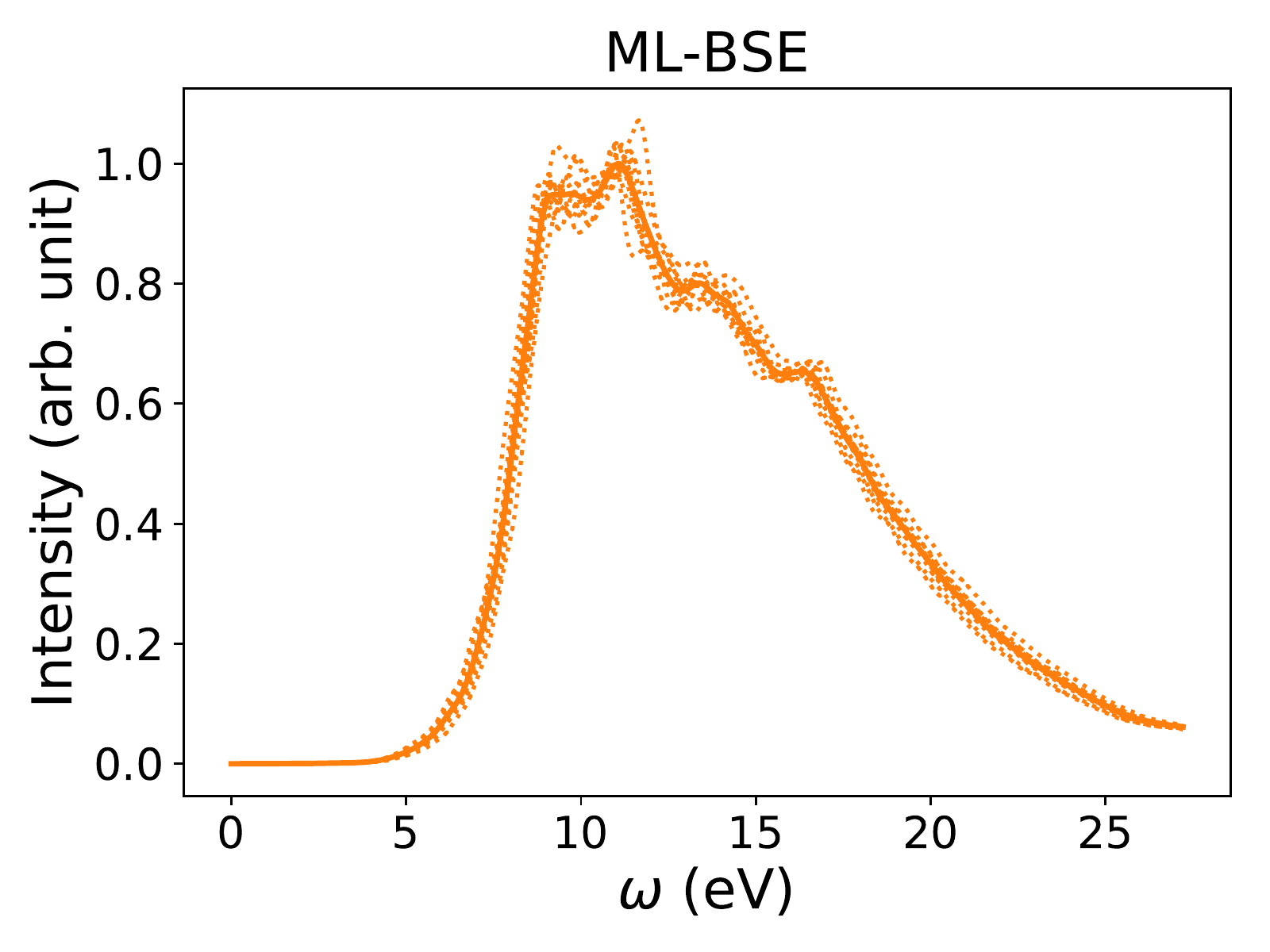}
    
    \end{subfigure}
    
    \caption{The absorption spectra of 10 individual snapshots of Si$_{35}$H$_{36}$ (dotted lines) and their averaged spectrum (solid line) from (a) FF-BSE and (b) ML-BSE calculations.}
    \label{fig:si35h36_ave_ind}
\end{figure}

The variations of individual spectra used to calculate the averaged spectrum of Si$_{35}$H$_{36}$ are shown in Figure~\ref{fig:si35h36_ave_ind}. The comparison between FF-BSE and ML-BSE spectra of each snapshot is in Figure~\ref{fig:si35h36_10}. The model used in ML-BSE uses a convolutional model $(7,7,7)$ trained on data obtained for the 0~K snapshot, the same model used in Figure~\ref{fig:si35h36_25ang_0K_ml}.

\begin{figure}[H]
\centering

\begin{subfigure}{0.45\textwidth}
\includegraphics[width=\linewidth]{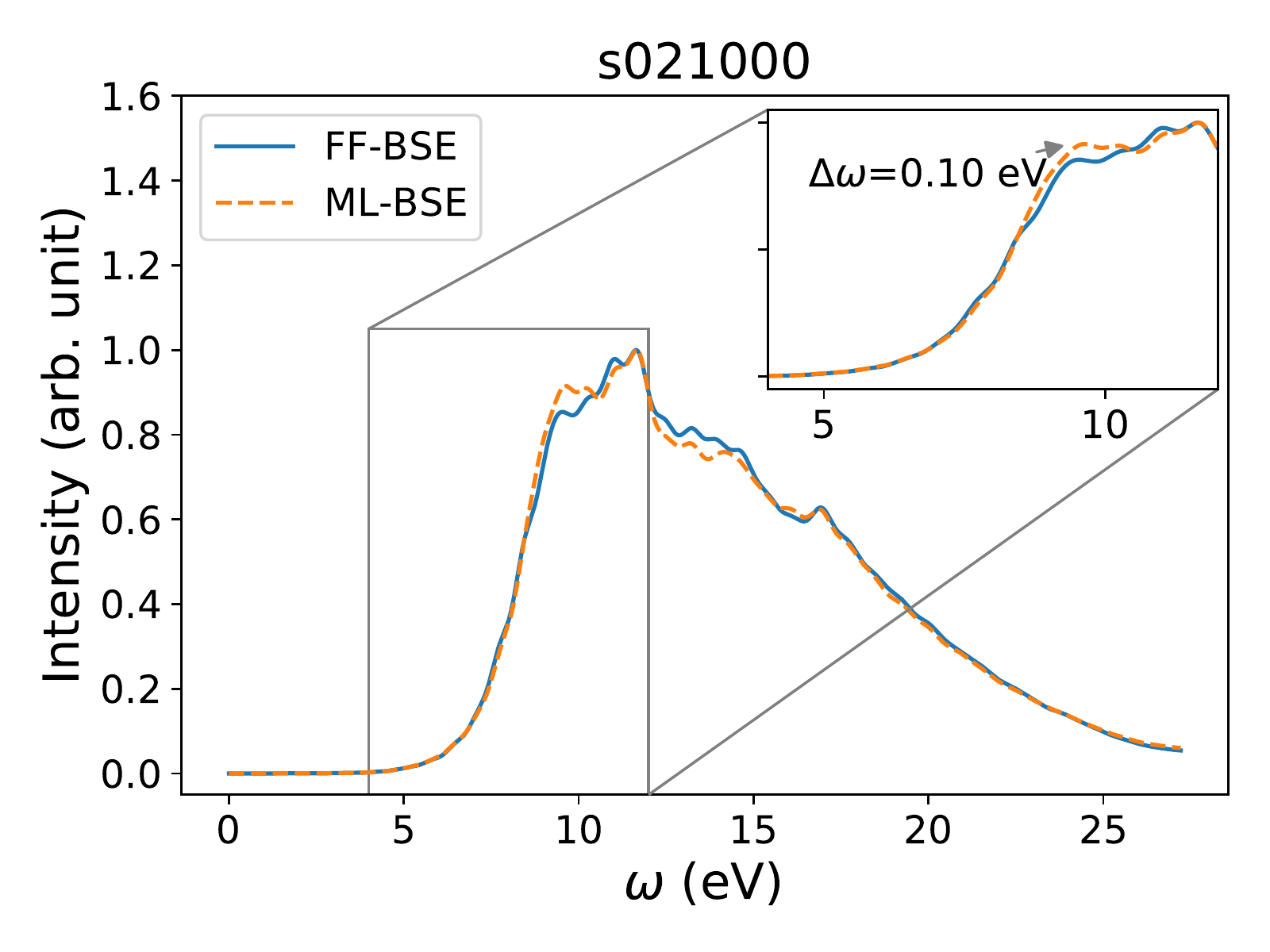}
\end{subfigure}
\hfill
\begin{subfigure}{0.45\textwidth}
\includegraphics[width=\linewidth]{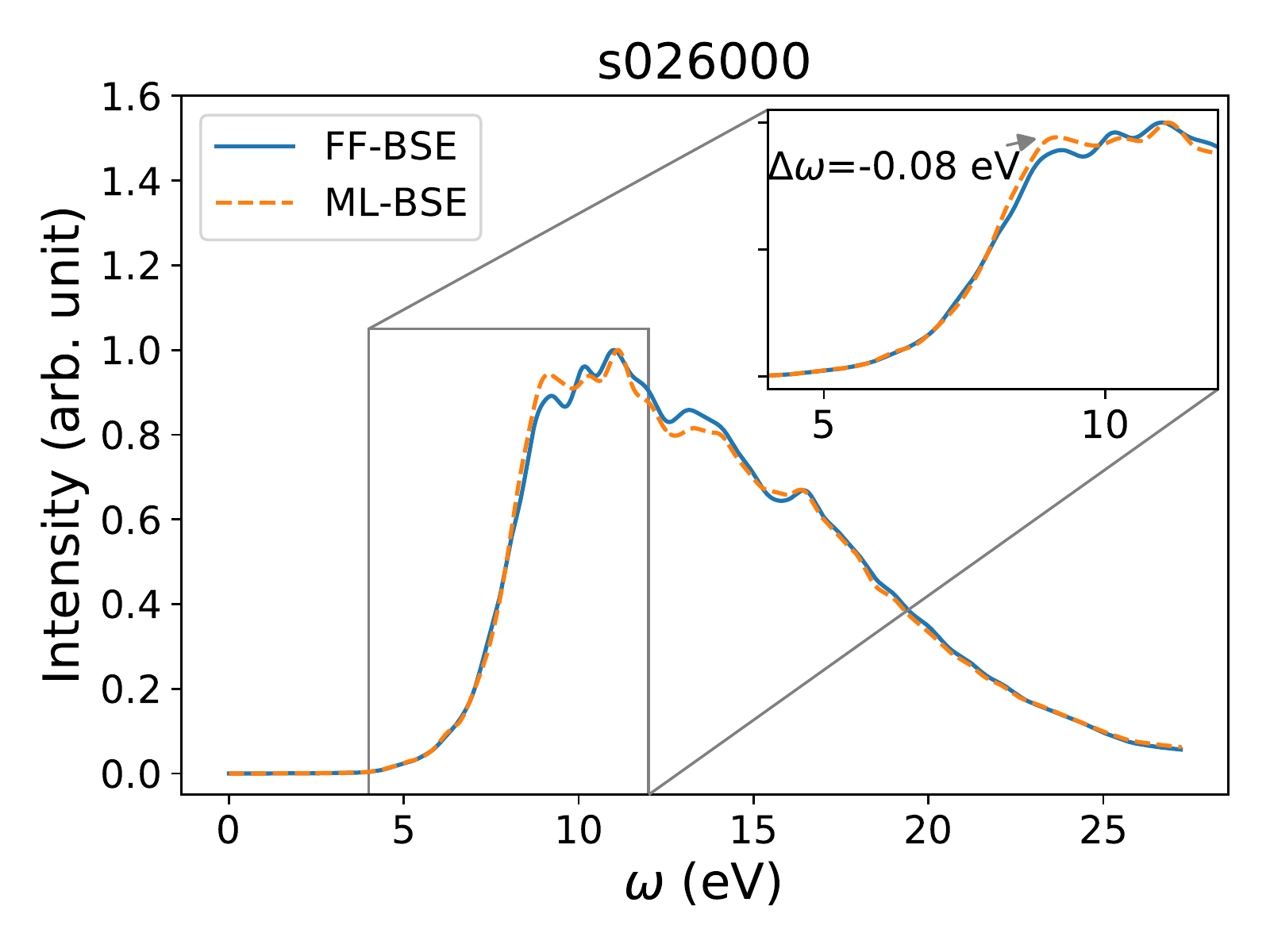}
\end{subfigure}
\hfill
\begin{subfigure}{0.45\textwidth}
\includegraphics[width=\linewidth]{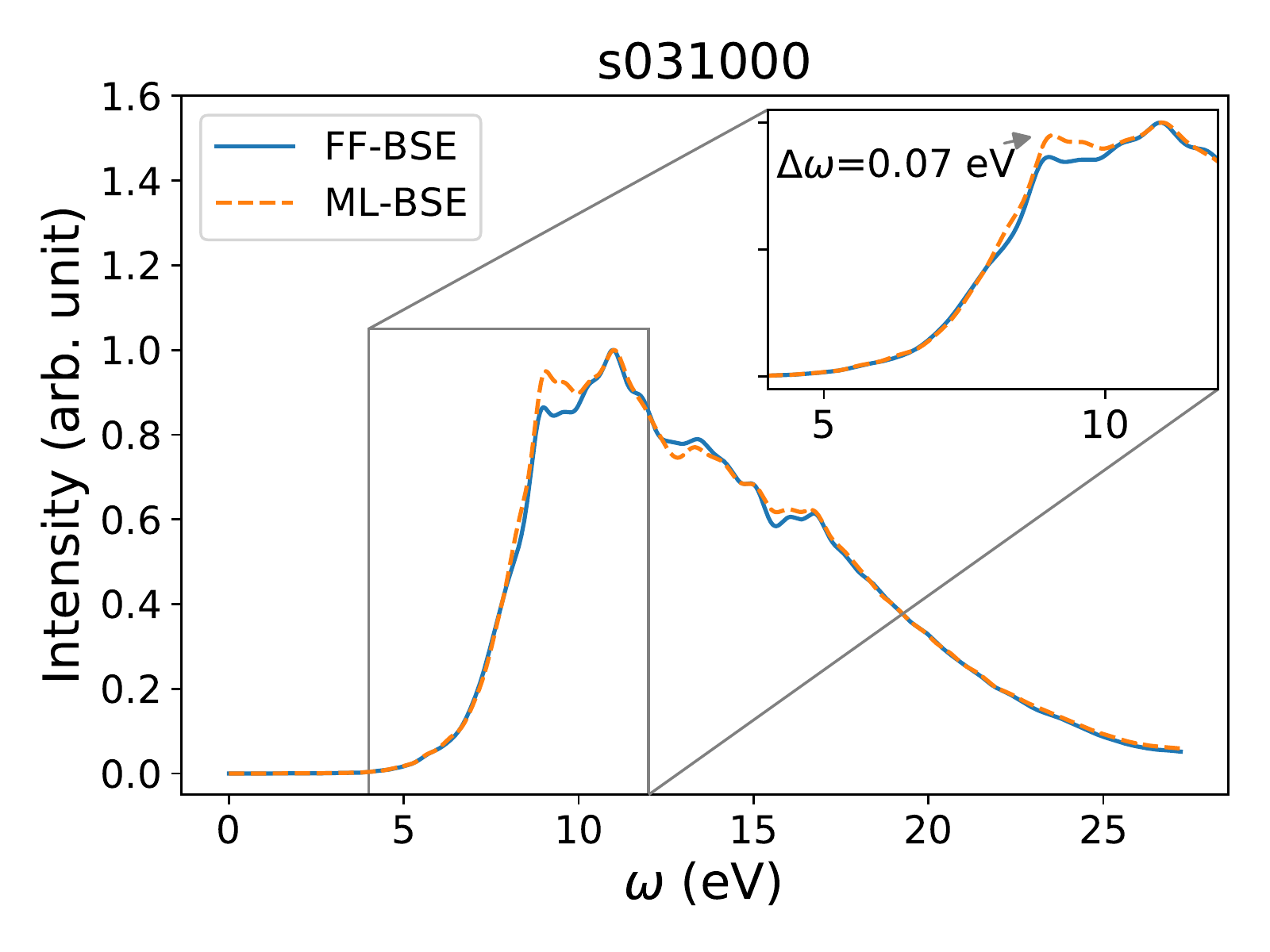}
\end{subfigure}
\hfill
\begin{subfigure}{0.45\textwidth}
\includegraphics[width=\linewidth]{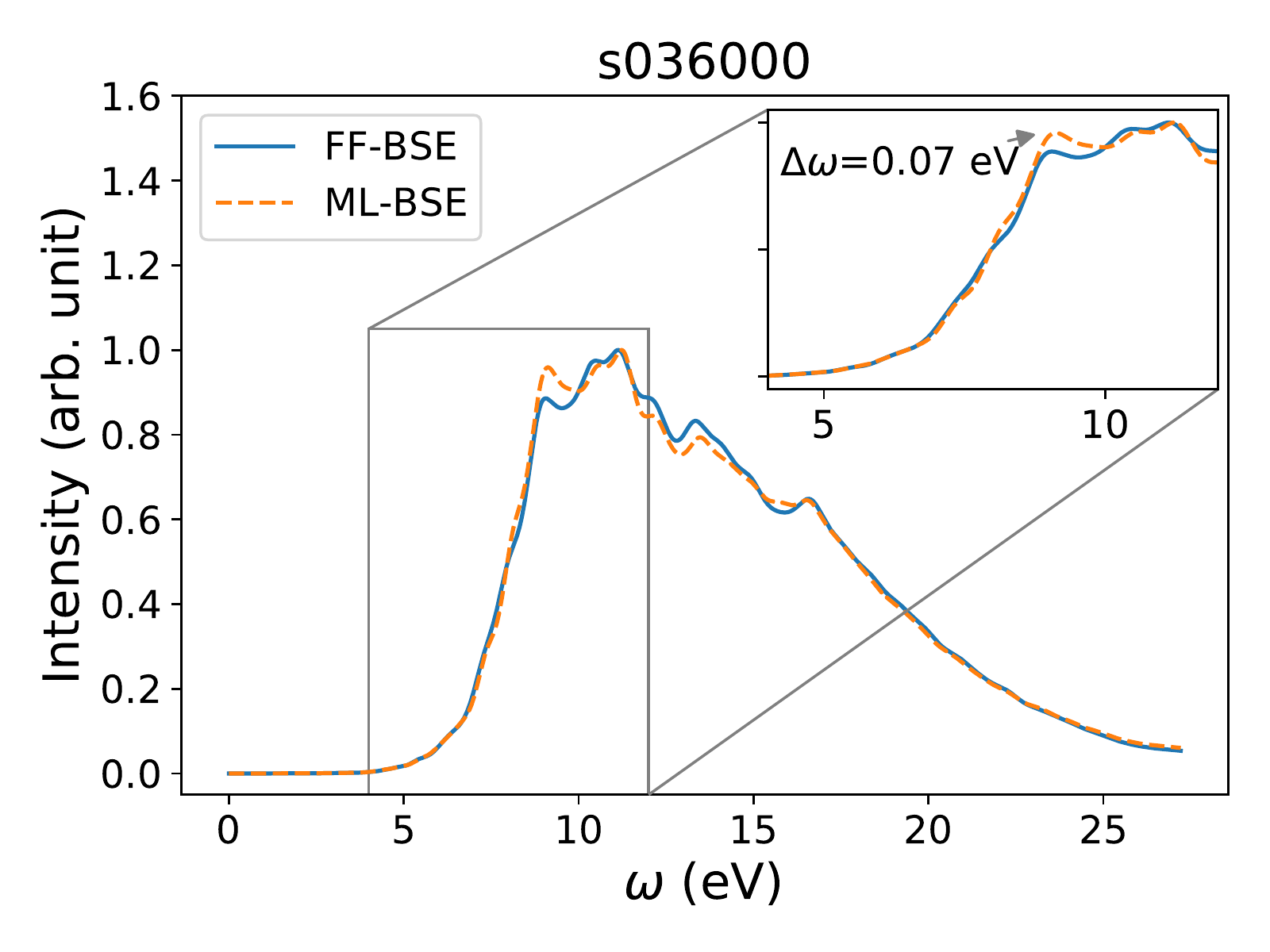}
\end{subfigure}
\hfill
\begin{subfigure}{0.45\textwidth}
\includegraphics[width=\linewidth]{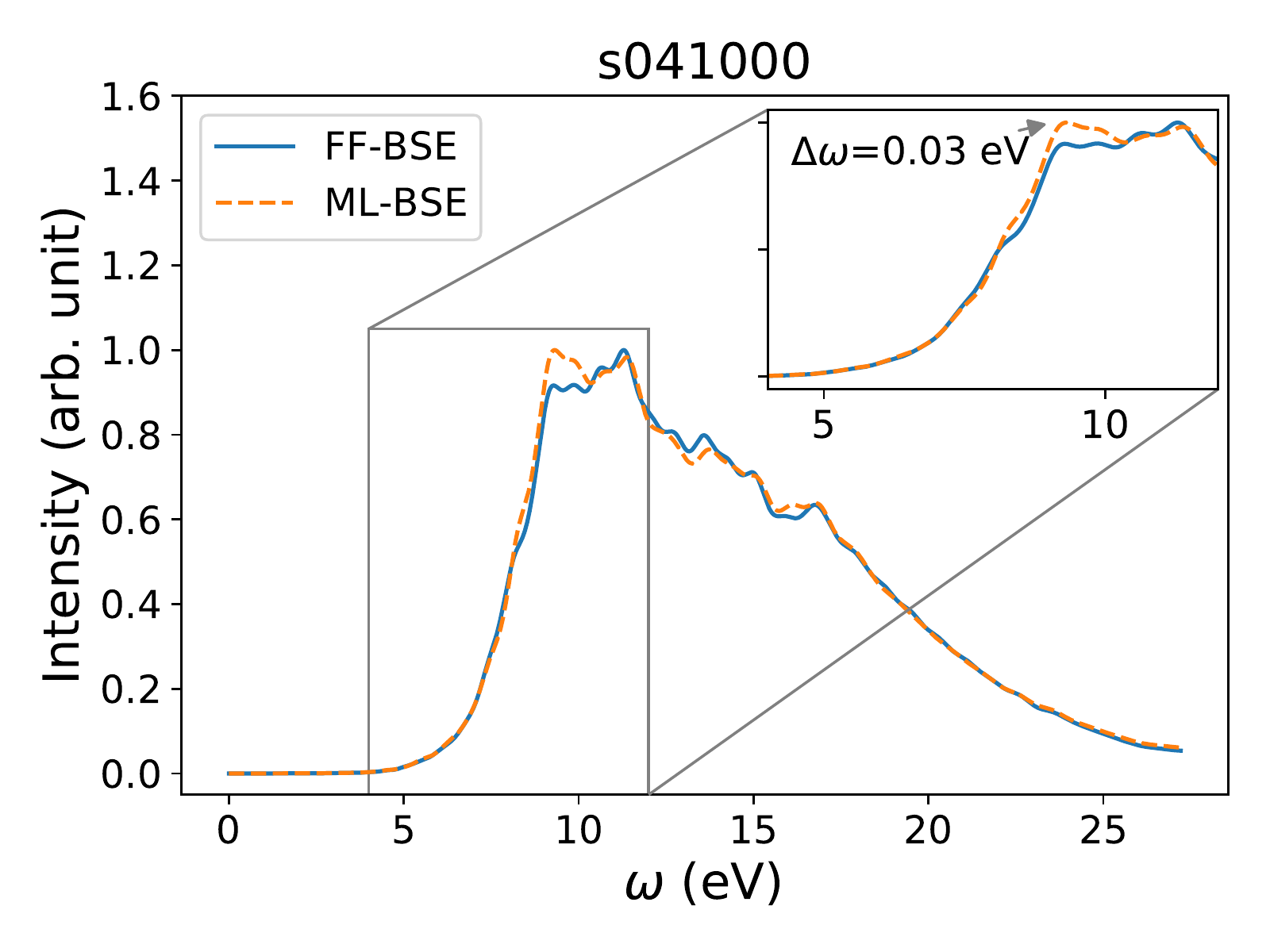}
\end{subfigure}
\hfill
\begin{subfigure}{0.45\textwidth}
\includegraphics[width=\linewidth]{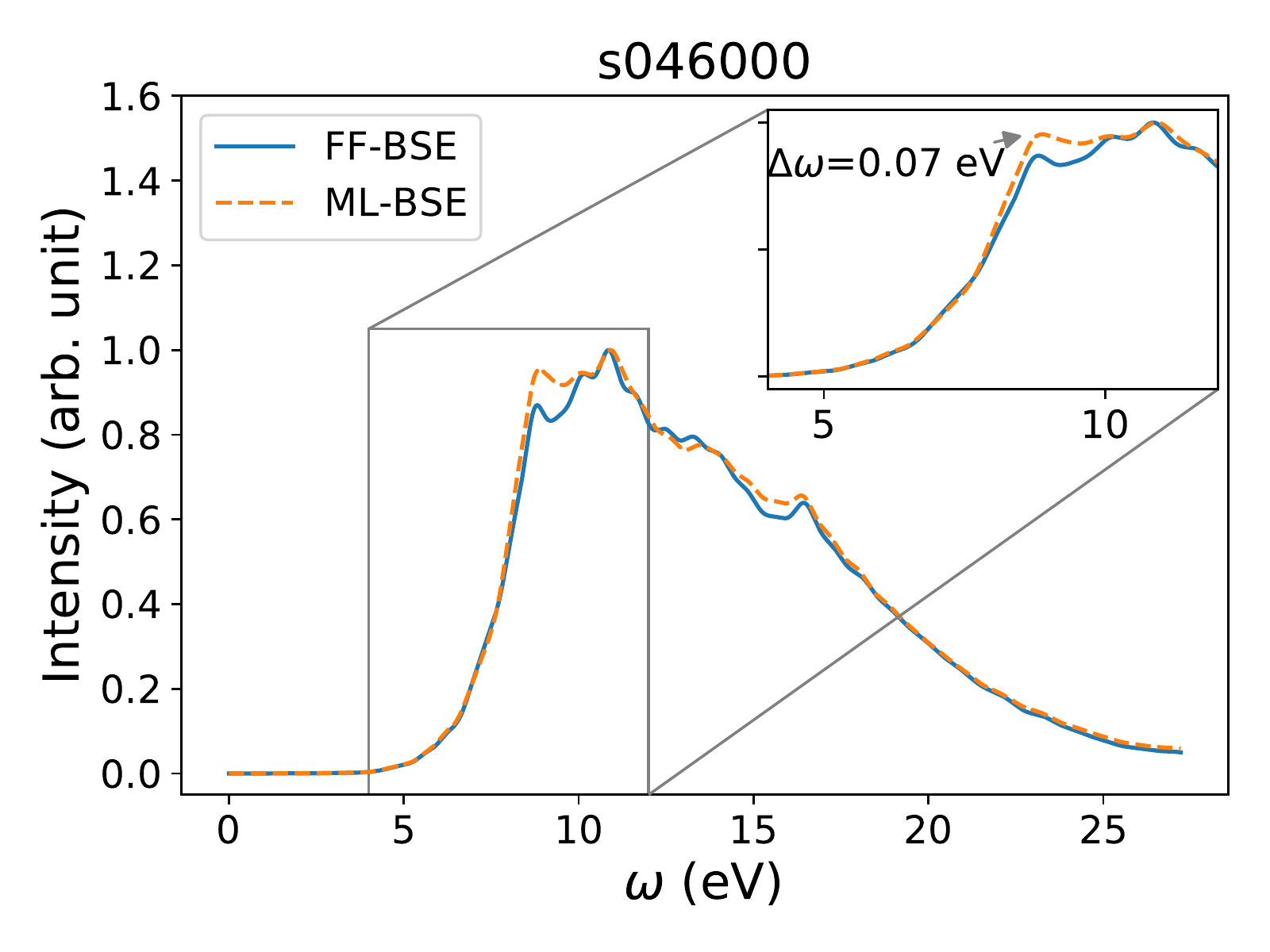}
\end{subfigure}

\caption{ML-BSE and FF-BSE spectra of 10 snapshots of Si$_{35}$H$_{36}$ snapshots extracted from FPMD trajectories at 500~K. The model used in ML-BSE obtained using a convolutional model $(7,7,7)$ trained with data obtained for the 0~K snapshot, the same model used in Figure~\ref{fig:si35h36_25ang_0K_ml}. The 10 snapshots were extracted every 5000 time steps from the FPMD trajectory, starting from the 21000th time step (10.16 ps onwards). We label these snapshots from s021000 to s066000.}
\label{fig:si35h36_10}

\end{figure}

\begin{figure}[H]
\ContinuedFloat
\centering

\begin{subfigure}{0.45\textwidth}
\includegraphics[width=\linewidth]{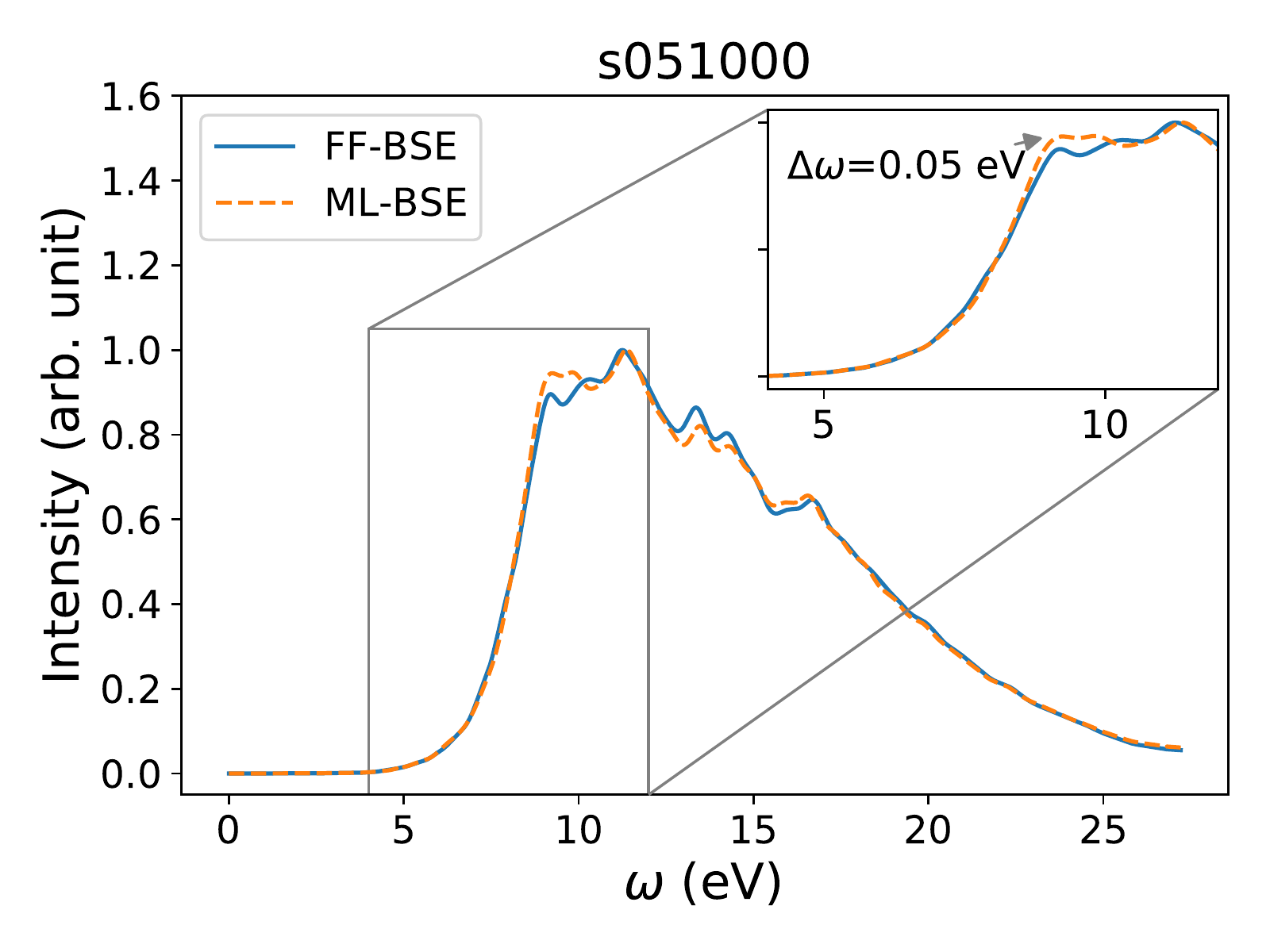}
\end{subfigure}
\hfill
\begin{subfigure}{0.45\textwidth}
\includegraphics[width=\linewidth]{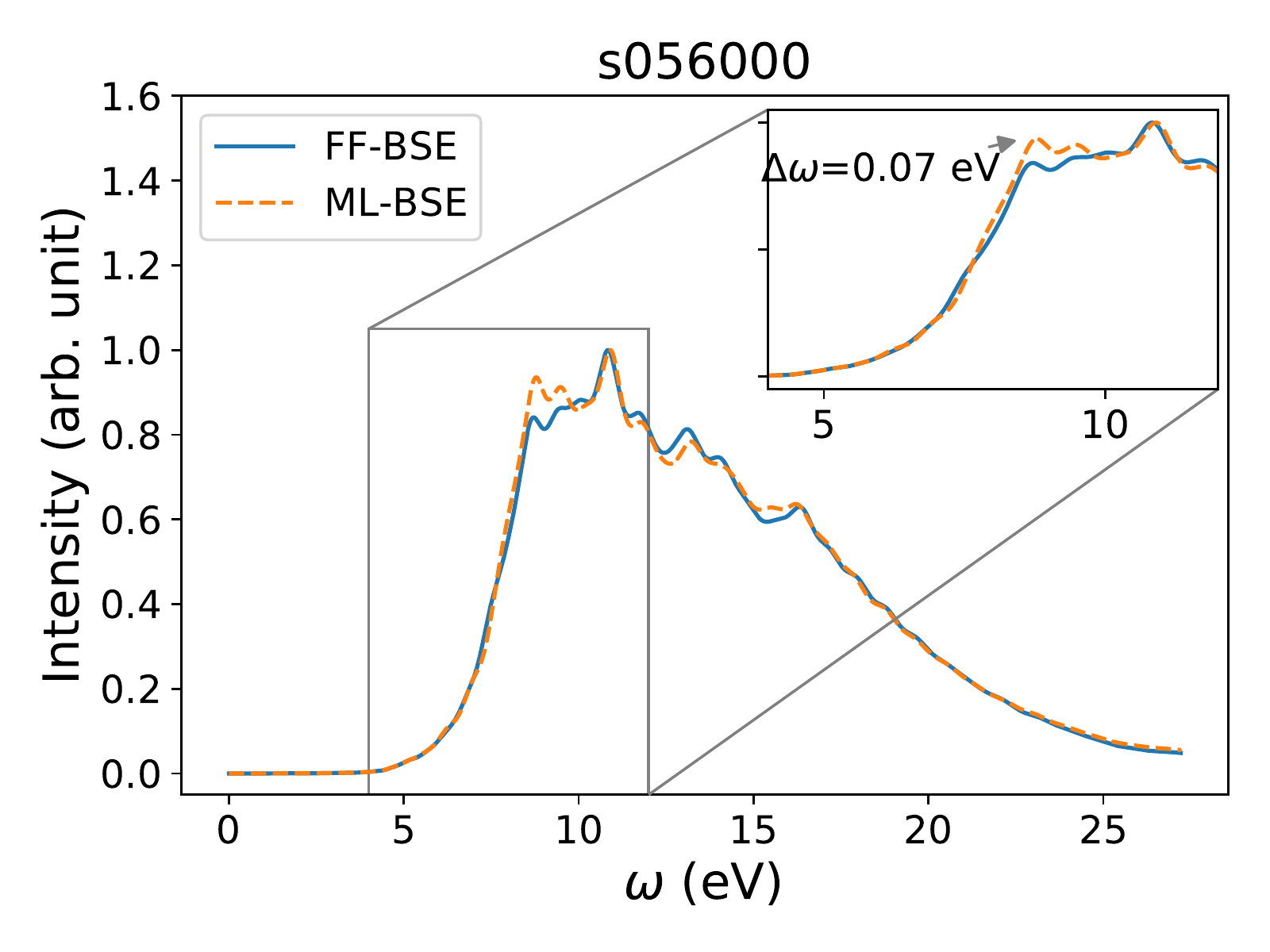}
\end{subfigure}
\hfill
\begin{subfigure}{0.45\textwidth}
\includegraphics[width=\linewidth]{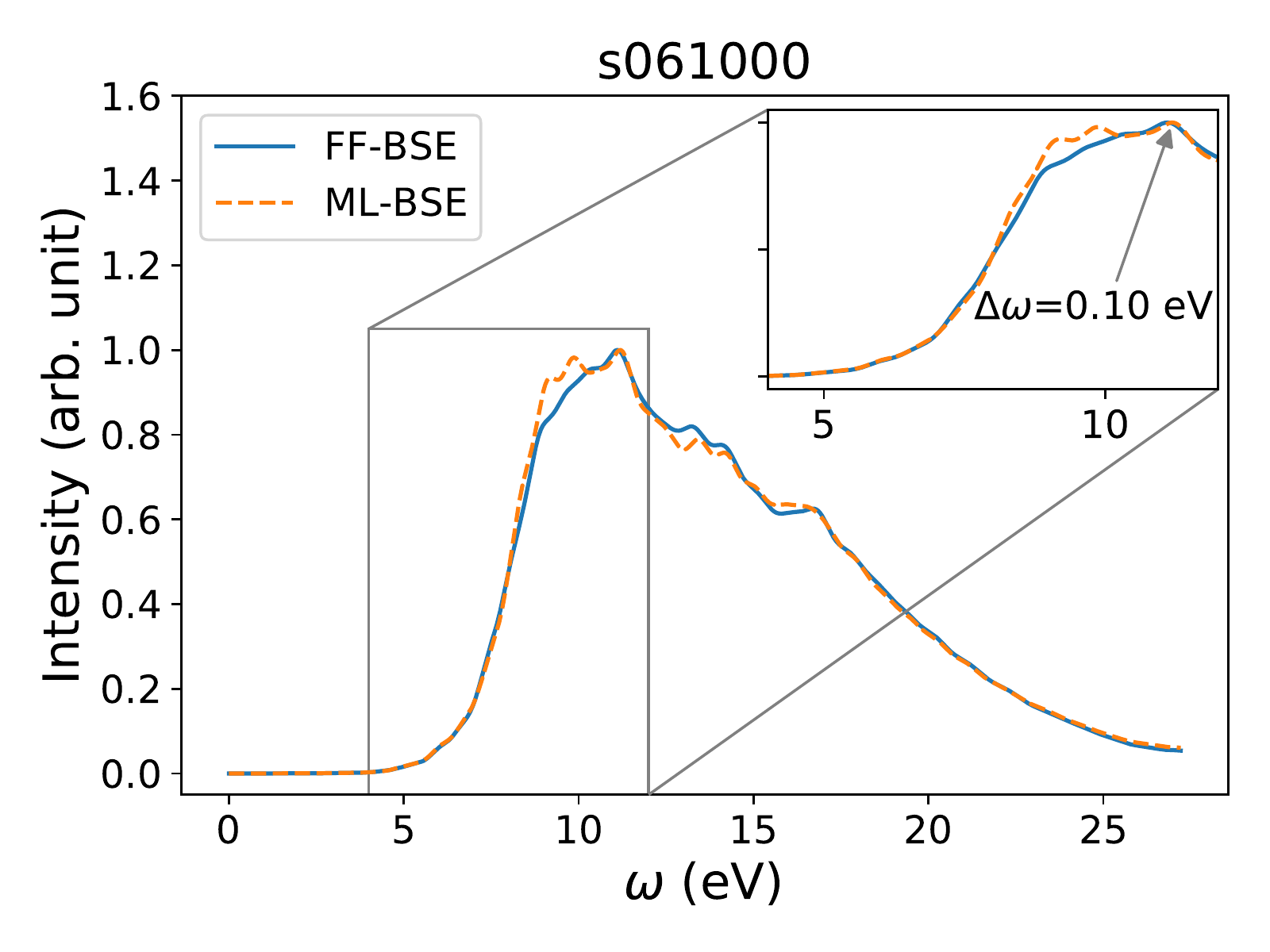}
\end{subfigure}
\hfill
\begin{subfigure}{0.45\textwidth}
\includegraphics[width=\linewidth]{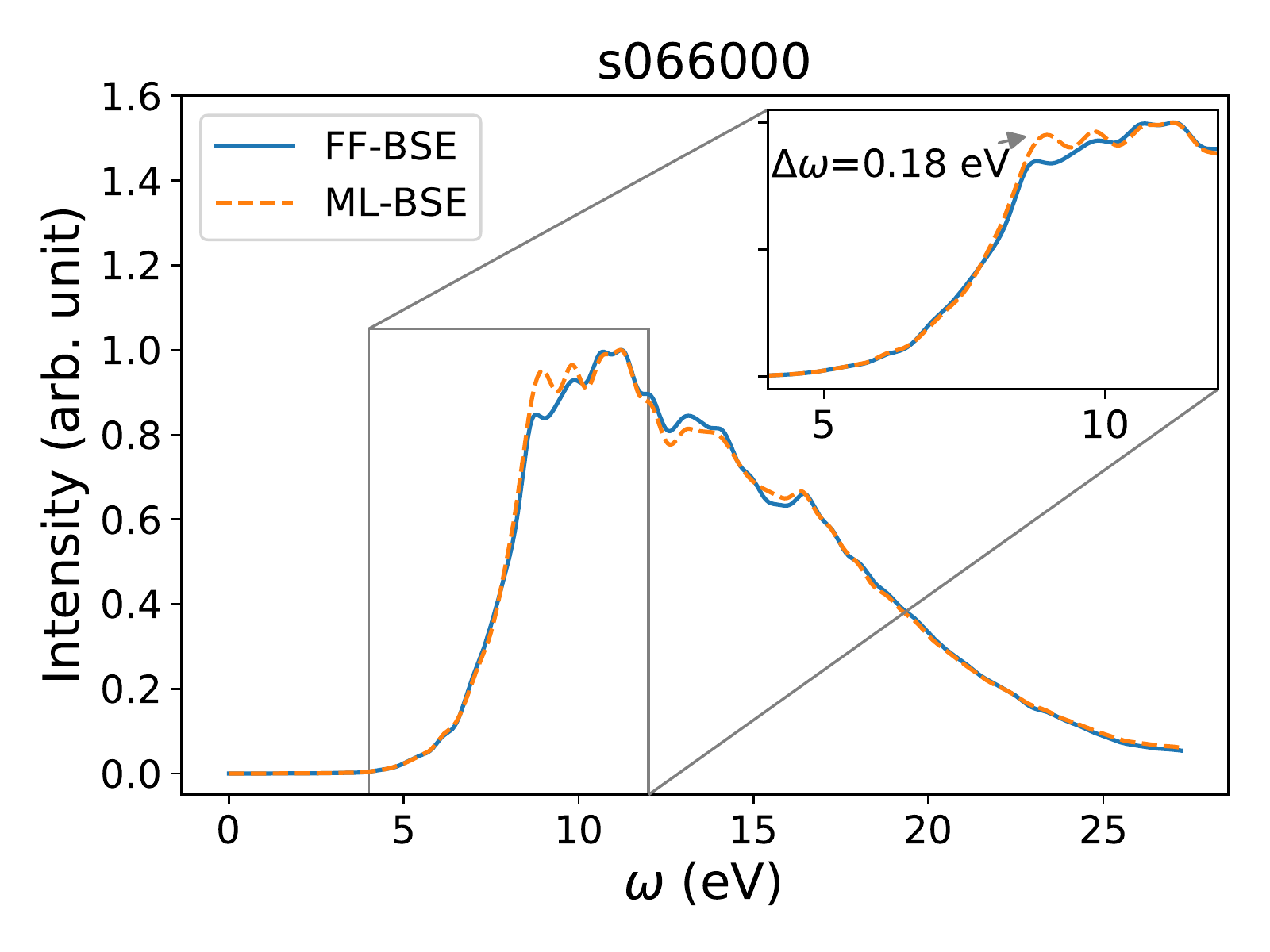}
\end{subfigure}

\caption{(Continued from the previous page) ML-BSE and FF-BSE spectra of 10 snapshots of Si$_{35}$H$_{36}$ snapshots extracted from FPMD trajectories at 500~K. The model used in ML-BSE obtained using a convolutional model $(7,7,7)$ trained with data obtained for the 0~K snapshot, the same model used in Figure~\ref{fig:si35h36_25ang_0K_ml}. The 10 snapshots were extracted every 5000 time steps from the FPMD trajectory, starting from the 21000th time step (10.16 ps onwards). We label these snapshots from s021000 to s066000.}
\label{fig:si35h36_10}

\end{figure}

\subsubsection*{Transferability of ML models for Si$_{35}$H$_{36}$}

\begin{figure}[H]
\centering

    \includegraphics[width=0.45\linewidth]{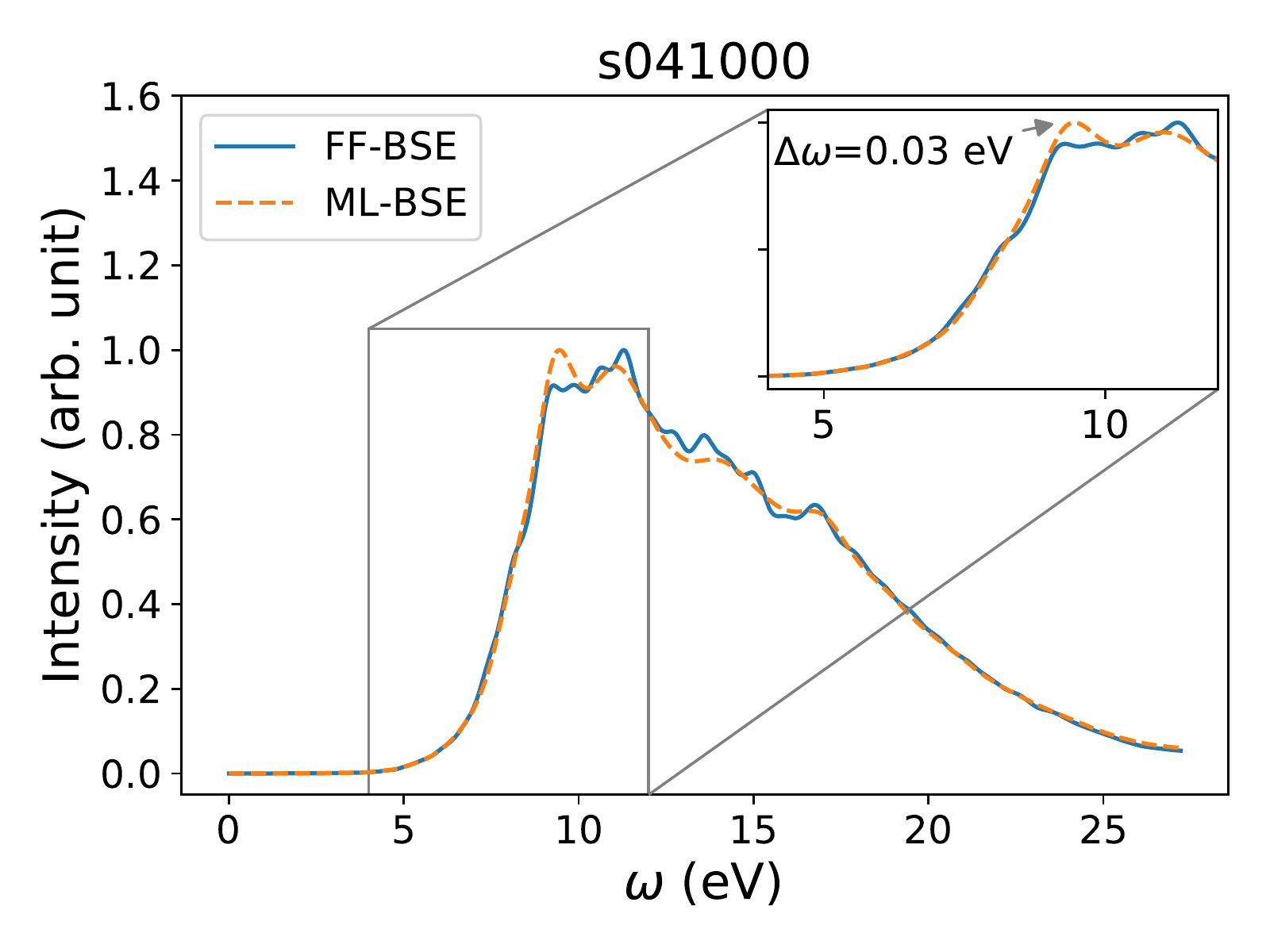}
\caption{Accuracy of the ML-BSE spectrum of Si$_{35}$H$_{36}$ (a 500 K snapshot s041000) obtained by using a convolutional model $(7,7,7)$ derived from the data for snapshot 041000. The RMSE value between the ML-BSE and FF-BSE spectra in this figure is 0.019, to be compared with RMSE=0.020 for the same snapshot's spectrum in Figure~\ref{fig:si35h36_10}.}
\label{fig:si35h36_0K_500K}
\end{figure}

As shown in Figure~\ref{fig:si35h36_10} (snapshot s041000) and Figure~\ref{fig:si35h36_0K_500K}, the accuracy of the convolutional model derived from data for the 0~K and 500~K snapshots is similar in predicting the absorption spectrum of the 500 K snapshot. This suggests that the convolutional model is indeed transferable from the 0 K geometry to 500 K geometries for this Si cluster. 

\subsubsection*{Si$_{87}$H$_{76}$}
\begin{figure}[H]
\centering
    \begin{subfigure}{0.45\textwidth}
    \caption{}
    \includegraphics[width=\linewidth]{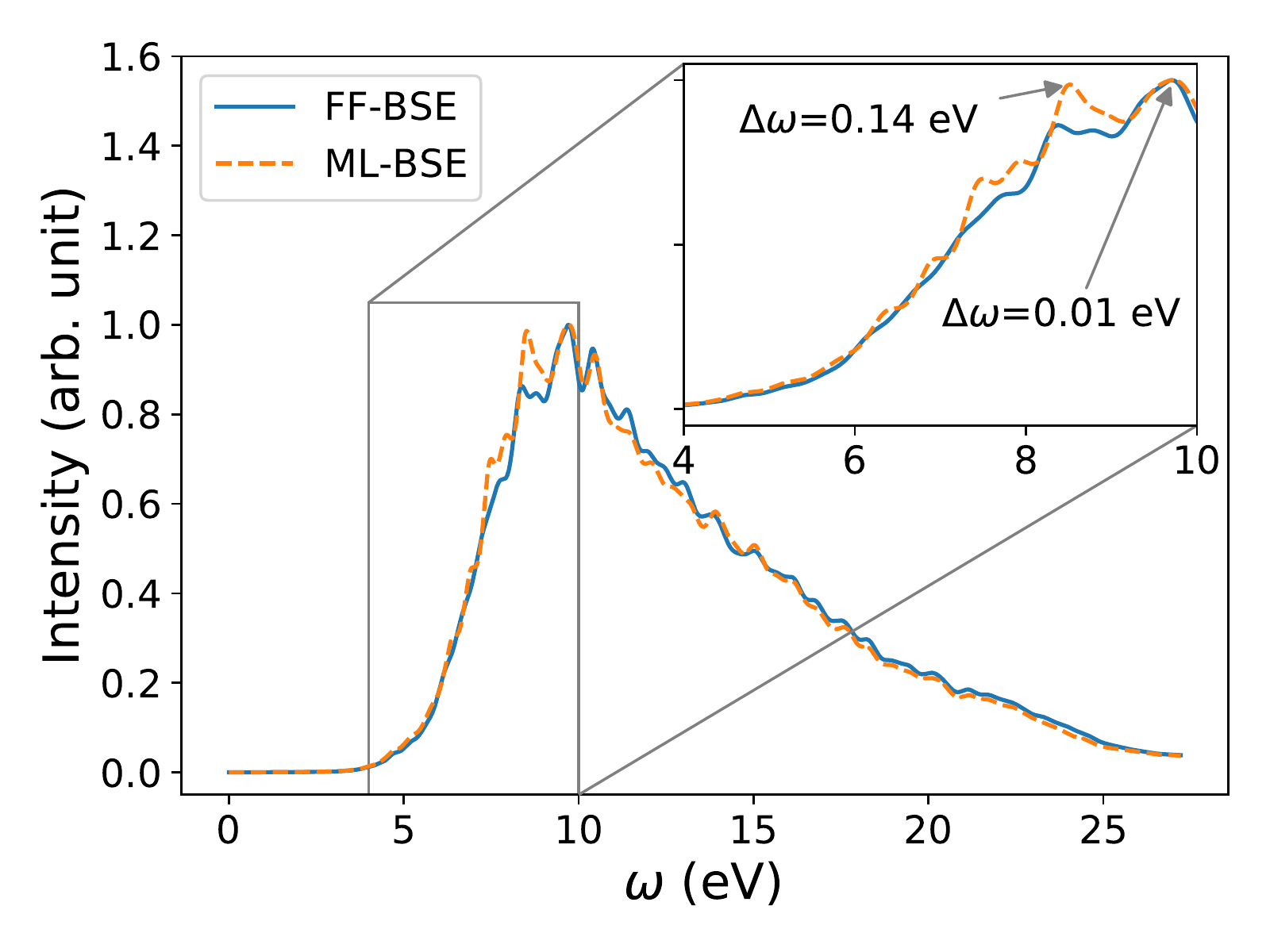}

    \end{subfigure}
    \hfill
    \begin{subfigure}{0.45\textwidth}
    \caption{}
    \includegraphics[width=\linewidth]{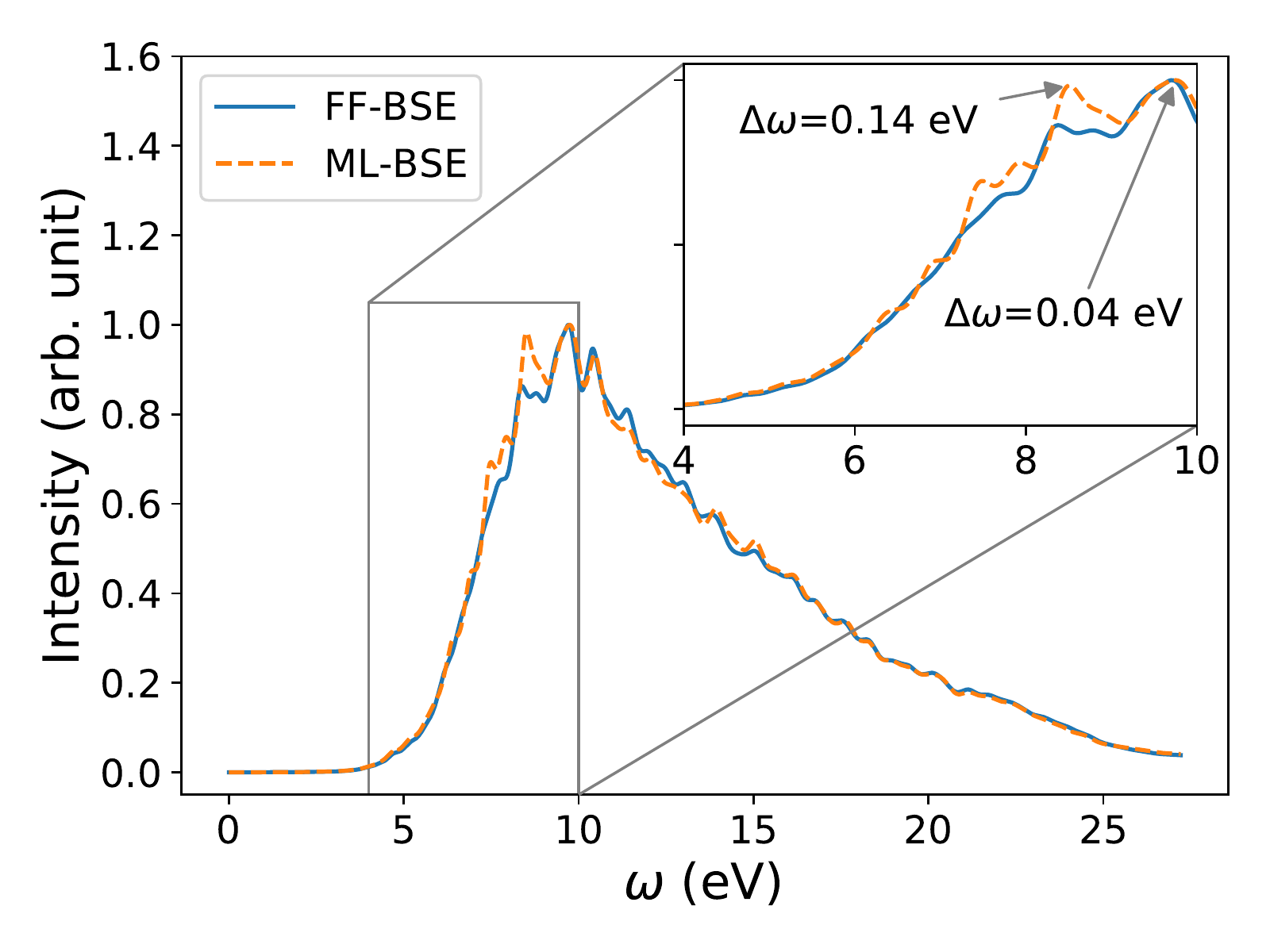}

    \end{subfigure}
    \begin{subfigure}{0.45\textwidth}
    \caption{}
    \includegraphics[width=\linewidth]{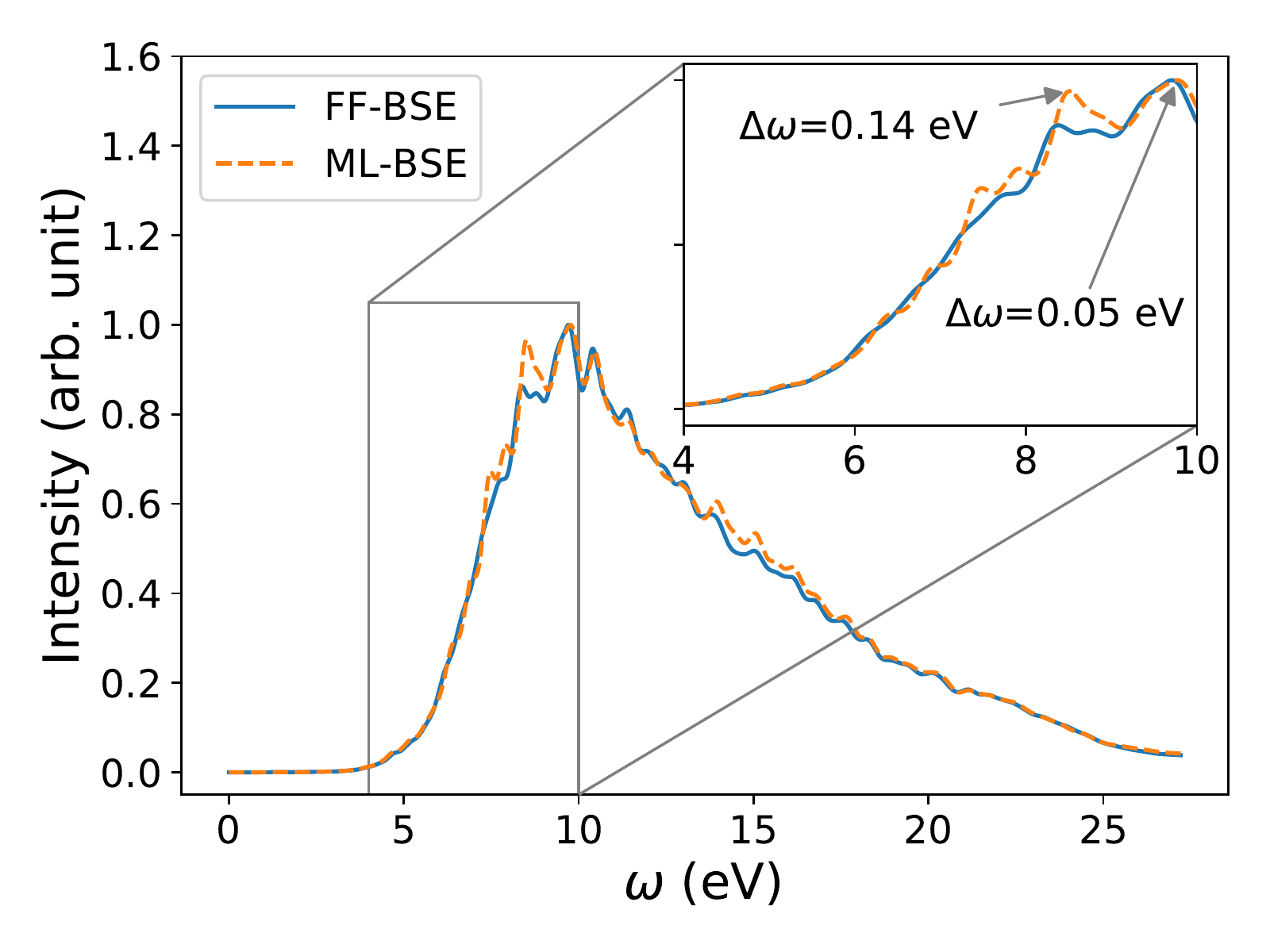}

    \end{subfigure}
\caption{Accuracy of the ML-BSE spectrum of Si$_{87}$H$_{76}$ obtained by using (a) a global scaling factor, (b) a convolutional model $(5,5,5)$, and (c) a convolutional model $(7,7,7)$. The RMSE of the spectra are 0.025 for (a), 0.024 for (b), and 0.020 for (c).}
\label{fig:si87h76_ml}
\end{figure}

The ML-BSE spectrum of Si$_{87}$H$_{76}$ obtained by applying various models obtained from Si$_{87}$H$_{76}$ is similar to the one obtained by applying the convolutional model obtained from Si$_{35}$H$_{36}$, as shown in Figure~\ref{fig:si87h76_ml} and Figure~\ref{fig:si87h76_si35model}. 

\subsubsection*{Computational time savings}
\begin{figure}[H]
\centering
    
    \begin{subfigure}{0.48\textwidth}
    \includegraphics[width=\linewidth]{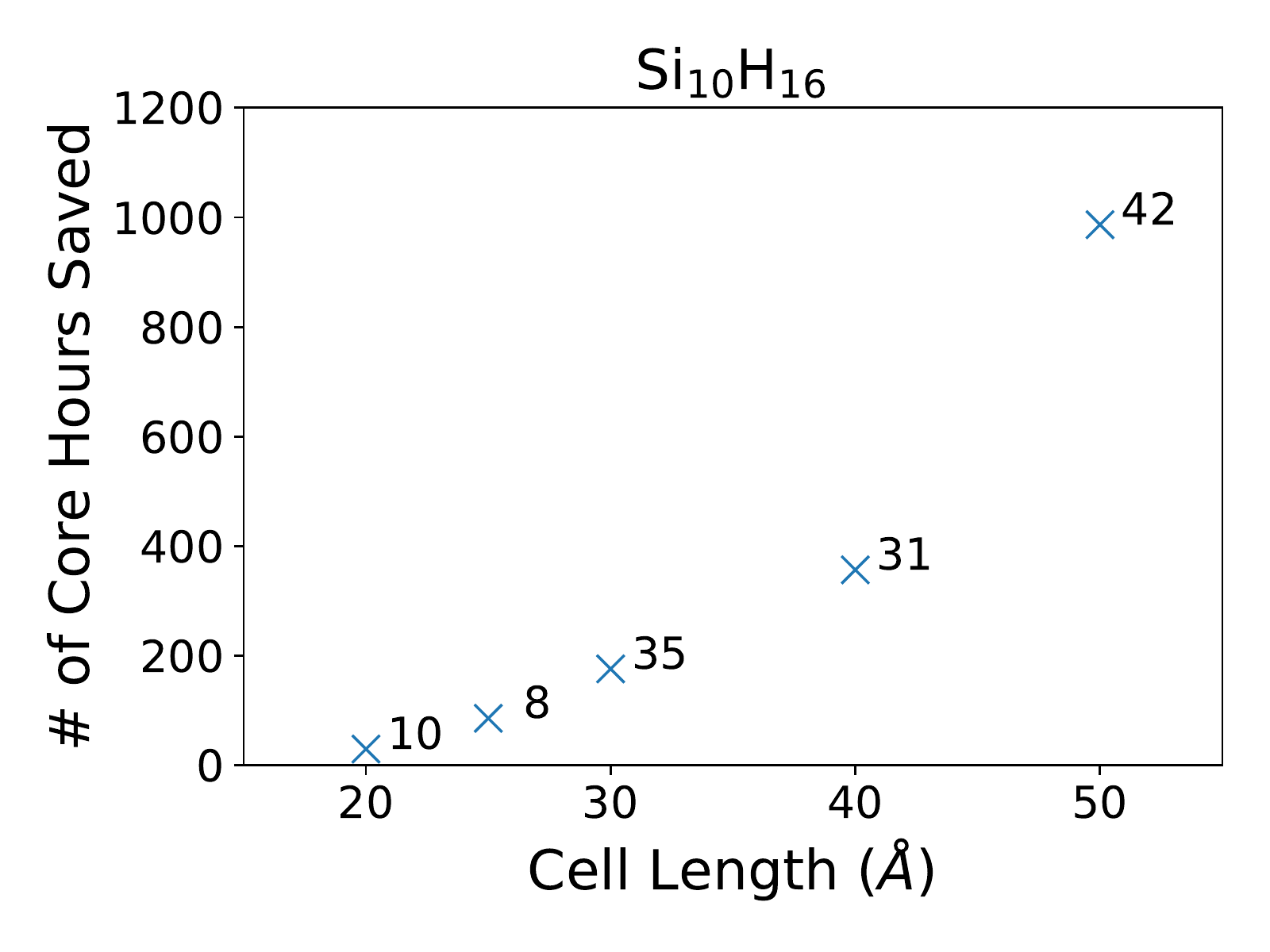}
    \end{subfigure}
    \hfill
    \begin{subfigure}{0.48\textwidth}
    \includegraphics[width=\linewidth]{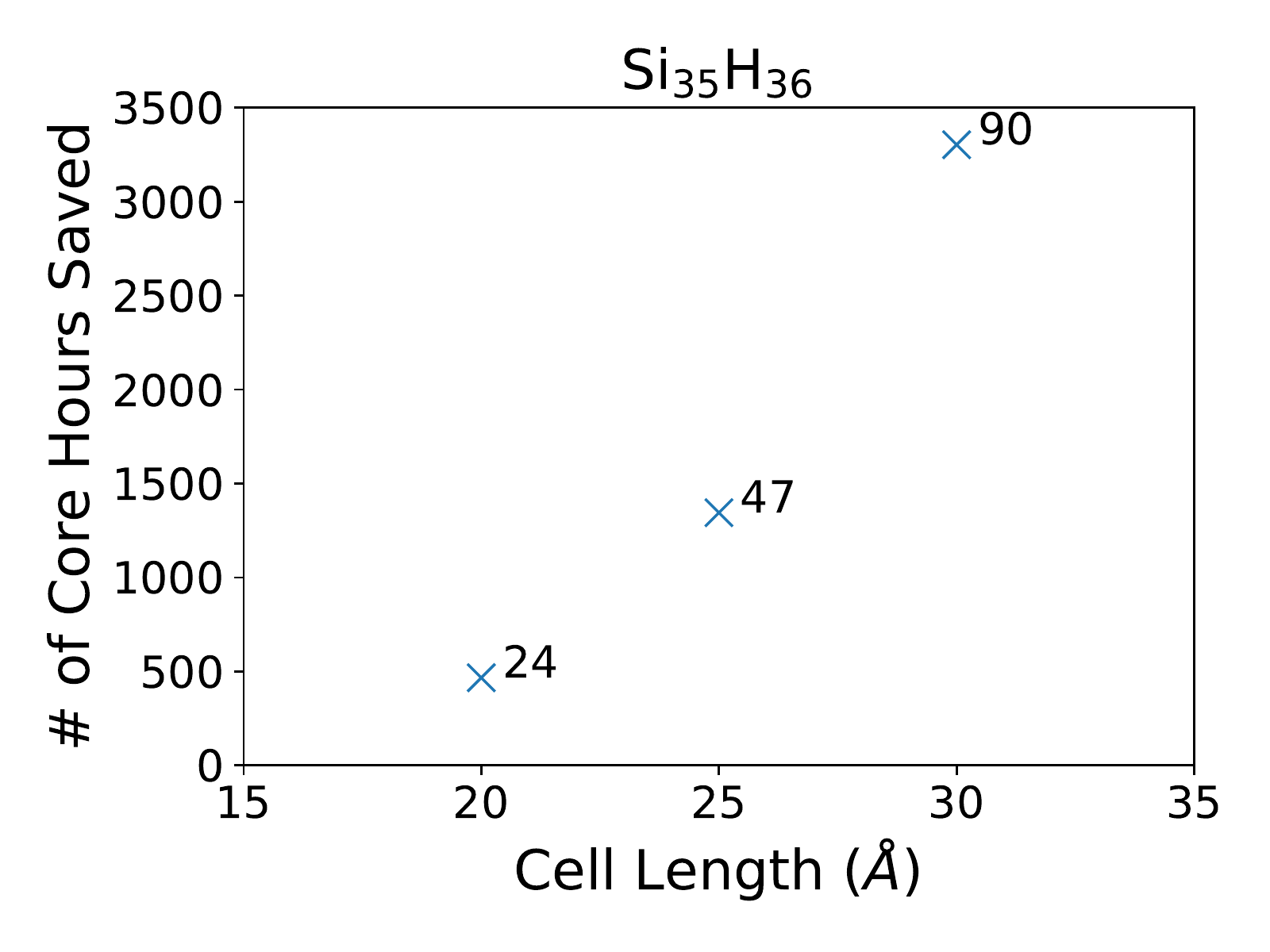}
    \end{subfigure}
\caption{Number of core hours saved for the Si$_{10}$H$_{16}$ and Si$_{35}$H$_{36}$ clusters as a function of the size of the simulation cell when using ML-BSE instead of FF-BSE. The numerical values beside each data point is the corresponding $\alpha_d$. Because of the relatively small computational cost of ML-BSE for Si$_{10}$H$_{16}$, its $\alpha_d$ is more sensitive to the machine status than that of Si$_{35}$H$_{36}$ and is not necessarily monotonic as a function of the size of the simulation cell.}
\label{fig:si10_si35_timing}
\end{figure}

As demonstrated in Figure~\ref{fig:si10_si35_timing}, the greater the simulation cell is (the larger the number of plane waves $n_{\text{pw}}$ is), the more substantial are the savings in the computational time.

\section{Timing and scaling of ML-BSE}

The calculation of absorption spectra can be decomposed into the following three steps:
\begin{enumerate}
    \item Calculation of unscreened integrals using Eq.~\ref{eq:tauu};
    \item Calculation of screened integrals using Eq.~\ref{eq:delta_tau_delta_rho} (in FF-BSE) or Eq.\ref{eq:CNN} (in ML-BSE);
    \item Calculation of the frequency-dependent spectrum using Eq.~\ref{eq:spectrum}.
\end{enumerate}

Here we have replaced Step 2 with a machine learning model. For the largest system considered in this work (COOH-Si/water interface), at least $\sim$50\% of the total computational time is spent in Step 2. We define the speed-up $\alpha_d=t_d^{\text{FF-BSE}}/t_d^{\text{ML-BSE}}$ as the ratio of the core hours needed to carry out Step 2 with FF-BSE or ML-BSE. We found that $\alpha_d$ increases as $n_{\text{int}}$ and the number of plane waves $n_{\text{pw}}$ increase. 

\bibliography{mlbse} 
\bibliographystyle{rsc} 
\makeatletter\@input{mainaux.tex}\makeatother